\documentclass{emulateapj}

\usepackage{lscape}
\usepackage{float}

\def\simless{\mathbin{\lower 3pt\hbox
        {$\,\rlap{\raise 5pt\hbox{$\char'074$}}\mathchar"7218\,$}}} 
\def\simgreat{\mathbin{\lower 3pt\hbox
        {$\,\rlap{\raise 5pt\hbox{$\char'076$}}\mathchar"7218\,$}}} 

\slugcomment{Accepted in \aj}

\shorttitle{The C4 SDSS Cluster Catalog}
\shortauthors{Miller et al.}

\begin{document}
 
\title{The C4 Clustering Algorithm: Clusters of Galaxies in the Sloan Digital Sky Survey}

\author{
Christopher J. Miller,\altaffilmark{1,2}
Robert C. Nichol,\altaffilmark{3}
Daniel Reichart,\altaffilmark{4}
Risa H. Wechsler,\altaffilmark{5,6} 
August E. Evrard,\altaffilmark{7,8}
James Annis,\altaffilmark{9}
Timothy A. McKay,\altaffilmark{7}
Neta A. Bahcall,\altaffilmark{10}
Mariangela Bernardi,\altaffilmark{12}
Hans Boehringer,\altaffilmark{11}
Andrew J. Connolly,\altaffilmark{12}
Tomotsugu Goto,\altaffilmark{13}
Alexie Kniazev,\altaffilmark{17,18,19}
Donald Lamb,\altaffilmark{16}
Marc Postman,\altaffilmark{14}
Donald P. Schneider,\altaffilmark{15}
Ravi K. Sheth,\altaffilmark{12}
Wolfgang Voges\altaffilmark{11}
}
\altaffiltext{1}{Cerro-Tololo Inter-American Observatory, NOAO, Casilla 603, La Serena, Chile}
\altaffiltext{2}{
email: cmiller@noao.edu }
\altaffiltext{3}{Institute of Cosmology and Gravitation,
University of Portsmouth,
Portsmouth, PO1 2EG, UK}
\altaffiltext{4}{
Department of Physics and Astronomy, University of North Carolina,
Chapel Hill, NC 27599}
\altaffiltext{5}{
Center for Cosmological Physics, Dept. of Astronomy
\& Astrophysics, \& Enrico Fermi Institute,
University of Chicago,
Chicago, IL 60637}
\altaffiltext{6}{
Hubble Fellow}
\altaffiltext{7}{
Department of Physics, University of Michigan,
Ann Arbor, MI 48109}
\altaffiltext{8}{
Department of Astronomy, University of Michigan,
Ann Arbor, MI 48109}
\altaffiltext{9}{
Fermi National Accelerator Laboratory, Batavia, IL 60510}
\altaffiltext{10}{
Princeton University Observatory, Princeton, NJ 08544}
\altaffiltext{11}{
Max-Planck-Institut f\"{u}r Extraterrestrische Physik,
Garching, Germany}
\altaffiltext{12}{
Department of Physics and Astronomy, University of Pittsburgh, PA 15260}
\altaffiltext{13}{
Institute for Cosmic Ray Research, University of
Tokyo, Kashiwanoha, Kashiwa, Chiba 277-0882, Japan}
\altaffiltext{14}{
Space Telescope Science Institute, Baltimore, MD 21218}
\altaffiltext{15}{Pennsylvania State University, 
                  University Park, PA 16802}
\altaffiltext{16}{University of Chicago, 
                  Chicago, IL 60637}
\altaffiltext{17}{Special Astrophysical Observatory, Nizhnij Arkhyz, Karachai-Circassia 369167, Russia}
\altaffiltext{18}{MPA, Königstuhl 17, 69117 Heidelberg, Germany}
\altaffiltext{19}{Isaac Newton Institute of Chile, SAO Branch}

\begin{abstract} 
We present the ``C4 Cluster Catalog", a new sample of 748 clusters of
galaxies identified in the spectroscopic sample of the
Second Data Release (DR2) of the Sloan Digital
Sky Survey (SDSS). The C4 cluster--finding algorithm 
identifies clusters as overdensities in a
seven-dimensional position and color space, thus minimizing
projection effects that have plagued previous optical cluster selection. The
present C4 catalog covers $\sim 2600$ square degrees of sky and ranges in
redshift from $z = 0.02$ to $z=0.17$.  The mean cluster membership is 36
galaxies (with redshifts) brighter than $r =17.7$, but the catalog
includes a range of systems, from groups containing $10$ members to massive
clusters with over $200$ cluster members with redshifts.  The catalog
provides a large number of measured cluster properties including sky
location, mean redshift, galaxy membership, summed r--band
optical luminosity ($L_r$), velocity dispersion, as well as quantitative
measures of substructure and the surrounding large-scale
environment. We use new, multi-color mock SDSS galaxy catalogs,
empirically constructed from the $\Lambda$CDM Hubble Volume (HV) Sky
Survey output, to investigate the sensitivity of the C4 catalog to the
various algorithm parameters (detection threshold, choice of passbands
and search aperture), as well as to quantify the purity and
completeness of the C4 cluster catalog.  These mock catalogs indicate
that the C4 catalog is $\simeq 90\%$ complete and 95\% pure above
$M_{200} = 1 \times 10^{14} \, h^{-1}\,{\rm M}_\odot$ and within $0.03 \le z \le 0.12$. Using the SDSS
DR2 data, we show that the C4 algorithm finds 98\% of X-ray identified clusters and 90\%
of Abell clusters within $0.03 \le z \le 0.12$.
Using the mock galaxy catalogs and the full HV dark matter simulations,
we show that the $L_r$ of a cluster is a more
robust estimator of the halo mass ($M_{200}$) than the galaxy
line-of-sight velocity dispersion or the richness of the cluster. However,
if we exclude clusters embedded in complex large-scale environments,
we find that the velocity
dispersion of the remaining clusters is as good an estimator of $M_{200}$
as $L_r$. The final C4 catalog will contain $\simeq2500$ clusters using
the full SDSS data set and will represent one of the largest and most
homogeneous samples of local clusters.
\end{abstract}

\section{Introduction}
\label{intro}

Catalogs of clusters and groups of galaxies are used extensively
throughout extragalactic astronomy and cosmology, from constraining
the cosmological parameters (e.g., Oukbir \& Blanchard 1992; Henry \&
Arnaud 1991; Viana \& Liddle 1996; Bahcall, Fan \& Cen 1997; Reichart
et al. 1999; Miller et al. 2001a,b), to magnifying the most distant
galaxies in the universe (Sand et al. 2002). Considerable effort has
been invested over the last half-century in constructing catalogs of
clusters and groups of galaxies ({\it e.g.}, Zwicky 1952; Abell 1958;
Abell et al. 1989; Gioia et al. 1990; Lumsden et al. 1992; Dalton et
al. 1992; Henry et al. 1995; Postman et al. 1996; Romer et al. 2000;
Boehringer et al. 2000; Gladders 2000; Postman et al. 2002).  In this
paper, we present one of the first catalogs of clusters and groups
constructed directly from the spectroscopic data of the Sloan Digital
Sky Survey (SDSS). This is now possible because of the present size of
the SDSS dataset (see Section \ref{data}), and is complementary to the
SDSS cluster catalogs selected using the SDSS photometric data ({\it
e.g.}, Annis et al. 2000; Goto et al. 2002; Kim et al. 2002; Bahcall
et al. 2003; Lee et al. 2003).

The distribution of matter in the Universe is described by the
statistics of overdensities. When these overdensities are small, the
equations of motion that follow the evolution of matter can be
linearized and solved.  As gravitational clustering is amplified into
the non-linear regime, a description of the matter distribution as a
point set of extended dark matter halos becomes more appropriate
(e.g., Cooray \& Sheth 2002).  How halos are populated with galaxies
of specific colors and luminosities (typically referred to as the halo
occupation) is not known precisely, and remains a serious challenge in
cosmology and astrophysics.  The details of how galaxies occupy halos
will clearly have an affect on attempts to identify clusters in
optical catalogs.  But progress can be made on both fronts: gaining
understanding about how galaxies populate clusters will be invaluable
to those who study structure and galaxy formation/evolution.
Vice-versa, mock catalogs that are representative of the real Universe
would be invaluable for accurately measuring the selection function of
any clustering algorithm, which specifies the contamination and
completeness of a dataset and is a prerequisite to many scientific
analyses.

The challenge in constructing a cluster catalog from galaxy data is to
minimize projection effects (or false--positive detections), while
maximizing completeness, i.e. controlling the selection function.
Previous analyses of large optical catalogs of clusters (Lucey et
al. 1983; Sutherland 1988; Frenk et al. 1990) have claimed various
levels (10-25\%) of contamination (see also Miller et al. 1999 and 2002).
The next generation of cluster catalogs must have very little
contamination and precisely known selection functions in order to
compete with the increasingly precise cosmological constraints from
other methods (e.g. Perlmutter and Schmidt 2003; Mandolesi, Villa, \&
Valenziano 2002).  Because of concern about the presence of projection
effects in optical cluster catalogs, there has been renewed emphasis
over the past decade on new ways of finding clusters of galaxies in
wavebands other than the optical. For example, many authors have
constructed catalogs of clusters from X--ray surveys of the sky (Edge
\& Stewart 1991; Gioia et al. 1990; Boehringer et al. 2000; Romer et
al. 2000 among others) as this is believed to be more robust for
selecting mass-limited samples than optical methods (see Ebeling et
al. 1997). An X-ray selected SDSS cluster sample has also been
presented by Popesso, Boehringer \& Voges (2004).  Additionally, many
authors have proposed the construction of catalogs of clusters using
the Sunyaev--Zel'dovich Effect (Carlstrom, Holder \& Reese 2002, Romer
et al. 2004) and weak gravitational lensing (Wittman et
al. 2002). Furthermore, Kochanek et al. (2003) recently presented a
new catalog of clusters derived from the 2MASS infra--red photometric
data.

As we outline below, we have now mitigated the problem of projection
effects in optical cluster catalogs by simultaneously using both SDSS
photometric and spectroscopic data to find clusters. The details of
our cluster finding algorithm, which we refer to as the ``C4''
algorithm, are presented in Section \ref{overview}.  The premise of this
algorithm is that optical clusters and groups of galaxies are
dominated, at their cores, by a single, co--evolving population of
galaxies which possess similar spectral energy distributions, {\it
e.g.}, the ``E/S0 ridge-line'' or ``red envelope'' (Baum 1959; McClure
and van den Bergh 1968; Lasker 1970; Visvanathan and Sandage
1977). The evidence for such a co--evolving population of galaxies in
the cores of clusters has been presented by many authors; see Gladders
(2002) for a detailed review of this evidence. Blakeslee et al. (2003)
provide evidence that this co--evolving population extends beyond
redshift one. As such, Ostrander et al. (1998), Gladders (2000) and
Goto et al. (2002) have all used the existence of a co--evolving
population of galaxies in the cores of clusters as a basis for their
cluster--finding algorithms.

\begin{figure*}[hb]
\caption{Due to size limitations, this figure only appears in the accepted version
or online at www.ctio.noao.edu/~chrism/c4. \\
  Color--magnitude relation of galaxies in all four SDSS colors
  for a previously unknown cluster of galaxies identified in the SDSS DR2
  dataset.  Black dots are galaxies within a projected aperture of $1h^{-1}$Mpc
  around the cluster center. Red and green
  dots are cluster members;
  the red dots have low $H\alpha$ emission and the green dots have
  high $H\alpha$ emission, indicative of ongoing star formation.
  Error bars on the colors of the red galaxies indicate the
  typical color errors in the spectroscopic sample.}
\label{fig:fourplot}
\end{figure*}

In Figure \ref{fig:fourplot}, we present the color--magnitude diagram,
in all four of the SDSS colors ($u-g$, $g-r$, $r-i$, $i-z$), for a
newly discovered cluster of galaxies at $z = 0.06$ in the SDSS Early
Data Release (Stoughton et al. 2002). The nearest cluster in the
literature is a Zwicky group which is $\sim 9$ arcminutes to the
southeast. All galaxies in this figure are within a projected $1h^{-1}$Mpc radius of the
cluster center (we use $h$ = H$_0$/100 km/s/Mpc).  The figure
demonstrates the existence of a tight relationship between the colors
of cluster galaxies, which can have an observed scatter of $\sim 0.05$
(see Bower et al. 1992).  The figure also illustrates that this ``red
sequence'' of galaxies is present in all four SDSS colors and not just
in the one color system used by Gladders \& Yee (2000). Note also the
existence of a ``blue ridge-line'' in Figure \ref{fig:fourplot}
(top-left), which has been highlighted previously by several authors
(see Chester \& Roberts 1964; Tully, Mould, \& Aaronson 1982; Baldry
et al. 2004).  In this one example (at z = 0.06), the 4000 Angstrom
break sits between the $u$ and $g$ filters. Thus, the $u - g$
color-magnitude diagram can separate the blue star-forming galaxies
from the older red passive population. As one moves towards redder
colors, the star-forming and passive populations begin to overlap in
color. In the reddest colors, the ``red sequence'' contains a mixture
of old passive ellipticals as well as young star-forming galaxies. So
instead of using an algorithm that attempts to model the ``red
sequence'', we simply allow for galaxies in clusters to have similar
colors.  Our clusters will be detected via a mixture of galaxy types,
both passively evolving and star-forming.

In this paper, we first present an outline of our algorithm based on
the premise of galaxies clustering in both position and color.  We
then spend a significant amount of effort analyzing how the C4
algorithm's free parameters affect the completeness and contamination
of the final cluster catalog. 
We do this using a new breed of mock galaxy catalogs which populate
N-body cosmological simulations with realistic galaxy populations.
Previous authors have realized the necessity of such a detailed
understanding of their clustering algorithms (e.g., Diaferio et
al. 1999, Adami et al. 2000, Postman et al. 2002, Kim et al. 2002,
Goto et al. 2002, and Eke et al. 2004). In most of these cases, the
authors have embedded fake clusters of various forms into real or
simulated data, which are then searched for to characterize the
contamination and completeness of the algorithm.  In this work, we
have gone one step further by using mock galaxy catalogs generated
from full N-body simulations.  A similar technique was used in recent
work by Eke et al. 2004 on 2dF groups, using N-body simulations
populated with semi-analytic models.  The catalogs we use, developed
by Wechsler et al (2005),  embed galaxies
with realistic luminosities and colors into cosmological simulations
which contain all of the messiness of structure formation, including
merging systems, systems with lots of substructure, systems with
ill-defined E/S0 ridgelines, systems that nearly overlap in redshift
space, etc.

In Section \ref{overview}, we provide details of the C4
algorithm.
In Section \ref{sim}, we describe the use of novel mock
catalogs --- based on populating large cosmological simulations with
realistic galaxy properties --- for calibrating the C4 algorithm and
determining the completeness and purity of our SDSS C4 cluster
catalog.  
In Section \ref{prop}, we introduce the observables that we
measure for each cluster, and then discuss the scaling relations
between those observables and halo mass in Section \ref{scaling}. The
SDSS data and the SDSS C4 cluster catalog are presented in Section
\ref{data} and we conclude in Section \ref{summary}.  Where
appropriate, we have used ${\rm h = H_o}/100\,{\rm km\, s^{-1}}\,{\rm
Mpc^{-1}}$, $\Omega_m=0.3$ and $\Omega_{\Lambda}=0.7$ throughout this
paper.

\section{The C4 Algorithm}
\label{overview}

\begin{figure*}
\plotone{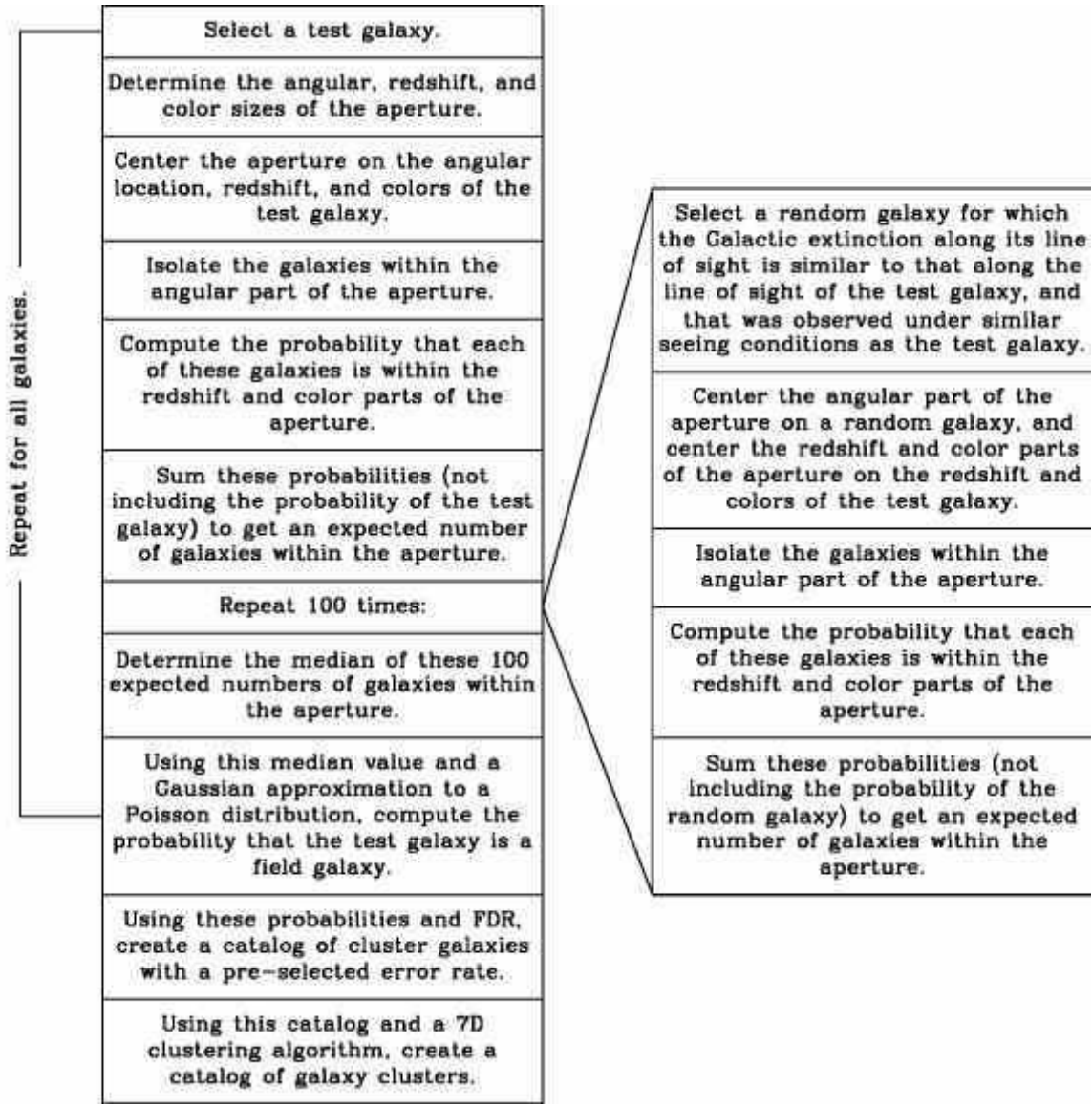}
\caption[]{Flow chart describing the algorithm}
\label{fig:flow}
\end{figure*}

In this section, we present the C4 cluster--finding
algorithm. Details and tests follow. Many readers will only want a
brief explanation of the algorithm and we suggest they examine the
flowchart of the algorithm given in Figure \ref{fig:flow} and read
this overview section.  For those who desire more details, each step
is described in more detail throughout the rest of this section.  The application of
the C4 algorithm to the SDSS data is discussed in Section \ref{data}.

The C4 algorithm begins by
placing each galaxy in a seven dimensional space of
right ascension
(ra), declination (dec), 
redshift and four color dimensions ($u-g$, $g-r$, $r-i$,
$i-z$). On each such target galaxy, we then perform the following steps:

\begin{enumerate}
\item We place an aperture around each target 
to only include galaxies in a specified range of ra, dec, and redshift.
We then measure the probability that every
galaxy within this spatial aperture has colors equal
to the target galaxy. 
The probabilities
are summed to obtain a ``number count''.

\item Using the target galaxy's spatial and color aperture,
we then select 100 random galaxies and 
perform step (1).
These 100 random locations provide a ``number count'' distribution
for the target galaxy.

\item Using the number count distribution, we compute the
probability of obtaining at least the observed number count around the original
target galaxy. By definition, target galaxies with low probabilities will be in
clustered regions.

\item We repeat this exercise for all (`target') galaxies in our sample and then
rank all the target galaxy probabilities obtained from step 3.

\item Using the false discovery rate algorithm (FDR; Miller et al. 2001c), 
we determine a threshold in
probability below which target galaxies are removed;
our threshold choice typically results in the
eradication $\simeq90\%$ of all galaxies. 
The galaxies that remain are called ``C4 galaxies''.
By construction, these reside in high density regions with
neighbors that possess similar colors.

\item We determine the local surface density around
  all C4 galaxies, using only the C4 galaxies,
 We then rank order these measured densities and
  locate C4 cluster centers as peaks in this density field. 

\end{enumerate}

In summary, the C4 algorithm is a semi-parametric implementation of
adaptive kernel density estimation. The key difference of our
approach, compared to previous color-based cluster--finding
algorithms, is that we do not attempt to model either the colors of
the cluster galaxies ({\it e.g.}, Gladders \& Yee 2000, Goto et
al. 2002) or the properties of clusters ({\it e.g.}, Kepner et
al. 1998; Postman et al. 1996; Kim et al. 2002). Instead, we only
demand that the colors of nearby galaxies are
similar to those of the target galaxy. In this way
we are sensitive to a diverse range of cluster and group types {\it e.g.}, our
algorithm would detect a cluster dominated by a ``blue'' population
of galaxies (see Figure \ref{fig:clust_fig2}).

\subsection{Defining the 7-dimensional Search Aperture}
\label{definebox}

Every target galaxy in the dataset has a uniquely defined location in
a 7-dimensional data--space. For example, the position of the $i^{th}$
galaxy is defined as:
\begin{equation}
{\bf r^{i}} = [{\rm ra}^i, {\rm dec}^i, z^i, m^i_{u} - m^i_{g}, m^i_{g} - m^i_{r}, m^i_{r} - m^i_{i}, m^i_{i} - m^i_{z}],
\end{equation}
where $m^i_{X}$ are the five passband Petrosian magnitudes from the
SDSS PHOTO version 5.4 data reductions (typically abbreviated
$u,g,r,i,z$). No $k$--corrections are used herein.

To look for clusters in this 7-dimensional data--space, we need to define a
search aperture. Clearly the size of this aperture will have an effect on the
types of clusters we find in this data--space. We begin by using
a projected radius that is fixed in comoving coordinates, and specifies
the ra and dec aperture surrounding the target galaxy.
The exact cosmological model used makes little difference over the
redshift range we examine here, ($z \sim 0.1$). This
aperture can be tuned 
to find the size that optimizes completeness and purity in the
mock galaxy catalogs. 

We next define the redshift (or line--of--sight) dimension of the C4 search
aperture. For the spectroscopic SDSS sample, all galaxies have known
redshifts and we simply place a $z$-constraint around the target galaxy.
For the SDSS photometric sample, one would need estimated redshifts
or else this constraint must be dropped entirely. We have chosen to
convert redshift to co-moving distance under an assumed model, but one
could also simply let the length of the redshift dimension vary with redshift.

Finally, we must define the color part of our search aperture. The size of the
four color dimensions will be driven by the well-established intrinsic
color-magnitude relation (CMR) seen in clusters (see Figure
\ref{fig:fourplot}) and the expected errors on the SDSS magnitudes. The CMR is
known to have a linear relationship with a small negative slope (with
increasing magnitude) and small scatter (Bower et al. 1992). Therefore, the size of the
``color-box'' should be set to capture the full range of colors in the CMR,
from the brightest to the faintest cluster galaxies in any given cluster in
our data. We additionally include the known statistical ($1 \sigma$)
uncertainties in the individual galaxy magnitudes.
For the SDSS main galaxy
spectroscopic sample, these errors are minimal (less than 0.1\% at m$_{r}$ = 17.7).
We sum in quadrature these statistical errors and also a systematic uncertainty via:

\begin{equation}
\delta {\rm C_{xy}}
= \sqrt{{ \sigma^2_{xy}(stat) + \sigma^2_{xy}(sys)}},
\end{equation}
where
$\sigma^2_{xy}(stat)$ is the observed error for the two
magnitudes ($x,y$), summed in quadrature.  Here $\sigma^2_{xy}(sys)$
is a measure of the inherent scatter in the CMR (see below).  Therefore, 
for each $i$ galaxy the size of the color box is given by,
\begin{equation}
\delta {\bf C^i} = [\delta {\rm C^i_{ug}}, \delta {\rm C^i_{gr}},\delta {\rm C^i_{ri}},\delta {\rm C^i_{iz}}].
\end{equation}

We have used the Petrosian magnitudes reported by the SDSS throughout,
as it is better suited for the analyses of galaxies in the SDSS
spectroscopic sample (see Stoughton et al. 2001). However, our final
cluster catalog is robust against the use of Petrosian versus model
magnitudes. We do not apply evolutionary or k--corrections to our
data, as we are looking for galaxies clustered in both position and
color around another galaxy: for a given redshift and color of a
galaxy, any excess of neighboring galaxies with similar colors should
occur independent of any evolutionary effects and k--corrections.

Once we have defined the search aperture around a target galaxy, we
then ``count'' the number of neighboring galaxies within that
aperture. To do this, we demand that any neighboring galaxy fit
exactly within the spatial part of the aperture (ra, dec, and
redshift) as these dimensions are accurately known.

In the color dimensions, we allow for uncertainties in both the
color box of the target galaxy and the individual colors of
surrounding galaxies.  Specifically, we replace the color boxes with
Gaussians having widths specified by Equation (4), which ``softens'' 
the sides of the 4-d color box. We also treat the errors on
the individual galaxies as Gaussians. We then measure the the joint
probability that any galaxy falls within the color box of the
target galaxy.  We then sum these probabilities for all neighboring
galaxies and report this as the ``number count'' of neighboring
galaxies. 

\subsection{Building the Count Distributions}
\label{counts}

The next step in the C4 algorithm is to build a distribution of
expected number counts for each target galaxy, given that it was in a
random position.  We place the 7-d aperture of the target galaxy
around 100 randomly chosen galaxies and ``count'' the neighbors as
described above. We allow for the fact that our algorithm can be
run on the SDSS photometric data, in which case the seeing conditions
and galactic extinction can have a large effect on the selection function
of the SDSS photometric sample. The random galaxies
can be selected such that they have the same seeing and reddening as
the target galaxy. However, on the complete  SDSS spectroscopic sample, we ignore
this constraint.
From these 100 randomly chosen locations in
the data, we construct a distribution of counts for the 7-d aperture of
the target galaxy. 
So long as we expect no more than half of
the galaxies to be in clustered environments ({\it i.e.}, have higher counts
with respect to the mean), the medians of these
distributions are robust descriptors of the distributions.

\subsection{Determining Probabilities}
\label{probs}

By this stage, we have defined a unique aperture for the target galaxy. We have
measured the number of neighboring galaxies within its aperture 
and built a count distribution from 100
random locations at the same redshift of the target galaxy. We then
ask the question: how likely is the observed neighboring galaxy count given
the distribution of neighboring galaxy counts for a specified 7-d aperture?
The exact form of the
distribution of neighboring galaxy counts depends on the number of counts
measured. For example, in the photometric SDSS data, where there are millions
of galaxies, the distributions of neighboring galaxy counts is
Gaussian. However, in the spectroscopic data, the count distributions can sometimes be small and
Poissonian. As a compromise, we adopt the Gaussian approximation to the
Poisson.  In order to justify the basis of Poissonian statistics, we need to
meet the following requirements in the count distribution; {\it i)} the count
within an aperture of zero volume is zero, {\it ii)} each of the 100 randomly
chosen counts must be independent, {\it iii)} the count values depend only on
the size of the aperture, {\it iv)} the aperture size does not change when
building the count distributions, and {\it v)} no two counts come from the
same location.  Requirements {\it i,ii,iv} and {\it v} are already met in our
algorithm, while {\it iii} requires that the randomly selected points have an
underlying count distribution that is also random. This, of course, is not
true for all galaxies, as galaxies are known to cluster, and galaxies within
clusters have a higher neighbor count than those in the field. However, if
a majority of the randomly selected galaxies are ``field-like'', {\it i.e.},
less clustered then the older elliptical population in clusters and groups,
then we can expect {\it iii} to hold. At worst, this assumption produces a
small bias by slightly raising our probabilities, resulting in a loss of
statistical power (which would affect the C4 completeness), and so our Poisson
assumption is a conservative one.

The Gaussian approximation to the Poisson distribution has the convenient feature
that the width of the Gaussian is equal to the square root of the mean of the
Poisson distrubution. Thus, when we build the count distributions based on the
100 random locations, we only need to calculate the median, which then fully
describes the Gaussian approximation. Thus, the probablility that a target
galaxy looks like a field galaxy is determined solely from the count around
the target and the median of the counts around the 100 random locations.

\subsection{Repeat}
Once the above steps are performed on the first target galaxy, we then
repeat for all galaxies in sample. This is conducted
in no specific order. Once we have looped over the entire sample, every galaxy has
a probability that it is a ``field galaxy''. These probablities are ranked
so that a threshold can be applied to separate the cluster galaxies from
the field galaxies.

\subsection{Choosing a Threshold}
\label{threshold}
Miller et
al. (2001c) present a new thresholding technique known as the False Discovery
Rate (FDR) originally devised by Benjamini and Hochberg (1995).  This
technique allows one to choose a statistically meaningful threshold, in the
sense that the fraction of false positive detections over the total number of
detections is controlled.  We apply the techniques discussed in Miller et
al. here. Briefly, this involves choosing a priori the maximum fraction
of acceptable false discoveries ($\alpha$) one is willing to tolerate.  The
 p-values (probabilities)
are rank--listed (from lowest to highest) and a line of slope
$\alpha$ drawn. Where the two lines intersect for the first time defines the
threshold one must use to guarantee the fraction of false discoveries (see Miller et al. 2001c and
Hopkins et al. 2002 for 
some examples).
After applying the FDR technique, all galaxies above the threshold are called
``cluster-like'' or ``C4'' galaxies and are then used to identify C4 cluster centers.

\begin{figure*}[tp]
\plottwo{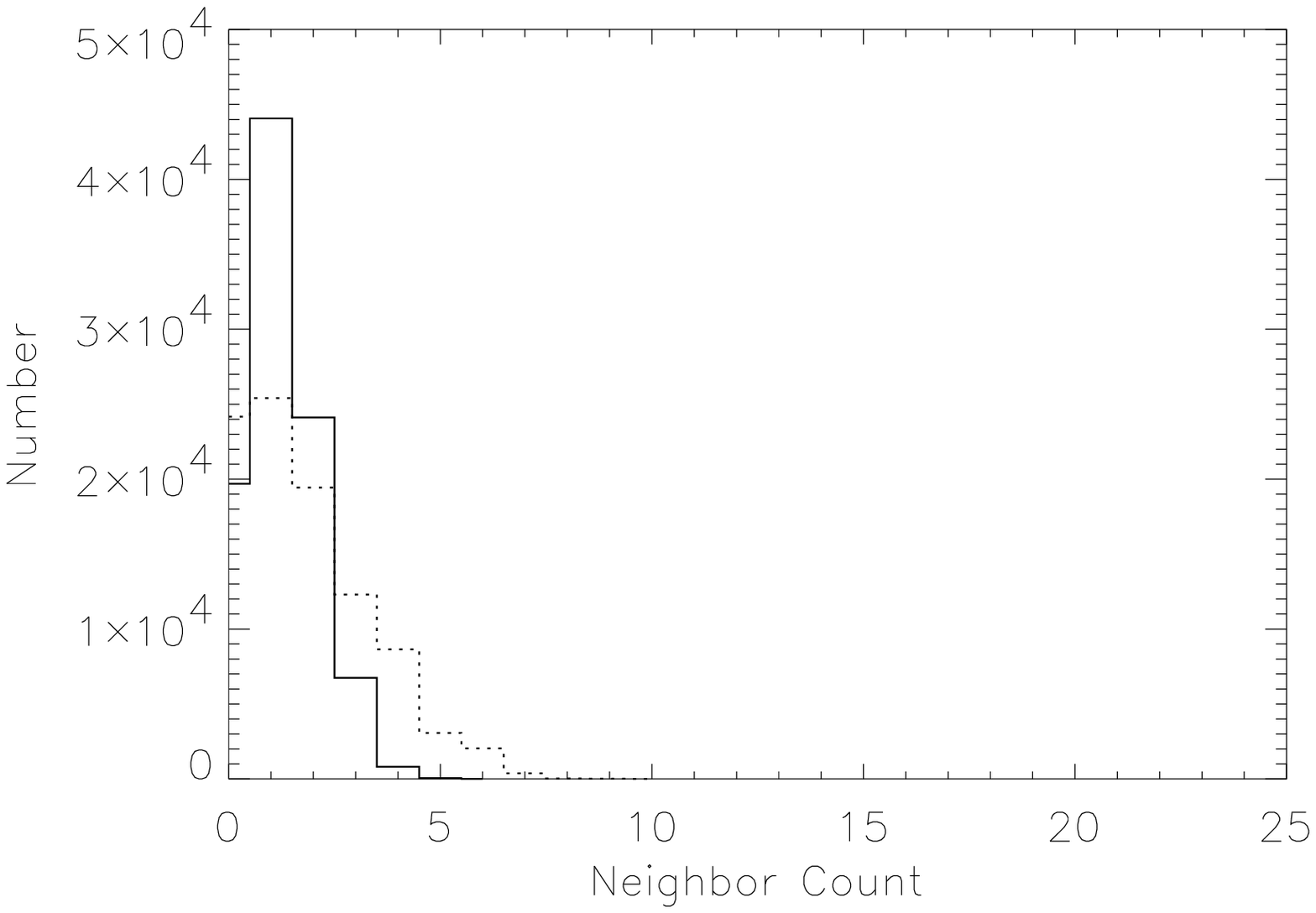}{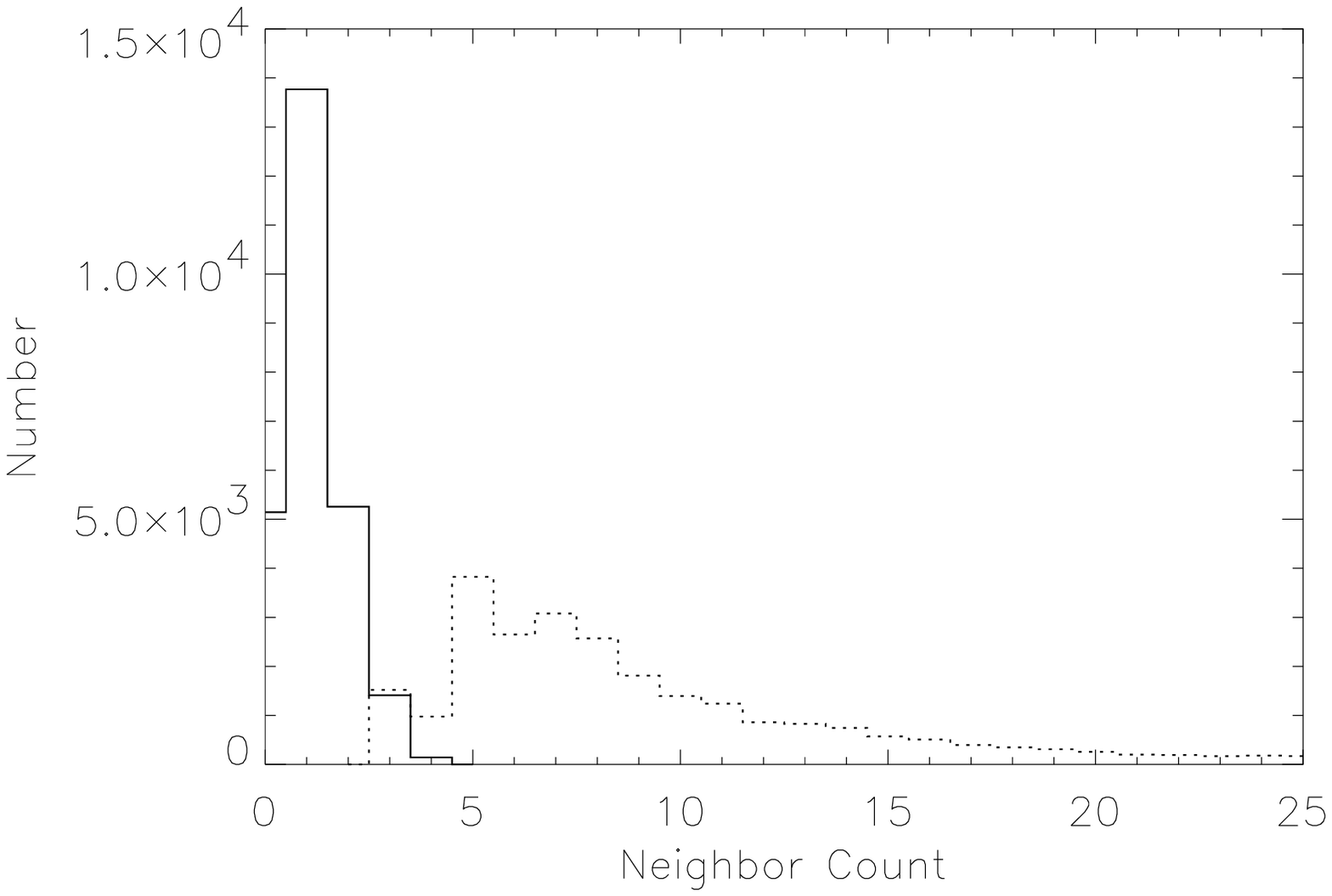}
\caption[]{We show the counts from the field based on the median of 100 random
locations for each galaxy (solid) and counts around each specific galaxy (dotted).
On the left we show for ``field-like'' galaxies, while on the right we show
``cluster-like'' galaxies. Notice that ``cluster-like'' galaxies have more neighbors
than the median of the field.}
\label{fig:counthistos}
\end{figure*}

We test our probability model and whether our FDR threshold can separate
``field'' and ``cluster-like'' galaxies on a galaxy-by-galaxy level.  In
Figure \ref{fig:counthistos}, we show the distribution of median counts for
the count distributions around all galaxies in the mock SDSS catalog (solid
line), compared to the individual counts around galaxies (dash line).  We have
split our sample into ``field-like'' and ``cluster-like'' as described above.
Note that the counts around ``cluster-like'' galaxies are significantly higher
than around ``field-like'' galaxies.  Also note that the median counts of the
field distributions, and also the counts around ``field-like'' galaxies, are
similar and Poisson. This shows that the medians of the random distributions
are representative of ``field-like'' galaxies. This justifies our use of the
median to represent the random distributions, and that the probability
threshold (discussed below) cleanly separates ``field-like'' and
``cluster-like'' galaxies.

\subsection{Identifying the clusters}
\label{clustids}

The C4 algorithm works by identifying galaxies clustered in a positional
and color space. Once the 7-d aperture and threshold are defined, 
galaxies with low probabilities of being field-like (as defined by
random positions on the sky) are identified as clustered C4 galaxies
(10\% of all galaxies using our fiducial parameters).
We then find centers of the clumps of C4 galaxies, and call these
C4 clusters. The algorithm (to this point) does not define the galaxy
membership of the clusters (see Section \ref{prop}).

We were motivated by the Spherical Overdensity method applied by
Evrard et al. (2002) to identify halos in the HV simulation. This adds
a level of consistency when comparing our observed C4 catalog to the
mock C4 cluster samples.  We begin by measuring the distance to the
sixth nearest projected neighbor for each C4 galaxy (using only C4
galaxies). We do this in redshift shells of $\Delta z = 0.02$.  The
nearest neighbor distances are ordered from the smallest to largest,
and the C4 galaxy with the smallest sixth nearest neighbor distance is
assigned as the center of the first C4 cluster. We then exclude all C4
galaxies from this list out to a projected radius corresponding to 15
times the background density of C4 galaxies centered on this first
cluster. This choice of enhancement is arbitrary. However, this same
overdensity is used when examining the real data or the mock
catalogs. We have checked to make sure that the distribution of
neighbor distances is the same in both the real SDSS data and in the
mock galaxy catalog.  We then move to the next highest density C4
galaxy that is not within any other C4 cluster and repeat {\it i.e.},
the C4 galaxy now with the smallest sixth nearest neighbor distance
becomes the center of the second cluster and so on.  The iterations
are terminated when all C4 galaxies are assigned to clusters or the
local densities fall below the threshold.  These initial centers are
then peaks in the C4 galaxy surface over-densities.

\begin{figure}[tp]
\plotone{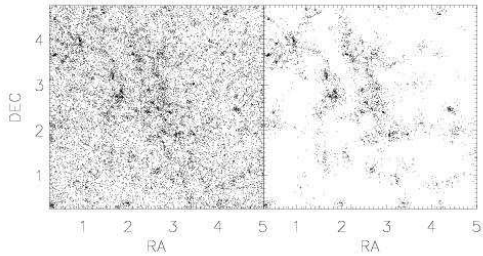}
\caption[]{Projected galaxy distribution of the simulations
before (left) and after (right) the C4 algorithm is run and a
threshold is applied to eliminate
field-like galaxies.}
\label{fig:before_after}
\end{figure}

\begin{figure}[tp]
\plotone{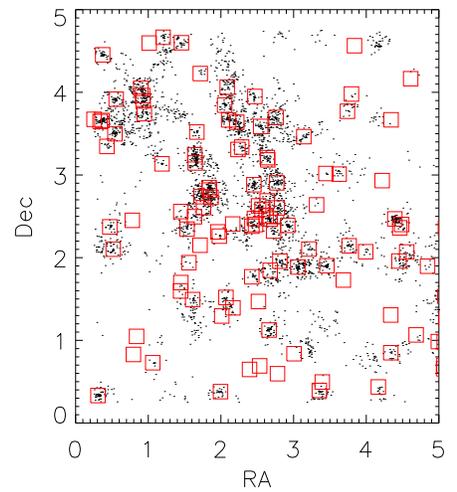}
\caption[]{The C4 galaxies, as in Figure \ref{fig:before_after} (right), with
the dark matter halo positions overplotted (red squares). The halos have masses greater than
$4.5\times 10^{13}$ $h^{-1}M_{\odot}$.}
\label{fig:c4_groups}
\end{figure}

This process is
shown visually in Figures \ref{fig:before_after} and
\ref{fig:c4_groups}.
We investigated other methods for finding the C4 cluster
centers ({\it e.g.,} ``friends--of--friends'' algorithm {\it etc.}),
and find this method to be the best in terms of accuracy,
completeness, and purity when compared to the actual halo catalog.

During this process, if there are fewer than three C4 neighbors around
any cluster center, we exclude it as a possible cluster. Likewise, if
we determine that less than 10\% of all galaxies in an
$1h^{-1}$Mpc aperture around the cluster are classified as C4
galaxies, we exclude it (the number of clusters excluded due to this
criteria is less than 2\% of the total found). Finally, we exclude
any cluster that has less than eight members within $1h^{-1}$Mpc so 
that we may measure a reliable velocity dispersion. We note that these
exclusions imply that C4 completeness (see Section \ref{uber})
is lower than it could be, since these excluded systems could often be real
systems.

\subsection{Other Algorithm Considerations}

\subsubsection{Survey Edges}
\label{edges}
Edge effects are taken into account during this procedure: C4 galaxies
are required to be further than 7.5 arcminutes from any edge of the
data sample. This allows use of all galaxies within a circle of 15
arcminute diameter when making our counts.
Edges cause their greatest affect on the catalog at the lowest
redshifts. To help minimize the problems of survey edges, we only
include cluster candidates whose initial redshift (estimated from
the C4 galaxies) is above $z =0.03$. Of course we do find clusters 
below this redshift (see Figure \ref{fig:clust_fig6}-- which is a known
X-ray cluster at $z = 0.027$), but the C4 algorithm
cannot be tuned to work statstically at such a low redshift.

\subsubsection{The Magnitude Limit}
\label{maglimit}
The current incarnation of the catalog is run on apparent
magnitude limited surveys. The SDSS spectroscopic Main sample was
targeted using galaxies brighter than $m_r = 17.77$, while the Luminous
Red Galaxy (LRG) sample includes additional elliptical-type galaxies
to $m_r = 19.5$.
While technically this is an input parameter that could be tuned, we
instead consider the effect on completeness, purity, and total number
of clusters after a single limit is chosen. When measuring the cluster
properties (see Section \ref{prop}), we apply an absolute magnitude limit as well.

\section{A Mock SDSS Galaxy Catalog}
\label{sim}
One advantage of the C4 catalog over many previous cluster
catalogs is our use of realistic cosmological N-body simulations to
refine the algorithm and to determine the completeness and purity of
the catalog.  Since the C4 algorithm is highly dependent on clustering
in {\em both} spatial and color space, any mock galaxy catalog must
have relations between galaxy color and density that mimic the those
found in real data. Similarly, the luminosity functions of the mock
catalogs and the data must be similar.  Simple bias schemes which
produce a population of galaxies above some luminosity cut (e.g., Cole
et al 1998) will not produce the information we need to find clusters.
Mocks created from more detailed semi-analytic models of galaxy
formation (e.g., Kauffmann et al. 1999 and Eke et al. 2004) do produce
colors and luminosities; unfortunately at this stage none of these
models are yet producing galaxies with properties that nearly enough
reproduce those seen in the SDSS data.  The mock catalog we use here
thus takes a very empirical approach, in which we aim to populate a
dark matter simulation with galaxies whose properties (especially, the
luminosity and color distributions as a function of environment)
closely match those seen in SDSS data.  The method for creating these
catalogs is described briefly in Wechsler (2004) and in detail in
Wechsler (2005, in preparation, hereafter W05); here we just give a
rough outline.  A complementary approach, using conditional luminosity
functions constrained by galaxy clustering, has been developed by Yang
et al. (2004).

The W05 catalogs are constructed using the distribution of dark matter
in the $\Lambda$CDM Hubble Volume simulation.  The simulation follows
$10^9$ particles of mass $2.25\times 10^{12} h^{-1}$ solar masses in a
periodic cubical volume with side length $3 h^{-1}$Gpc, in a flat
$\Lambda$ CDM universe with $\Omega_m=0.3$, $\sigma_8=0.9$, and
$h=0.7$. We use a subset of the MS sky survey output described by
Evrard et al. (2002) which mimics the collection of data on the past
light cone of an observer located at the center of the volume. We only
look at halos more massive than 4.5$\times{10}^{13} h^{-1}$ solar
masses (see below).  The large size of the simulation allows the
creation of a full-sky survey out to a depth of $z_{\rm max}=0.57$,
and thus can also be used to test cluster finding algorithms that use
only the SDSS photometric data and extend to higher redshift.

Briefly, the algorithm for creating the W05 catalogs consists of
constraining the relation between local dark matter density and
luminosity in the $r$-band, such that the mock galaxies match the
luminosity-dependent two-point correlation function measured in the
SDSS data (Zehavi et al. 2004).  We then measure local galaxy
density in the $r$-band, for both the data (see Gomez et al. 2003) and
the simulation, and assign the colors of real SDSS galaxies to mock
galaxies that have similar luminosities and local galaxy densities.
All relevant details of these new simulations are presented in W05,
including the prescriptions used to assign galaxy properties to each
dark matter particle and the extensive tests performed on these
simulations to ensure they closely mimic the real data.

The mass resolution of the simulation allows us to include galaxies
brighter than about 0.4 L$^{\star}$, and to resolve halos more massive
than $4.5\times10^{13}h^{-1}$ solar masses.  Therefore, these
simulations can only resolve bright galaxies in
intermediate--to--massive clusters of galaxies, {\it e.g.}, a cluster
like the Coma Cluster would contain about 500 galaxies in these mock
SDSS catalogs.  This makes them well-suited to the brighter (and more
massive) galaxies in the SDSS spectroscopic sample (which is what we
use in this work), which only includes dimmer galaxies at the lowest
redshifts $z<0.05$.  
As discussed above, we place a constraint on any C4 cluster that
there be at least eight galaxies within $1h^{-1}$Mpc of the cluster
centers.  For halos at the minimum well-resolved mass in the
simulation, only 1\% have fewer than
eight galaxy members, so nearly all are intrisically detectable by the
C4 algorithm.  This indicates that these simulations allow us to
reliably estimate the completeness of our catalog over the full mass
range of interest, but higher resolution simulations will be required
to fully characterize the purity of the smallest C4 systems.

Another advantage of using the HV simulation is that we can directly
relate any detected clusters in these mock SDSS catalogs to the
measured halo masses discussed by Jenkins et al. (2001) and Evrard et
al. (2002). In other words, we can directly relate the observables
(e.g., velocity dispersions and summed total cluster optical luminosities)
to the dark matter
halos used by Jenkins et al. (2001) for constructing the cosmological
mass function. There is no ambiguity in the definition of
mass between the theoretical models and the observables.

Since the real data will contain photometric errors, we
have added errors to each mock galaxy magnitude. However, we
note that for the spectroscopic sample, the SDSS Petrosian (1976) magnitude
errors are tiny (the largest fractional errors are $< 0.1\%$ in the $r$-band).
As described in Section \ref{definebox}, the size of the color-box is at least
an order of magnitude larger than the errors on the colors. Thus, this
step is not necessary for the spectroscopic data. However, when the full photometric
SDSS sample is used (to m$_{r}$ = 22), the photometric errors can in fact
dominate the color-box. Therefore, we have built in this mechanism at this stage.
We note that we use the true SDSS errors when the algorithm is run on the
real SDSS data. In the mock catalogs,
we use the median observed
error for galaxies in the SDSS main galaxy sample (see Strauss et
al. 2002). Our results are insensitive to changes ($<$50\%) in the
value of this constant error.

The mock galaxy catalog is complete down to an absolute magnitude of
$M_r = -19.6$, which corresponds to $m_r = 19.2$ at $z =0.17$ (which
is the highest redshift cluster we find in the real data).
We apply an apparent magnitude limit of m$_r \le 17.7$ to the SDSS
mock catalog to mimic the real magnitude--limited SDSS main galaxy
sample (see Strauss et al. 2002).  Once the magnitude limit is applied
and the errors are added, we are able to run the exact same C4
algorithm on the mock galaxy catalogs and identify clusters as
outlined in Section \ref{overview}.  For this work, we have used a volume
that is larger and more contiguous than the SDSS DR2. However, edge
effects are handled identically in the data as they are in the mock
catalogs (see Section \ref{edges}). While we have not applied the SDSS
targeting algorithm, in Section \ref{targeting} we study this issue in
detail and find that the effect on completeness and purity is
small.

\section{Completeness, Purity, and Tuning the C4 Algorithm}
\label{uber}

We use the SDSS mock galaxy catalogs to test the C4 algorithm, fine
tune the choice of parameters, and measure the completeness and purity
of the catalog (i.e., the selection function). We do this by running
the C4 algorithm on the mock galaxy catalog, and comparing the found
C4 clusters to the known halos from Evrard et al. (2002).  To make the
comparison, we apply a matching algorithm to associate C4 clusters
with halos.  We have investigated several prescriptions for matching
these two datasets and have found that our matches are robust against
the details of the matching algorithm. Here, we present results based
on matching a dark matter halo with any C4 cluster within a projected
distance corresponding to one virial radius and within $\Delta z =
0.005$. We discuss this matching in more detail in Section
\ref{false_matches}.  To estimate purity, we match clusters to
any simulated halo within the estimated $r_{200}$ of the ``observed''
C4 cluster, while for the completeness measurements we match each
``observed'' C4 cluster to the nearest dark matter halo within $\Delta
z= 0.005$ and the projected $r_{200}$ of the halo.

This method for matching allows for multiple matches. In other words,
when measuring completeness, multiple C4 clusters can be matched to
one HV halo, and similarly when measuring purity, multiple halos can be
matched to a single C4 cluster. There are many ways to deal with this
problem. For instance, when multiple halos match to one C4 cluster, we
could take the most massive halo as the fiducial match. Or we could
take the halo that has the most similar luminosity to the C4 cluster,
or any other method. We have chosen to simply take the match that is
closest in separation on the sky (and within $\Delta z= 0.005$).  We
have investigated a few of the other methods we mentioned and find no
clear winner. The C4 algorithm finds fewer clusters in the mock
catalog than there are real HV halos (i.e., the C4 algorithm is never
100\% complete). As seen and discussed in the following sections, this
completeness drops with halo mass such that the C4 algorithm can miss
up to 50\% of the halos at masses $\sim 5\times10^{13}h^{-1}$ solar
masses.  This means that there will always be more multiple halo
matches to the C4 clusters than vice versa. On average, 50\% of the C4
clusters have multiple halos within $\Delta z = 0.005$ and $r_{200}$
while only 5\% of halos have multiple C4 clusters within those same
constraints.

After the matching is done, we will plot the cumulative quantity:
\begin{equation}
{\rm Purity} (L_r) =\frac{{\rm Number} (>L_r){\rm ~ C4 ~ Matched ~ to ~ Halos}}{{\rm Number} (>L_r){\rm ~ C4 ~ Clusters ~ Found}}
\end{equation}
\begin{equation}
{\rm Completeness} (M_{200}) =\frac{\rm{Num} (>M_{200}){\rm~ Halos ~ Matched ~ to ~ C4}}{\rm{Num} (>M_{200}){\rm~ Total ~ Halos}}
\end{equation}
where $M_{200}$ is the mass within a radius that is 200 times the critical density and
$L_r$ is the summed luminosities of the cluster member galaxies as defined in detail
in Section \ref{prop}.
Since completeness is defined against the ``true'' halos from the mocks,
we plot completeness versus halo mass. On the other hand, purity is measured from the point of
view of the measured clustered catalog, and so purity is plotted against the observable: optical
luminosity. It is important to keep in mind that the high mass (or high luminosity) systems
are rare, and so the purity and completeness measurements can be noisy in these regimes.

\begin{figure*}[h]
\plotone{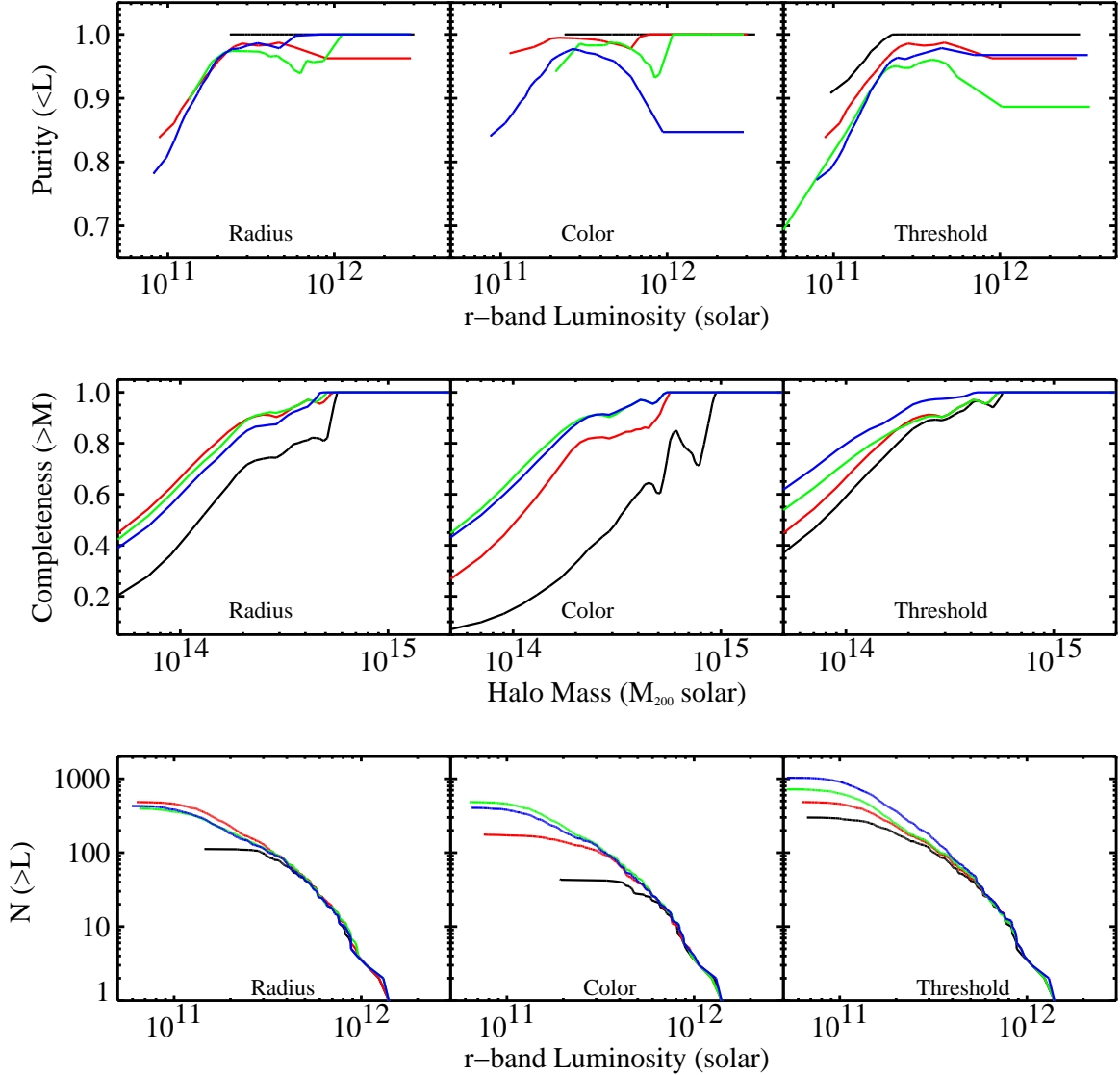}
\caption[]{
How the choice of box size, threshold, and
redshift-bin affect completeness and purity for clusters found using
the simulated mock galaxy data. Also show the cumulative number
plot ($N(>L)$) for the mock catalog clusters.
The variation of the box size is described in Section \ref{definebox}; the
variation of the FDR threshold is described in Section \ref{threshold}.
Generally, the size of the box
increases from black to red to green to blue.
The purity and
cumulative number counts are measured against the summed $r$-band
luminosity of the discovered clusters, while the completeness is
against the halo mass.}
\label{fig:uber}
\end{figure*}

In Figure \ref{fig:uber} (left), we present the completeness and purity of the mock C4
catalog as a function of different radii for
the search aperture: $500,1000,2000$, and $6000h^{-1}$kpc. In this
figure, the other dimensions of the search aperture are fixed at the final
values as discussed below. A radius of $500h^{-1}$kpc (black) appears to be
too small, as it significantly lowers the completeness of our sample for all
but the most massive systems (although it does produce the purest sample).
However, larger search-radii make little difference to the completeness
or purity of the algorithm. The highest completeness and purity occur when
a co-moving radius of $1h^{-1}$Mpc is used (the red line).

We varied the redshift dimension of the 7-d box to be a co-moving length of $25,50,100,200h^{-1}$Mpc. The
size of the aperture in the redshift direction must be large enough to allow
for significant (and unknown) peculiar velocities of galaxies within massive
clusters of galaxies and therefore, our 3-dimensional positional aperture is
shaped like a narrow cylinder. Using these tests, we find that our final
cluster catalog is {\it independent} of the length of the line-of-sight aperture.
We attribute this to the fact that there are not many clusters or groups that lie directly
along the line-of-sight that also have similar global colors.  Alternatively,
one could argue that by not using k--corrections for our SDSS colors, we have
already accounted for the redshift dimension in the ``color-box''.
We set the
redshift dimension of the search aperture to $50h^{-1}$Mpc.

In the middle panel of Figure \ref{fig:uber}, we show the completeness
and purity for the mock SDSS catalog as a function of the
``color-box'' size, holding constant the spatial part of the search
aperture. We examine only the effect of changing $\sigma_{xy}(sys)$,
using $\sigma_{ug}(sys) = \gamma \times 0.15, \sigma_{gr}(sys) =
\gamma \times 0.12,$ $\sigma_{ri}(sys) = \gamma \times 0.1$, and
$\sigma_{iz}(sys) = \gamma \times 0.1$. These values represent
reasonable widths for the color-magnitude relation, decrease with
increasing wavelength (as indicated in Figure \ref{fig:fourplot}), and
are motivated by the results of Goto et al. (2002). However, we note
that our algorithm is not attempting to model the color-magnitude
relation.  Thus, we allow the color-box size to be a free parameter in
our algorithm by varying $\gamma$ as $1,2,4,6$. We then use the mock
galaxy catalogs to tune this variable. We note that the median of
$\sigma_{xy}(stat) = 0.02$ for our data changes very little over our
magnitude-range (recall, these are the bright galaxies in the
spectroscopic SDSS data). Thus, $\sigma_{xy}(sys)$ is the dominant
term in Equation (4). 

As seen in Figure \ref{fig:uber} (middle), the smallest ``color-box''
dimension produces a very pure, but highly incomplete (black) sample,
as was the case for the smaller radial aperture.  As we increase the
size of the color box, we increase the completeness, while
decreasing the purity.  For the final algorithm, we choose $\gamma =
4$ which has the highest completeness (for $M \ge 10^{14}\,{\rm
M_{\odot}}$ systems), while still maintaining a high level of purity.

In the right panel of Figure \ref{fig:uber},
we show the completeness and purity of the C4 sample
as a function of the FDR threshold. We vary $\alpha$ from 0.05, 0.10, 0.15,
0.20, and 0.50.  We note that our least conservative threshold ($\alpha =
0.5$) produces the highest completeness, but at the expense of purity.
By lowering the FDR threshold, one
simply increases the number of C4 galaxies being selected, but these extra
galaxies either increase the detection likelihood of clusters already detected
at higher FDR thresholds, or form a background which decreases the
purity. Changing the threshold by a factor of four only improves the C4
completeness for ${\rm M200} \le 1\times 10^{14} {\rm M_{\odot}}$ systems by
$\sim 10\%$, but at the price of decreasing the purity for such systems by
$10\%$.
Based on this, we choose $\alpha = 0.1$ (the red line) which provides
both a high purity and completeness. This is preferred over maintaining a
higher completeness ({\it e.g.} $\alpha = 0.5$), as it provides users of the
C4 catalog the confidence to pick and choose real clusters
for any scientific analyses.  Also, as we discuss in section \ref{false_matches}, gains in
the measured completeness are as much a result of random matches as they are from a more
efficient algorithm.
When this final threshold is applied, approximately
90\% of all galaxies are excluded as not being in color and spatially clustered
environments.

\begin{figure*}
\epsscale{0.6}
\plotone{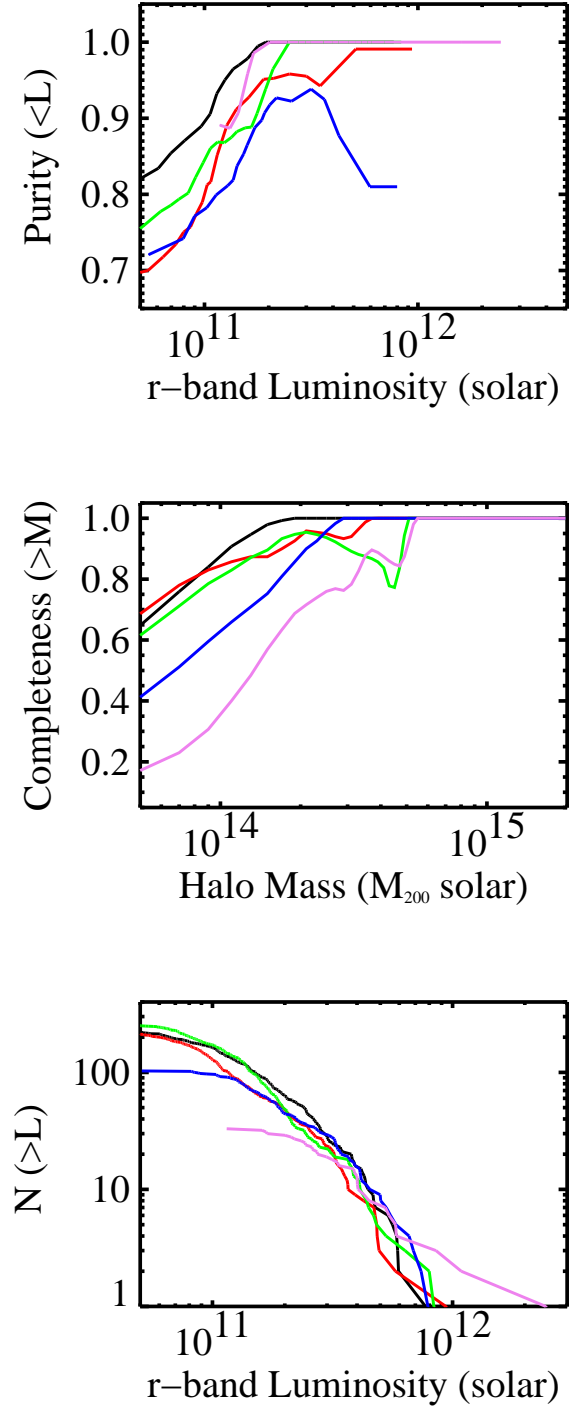}
\caption[]{The completeness, purity, and number of halos brighter than $L_r$ for
equal volume shells, each at increasing redshift. The shells are described in the
text. They go from the lowest to highest redshift as black,red, green, blue, violet.}
\label{fig:zlimit}
\end{figure*}

In Figure \ref{fig:zlimit}, we show purity, completeness, and number
for clusters in shells of equal volume, increasing in redshift from
black to red to green to blue to violet, covering redshift ranges of
[0.03,0.075], [0.075,0.093], [0.093,0.107], [0.107,0.118],
[0.118,0.128] respectively.  As with Figure \ref{fig:uber}, these
panels use the mock galaxy catalogs. However, unlike Figure
\ref{fig:uber} whose clusters numbered in the many hundreds, these
smaller volume bins contain  $\sim 100-200$ clusters and so
the results are noisier.
As expected, completeness decreases with increasing redshift, but
varies little out to $z = 0.107$ for all masses. Beyond that,
completeness drops steeply. Purity is fairly constant (to within 10\%)
over all redshift ranges, however the lowest redshift bin is the
purest.  The C4 catalog is $>90\%$ complete and $>95\%$ pure for
systems more massive than $\sim 2\times10^{14}M_{\odot}$ (or brighter
than $\sim 3\times10^{11}L_{\odot}$) and out to a redshift of $z \sim
0.12$. In the bottom panel of Figure \ref{fig:zlimit}, we see that the
number function is mostly dependent on the completeness in each
redshift shell.  The most complete bin (lowest redshift) shows the
highest number of low luminosity (or mass) clusters. As completeness
dwindles with redshift (and mass), the number of found halos decreases
similarly. The excess of $N(>L)$ in Figure \ref{fig:zlimit} (bottom)
for the highest redshift bin (violet) is due to a single bright,
massive halo in the simulations. 

\subsection{The Strength of Color Clustering}
\label{strength}

We have run a series of tests to determine whether our choice of all four
colors is necessary for our stated goals (high purity and known completeness)
compared to using just a subset of these colors. Specifically, we ran the C4
algorithm using each of the four colors separately, as well as using subsets
of the colors, {\it e.g.,} $g-r$ and $r-i$, but not $u-g$ or $i-z$.

\begin{figure*}[tp]
\plottwo{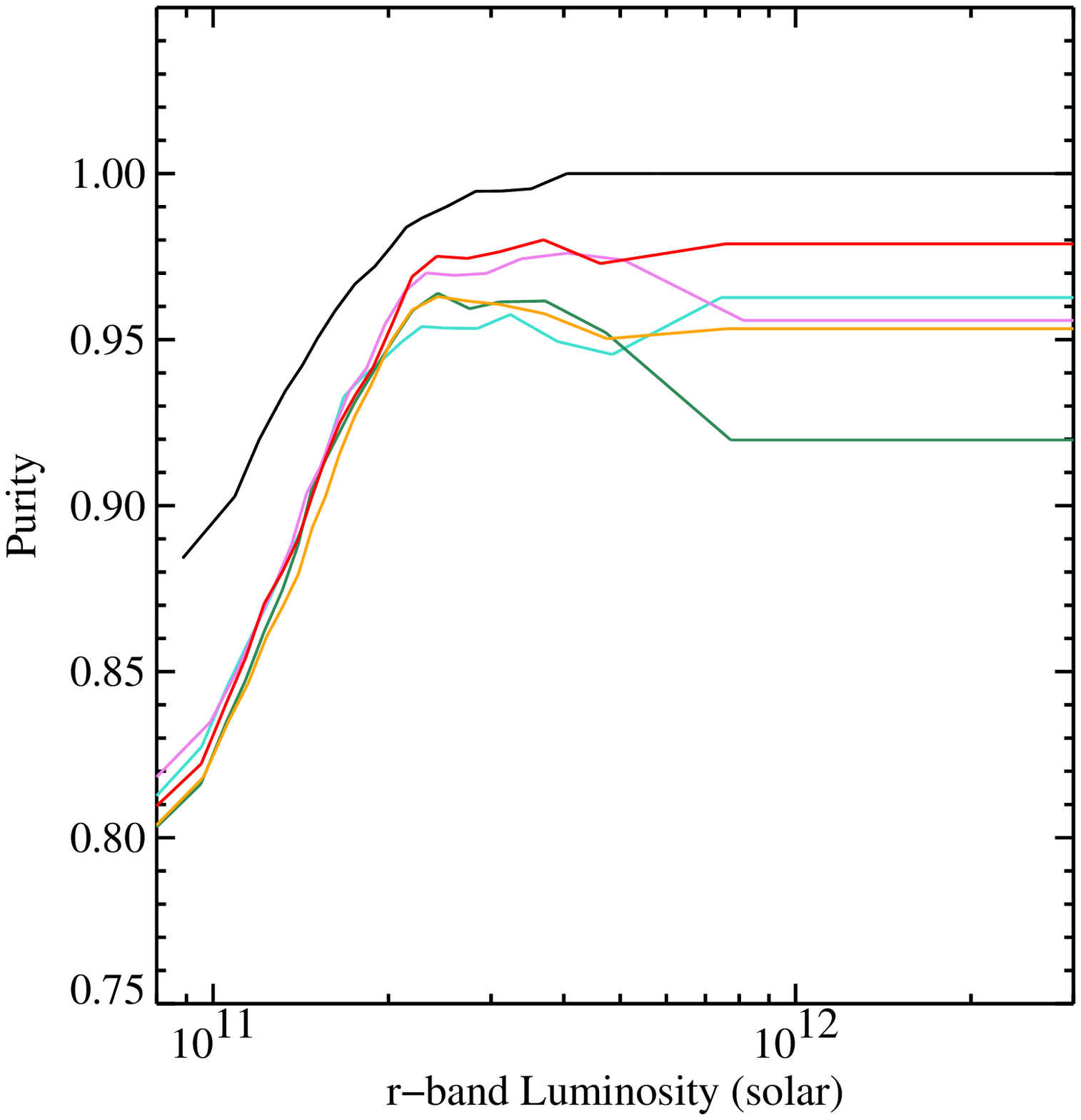}{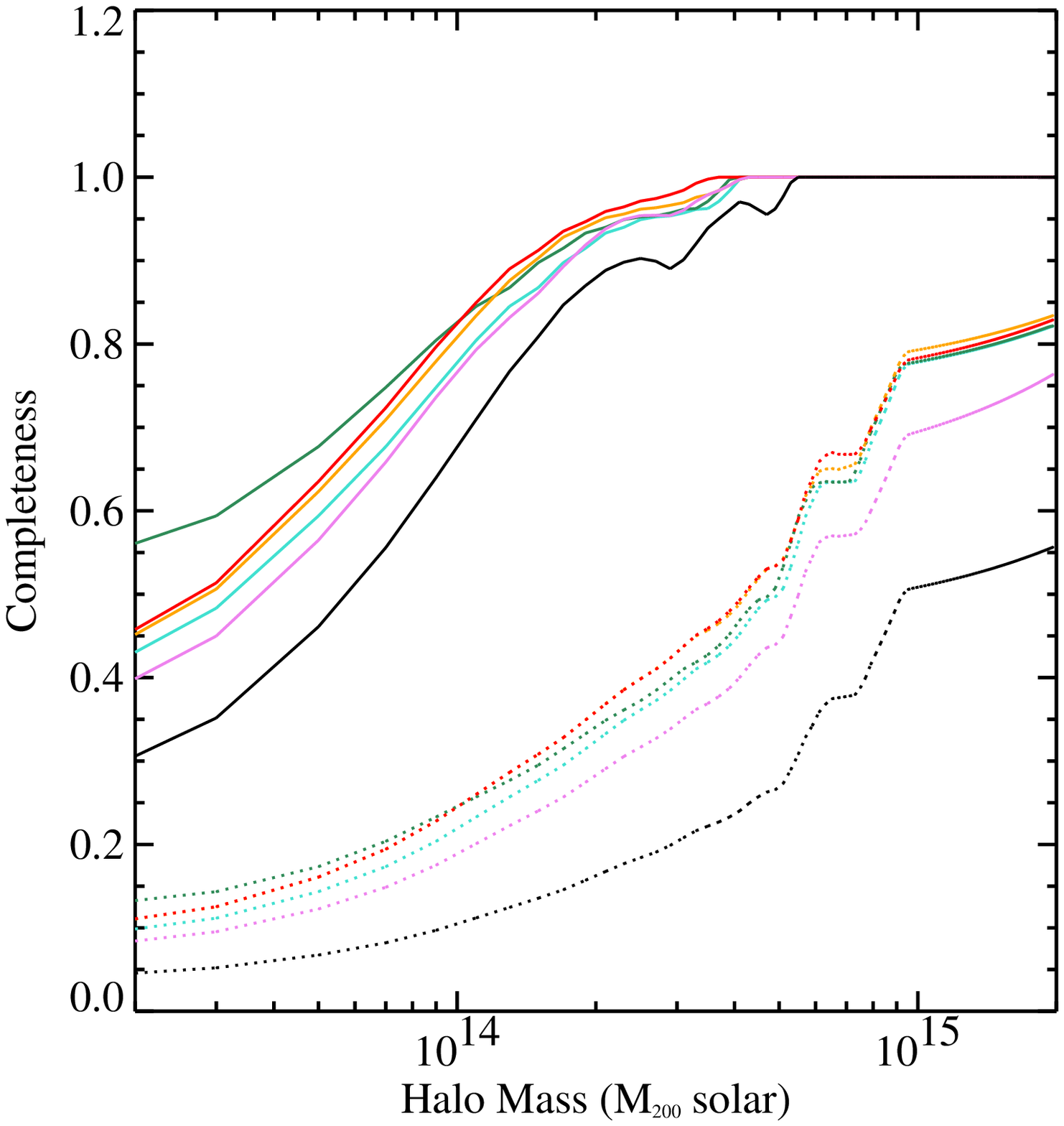}
\caption[]{{\bf Left:}
Purity of the C4 catalog as a function of different color selections.
The black-line (highest purity) comes from using all four colors.
In both panels, the green,
turquoise, orange, and red correspond to using only $u'-g', g'-r', r'-i'$, and
$i'-z'$ only. The purple corresponds to using both $g'-r'$ and $r'-i'$ only.
{\bf Right:}
Measured completeness of clusters found using the C4 algorithm (solid lines).
The dotted lines indicate the completeness measured for a random
selection of clusters of the same sample size as found by the algorithm.
The black lines
(the smallest sample with the lowest true and random completeness) comes from using all four colors.
}
\label{fig:pure_colorx}
\end{figure*}

In Figure \ref{fig:pure_colorx} (left), we show how the purity of the
C4 catalog changes as we add in more color information. The highest
purity comes from using all four colors. As expected, the reddest
color selection $i-z$ does well (even though the $z$-band magnitudes
have greater photometric uncertainties). Combining two colors does
reasonably well and is better than only using a single color. We
conclude that the choice of four colors gives us our highest purity.

We next examine the completeness as a function of our color choices.
In Figure \ref{fig:pure_colorx} (right), we present completeness as a
function of halo mass for the same color selections as used above in
the purity case (solid lines).  The highest purity case (using all
four colors) results in the lowest completeness. At first glance, it
appears that the use of a single or double color criteria produces
better results, with $~10\%$ higher completeness than the four color
case.  This is, however, misleading: the higher formal completeness is
in fact entirely due to random matches (the dotted lines in Figure
\ref{fig:pure_colorx}).  For example, at $M = 1\times10^{14}\,{\rm
M_{\odot}}$, the completeness increases from $\sim$ 70\% for the four
color criteria to 80\% for the single $i-z$ color selection.  At the
same time, the random matches (described below) increase by 10\% and
the purity decreases by $\sim$ 10\%. In other words, the single and
double color criteria have approximately twice as many detected
``clusters'' as the four color criteria, producing a much greater
chance for random matches. Of course, as seen in Figure
\ref{fig:pure_colorx}, a larger fraction of these detected
``clusters'' are spurious. We discuss in more detail the random
matches in the next section.

As a final test, we should mention here that we experimented with
shuffling the colors of the galaxies in the mock catalogs, while
keeping their positions fixed, and re-ran the C4 algorithm. We found
only a few of the closest richest clusters, which again demonstrates
the power of color clustering in 4--dimensions.

\subsection{Random Matches}
\label{false_matches}

These first results raise the issue of the number of random matches
one would expect given {\it any} sample. We quantify random matches by
selecting the same number of clusters as found by the C4 algorithm,
but centered on random galaxies in the mock catalog.  For example, if
we find 934 clusters in the mock catalog using the algorithm, we
select 934 galaxies at random from the same mock catalog and use them
as our cluster centers. We then match these with the dark matter halo
catalog using the same criteria as before.  We show the
``completeness'' from a sample of randomly-placed cluster centers as the
dotted line in Figure \ref{fig:pure_colorx} (right). As expected, the
``completeness'' of the random matches monotonically increases with
cluster mass because the number of clusters as a function of mass
monotonically decreases, while the number of matches remains fixed.
In other words, for a fixed number of random positions, a greater
fraction of rare rich clusters is recovered compared to the numerous
poor clusters.

\begin{figure*}[tp]
\plottwo{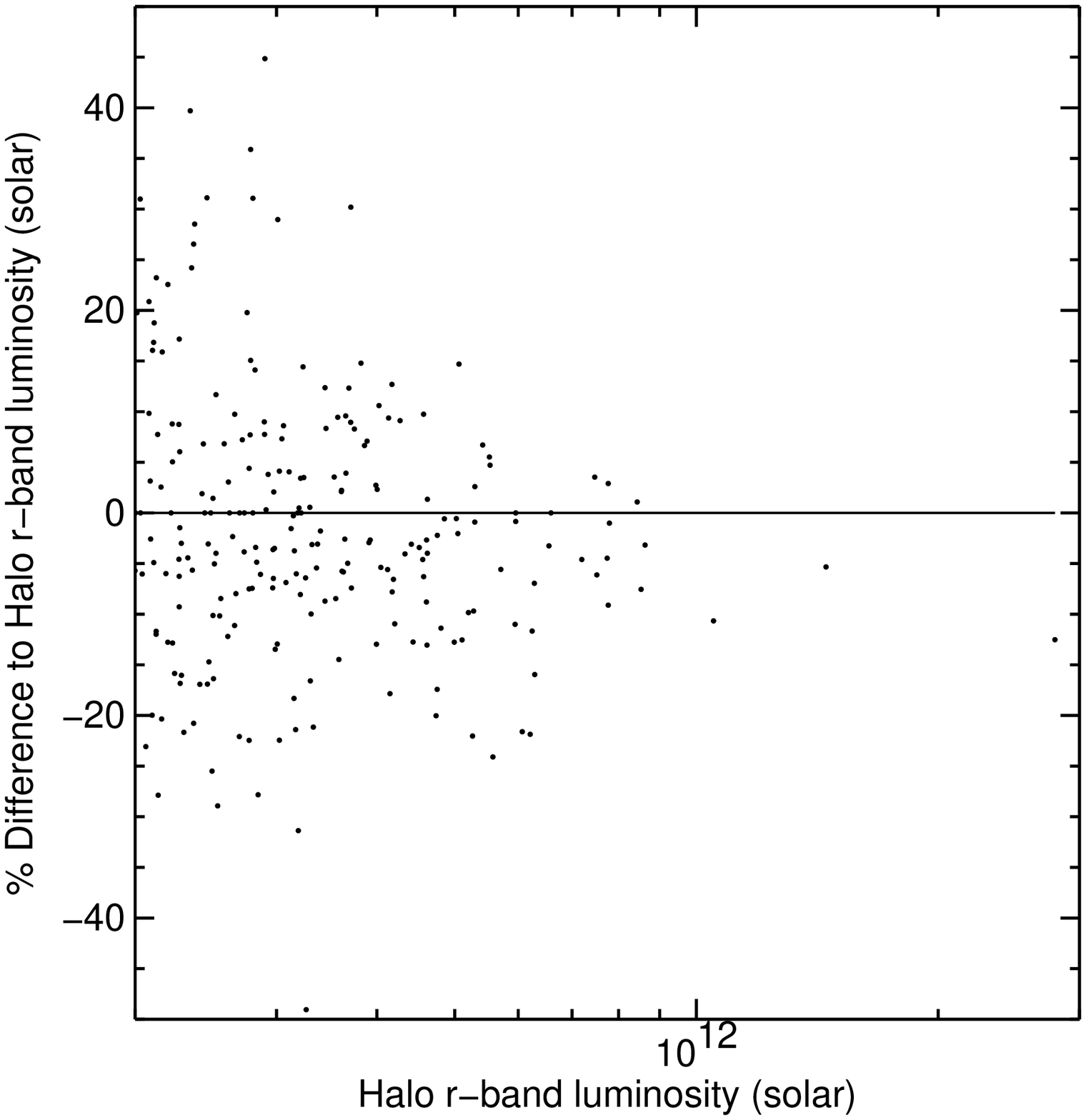}{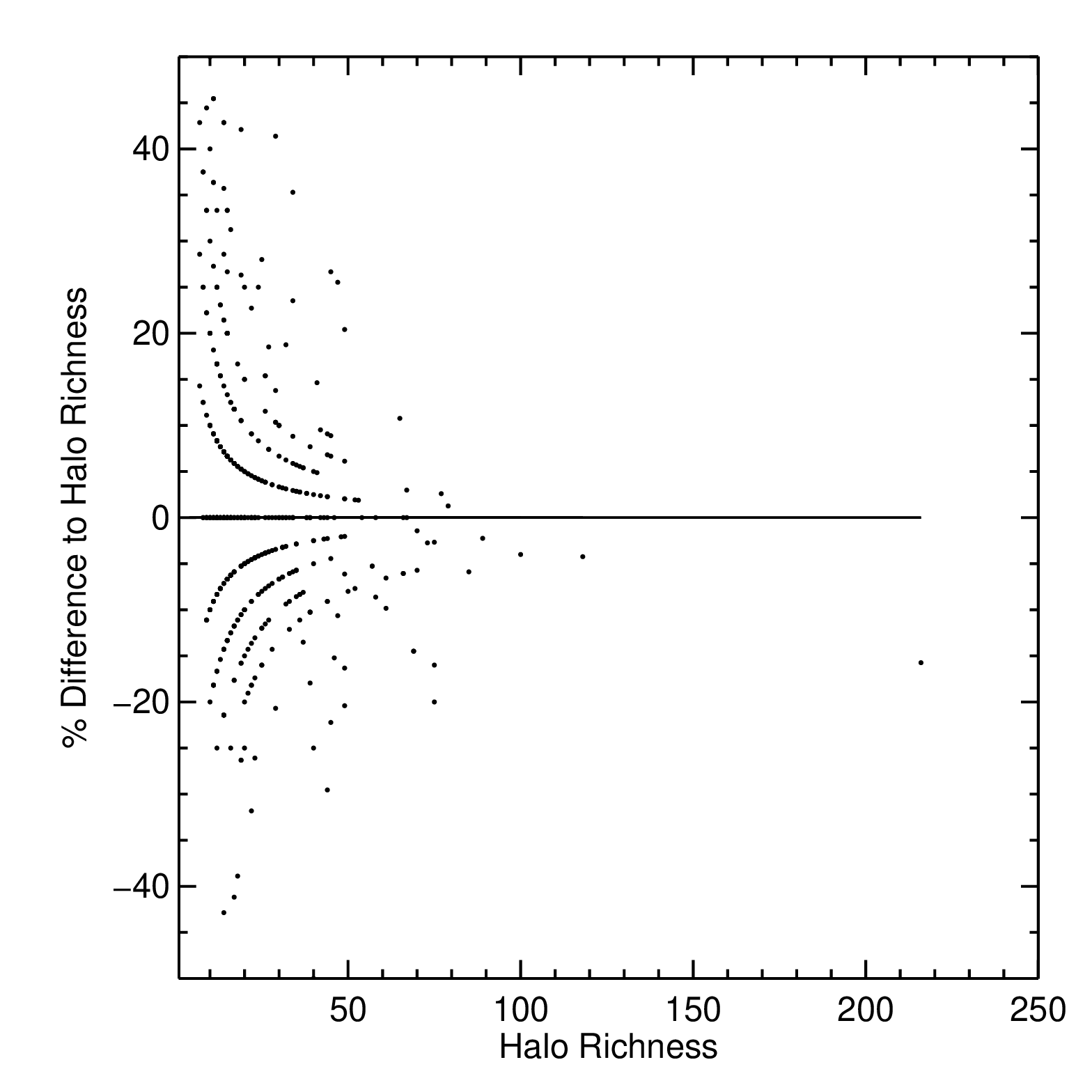}
\caption[]{{\bf Left:} The percentage difference between the measured cluster
optical luminosity compared to the ``true'' halo luminosity. {\bf Right:} Same
as {\bf Left} except for the Richness. The method of measuring
luminosities and richnesses is described in Section \ref{prop}.
For the C4 clusters, we use the
centroid of the found cluster, while for the halos we use the centroid as reported in
the halo catalog.}
\label{fig:lumr_v_lumr}
\end{figure*}

This conclusion is as much a statement of our matching criteria as it
is to one's ability to randomly find clusters. What does not appear in
this analysis is the scatter in the cluster observables at fixed halo
mass due to accidental (i.e. random) matches. In Figure
\ref{fig:lumr_v_lumr}, we show how well we recover the halo
observables after we find C4 clusters in the mock catalogs and match
to the halos.  We show difference between the recovered and the
``true" summed optical luminosities and richnesses (see Section
\ref{prop}). These figures show that we recover the true observables
to typically within 20\%.

Inherent in these analyses is our ability to match halos
to the C4 clusters.
Keep in mind that halos from simulations are
themselves messy, non-spherical systems whose boundaries are dependent
on the exact identification algorithm (Lacey \& Cole 1994; White
2002).  The halo sample we employ is based on a spherical overdensity
approach.  Details of the finding algorithm and the resultant halo
samples are published in Evrard et al. (2002).  A more detailed
exploration of matching clusters to dark matter halos is presented in
W05.

\subsection{The Final C4 Algorithm Parameters}
The parameters of our final algorithm are: (1) an aperture on the sky corresponding
to 1$h^{-1}$Mpc
projected at the redshift of the target galaxy; (2) a redshift box corresponding
to a fixed co-moving $\pm50h^{-1}$Mpc around the target galaxy ; (3) a 4-d color-box of width specified by Equation (5) and
\begin{equation}
[\sigma^{sys}_{ug}, \sigma^{sys}_{gr}, \sigma^{sys}_{ri}, \sigma^{sys}_{iz}]  = [0.6, 0.48, 0.4, 0.4]
\end{equation}
We apply a probability threshold that results in no more than 10\% contamination.

These parameters have been tuned to produce a cluster catalog with the
highest possible purity and similarly high completeness. We note that
Figure \ref{fig:uber} shows that the measured purity and completeness
are very robust to modest changes in the tunable parameters.  The
algorithm is demonstrably robust.

\subsection{Summary of C4 Catalog Purity and Completeness}
\begin{figure*}[tp]
\plottwo{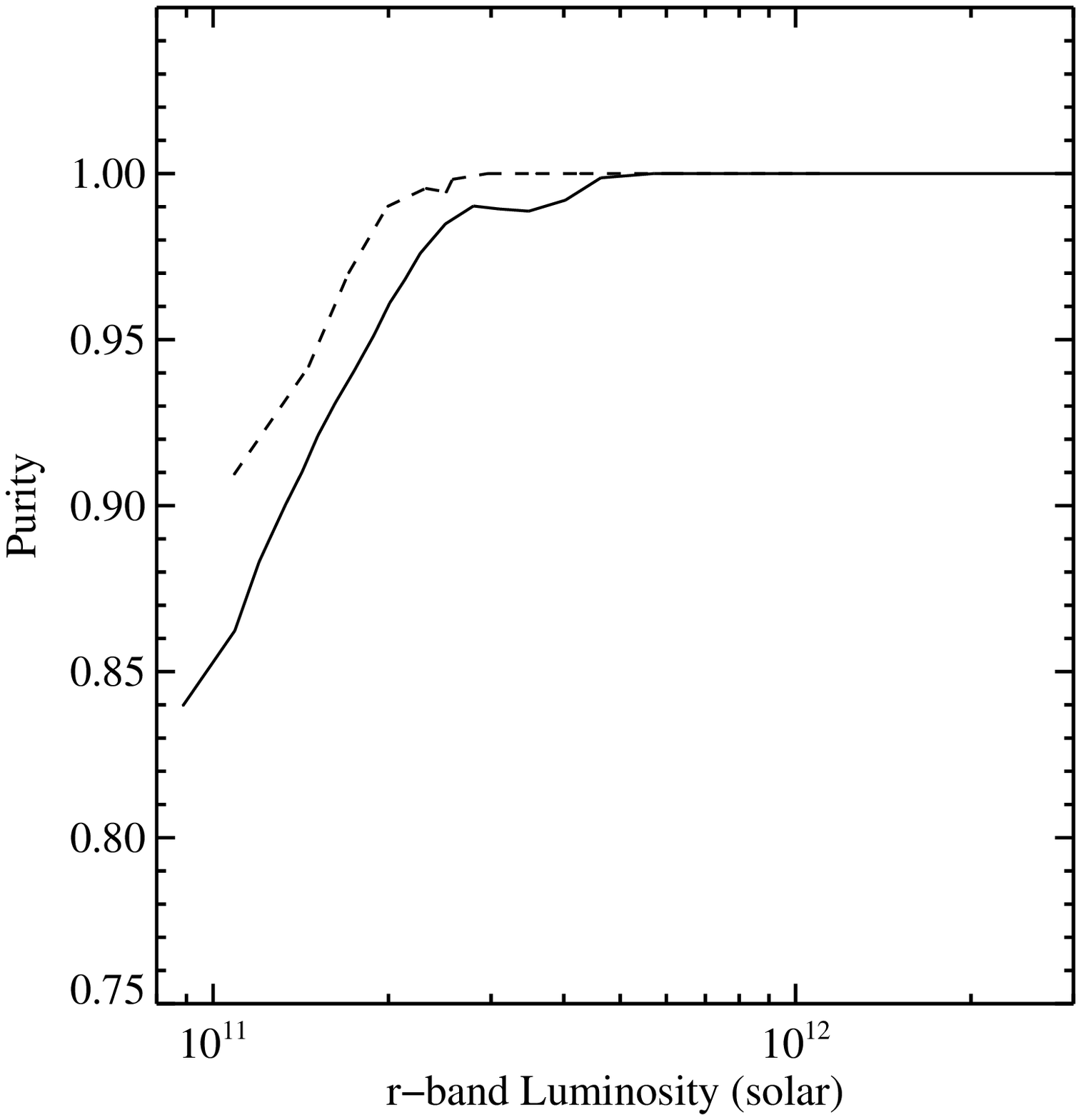}{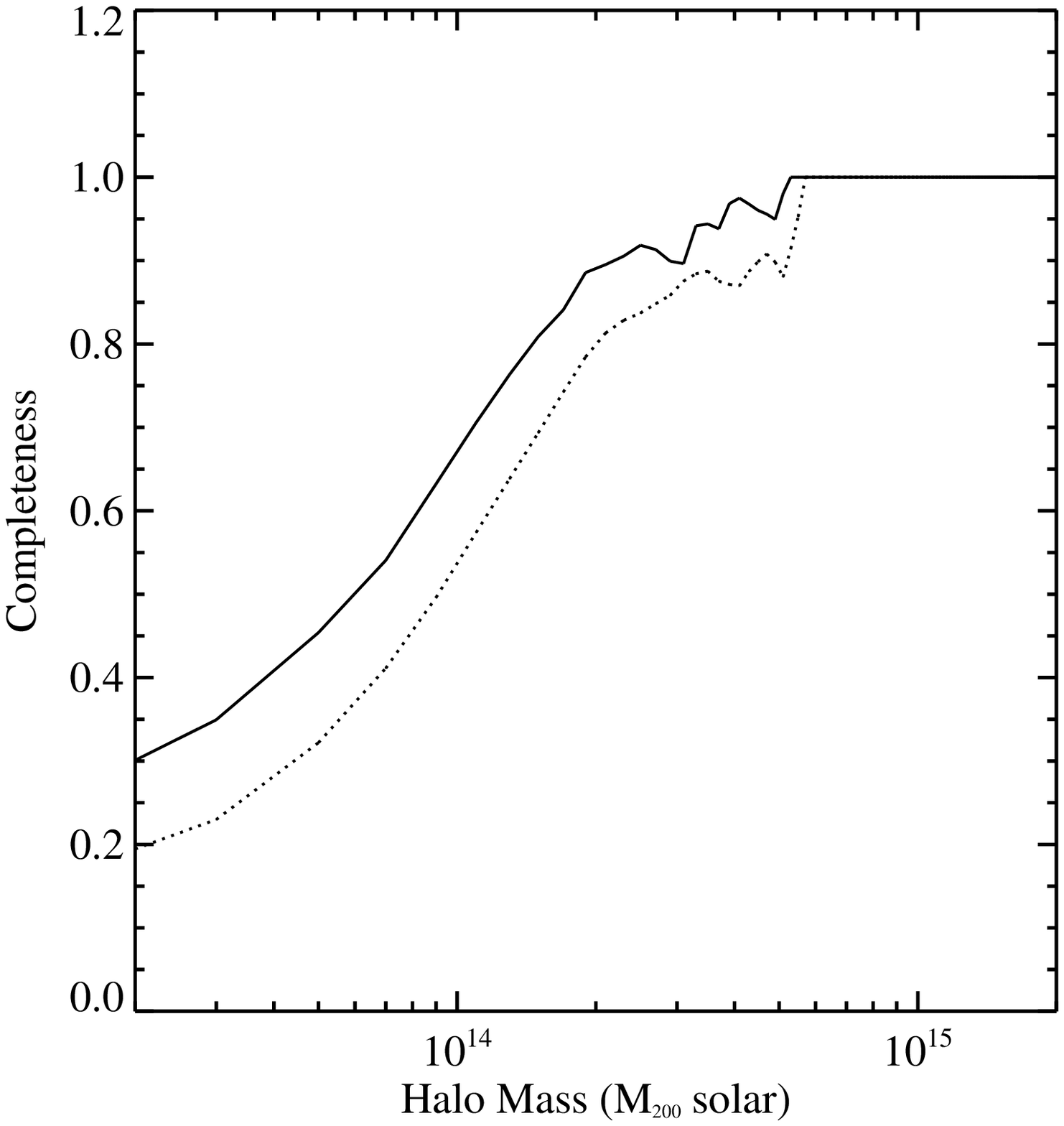}
\caption[]{The final measured purity and completeness of our C4 catalog using the mock SDSS catalog.
The solid is measured before the fiber collision algorithm is applied (see Text). The dotted line shows
the effect missing galaxies due to fiber collisions.}
\label{fig:purity}
\end{figure*}

In Figure \ref{fig:purity} we present the final purity and
completeness of the C4 catalog based on our optimal parameter choices
as discussed above (over all redshifts).  Recall, purity is defined to
be the percentage of systems detected in the mock SDSS catalog, using
the C4 algorithm, and matched to any dark matter halo (more 
massive than $4.5\times10^{13}$) in the HV
simulation.  We also measure the purity as a function of velocity
dispersion, using only those clusters which contain ten or more
galaxies.  We find that our C4 catalog is 100\% pure for such systems
and thus do not present this result in a figure.

As one can see in Figure \ref{fig:purity} (left--solid line), the purity of the C4 sample remains
at 100\% for the most massive systems, dropping down to $\sim 90\%$ for the
remainder. The C4 catalog is more than $99\%$ pure for luminosities larger
than $3\times10^{11}$L$_{\odot}$. 
The high purity of the C4 catalog is a direct
product of our search for clusters in a high--dimensional space.  

In Figure \ref{fig:purity} (right--solid line), we also show the completeness of the
C4 algorithm, as a function of halo mass ($M_{200}$), as selected from
the mock SDSS catalog. This figure demonstrates that the C4 catalog
remains more than $90\%$ complete for systems with ${\rm M_{200}}
\simgreat 2\times 10^{14} {\rm M_{\odot}}$.
Below this mass the
catalog becomes progressively more incomplete and is only 55\%
complete at ${\rm M_{200}} \simeq 1\times 10^{14} {\rm
M_{\odot}}$. The completeness is only 30\% for the lowest mass systems
probed here (${\rm M_{200}} \simeq 2\times 10^{13} {\rm M_{\odot}}$).

\subsection{Questions about the C4 Methodology}

We address here three common questions
raised about the C4 approach. These are:

\begin{enumerate}
\item{Why focus on the photometric data in the C4 algorithm, when the
redshifts (i.e., the 3-d positions in redshift-space) are known?}

\item{Why use all four SDSS colors ($u-g$,$g-r$,$r-i$,$i-z$)? Why not use the
spectra of the galaxies, instead of the broad-band filters?}

\item{Does the algorithm miss clusters with younger stellar
populations?}

\end{enumerate}

\begin{figure*}
\epsscale{0.8}
\plottwo{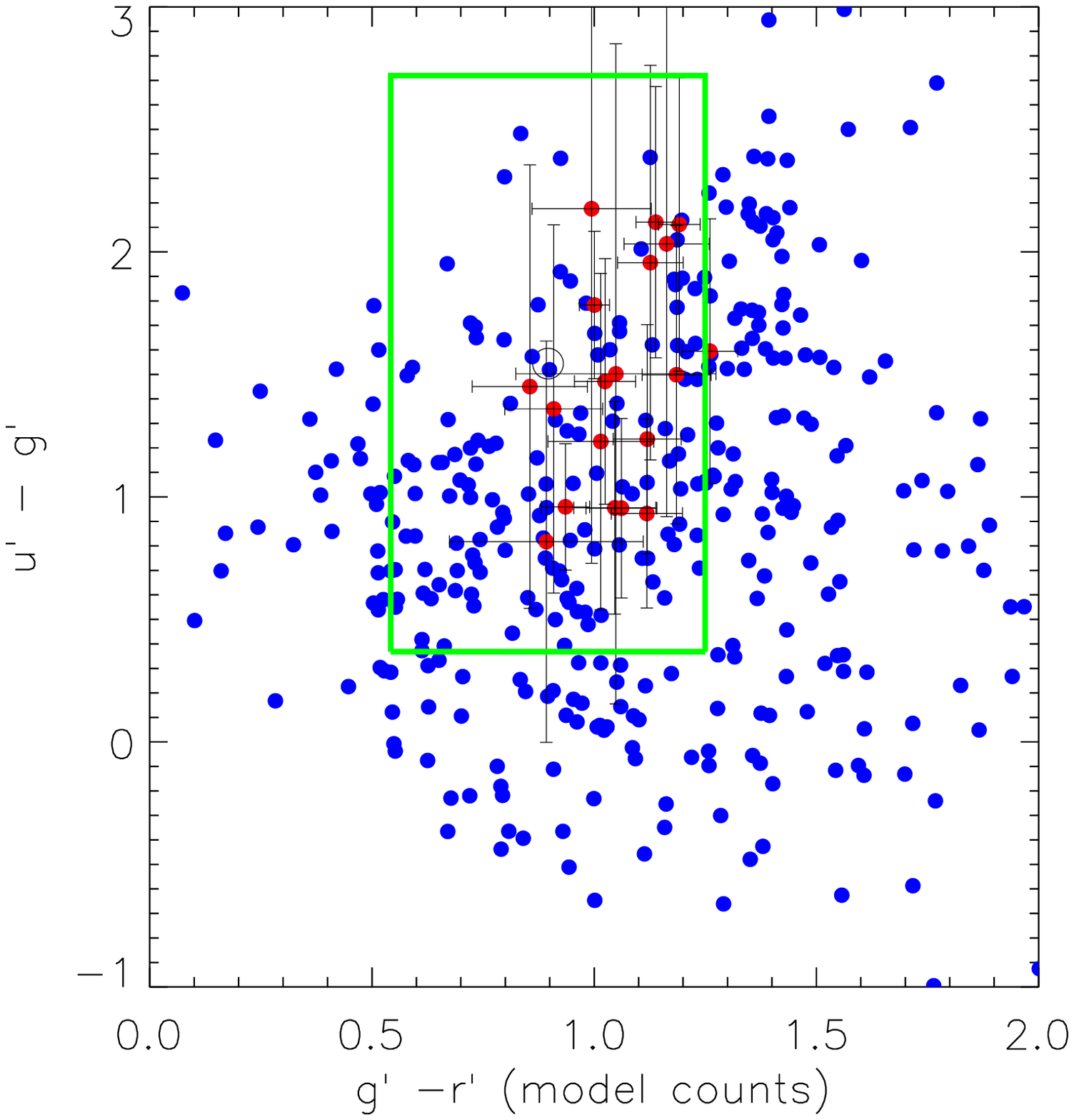}{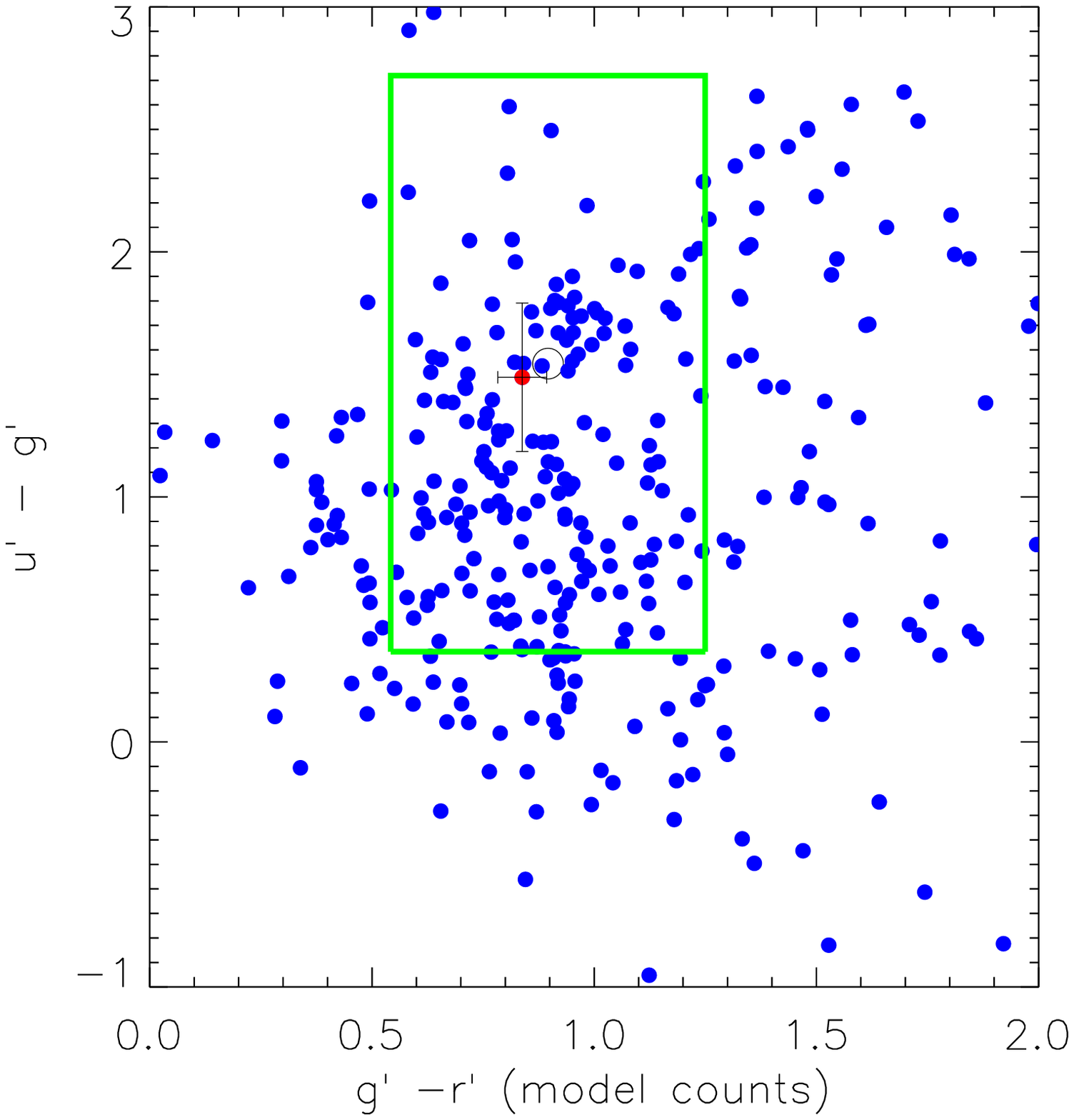} \\
\plottwo{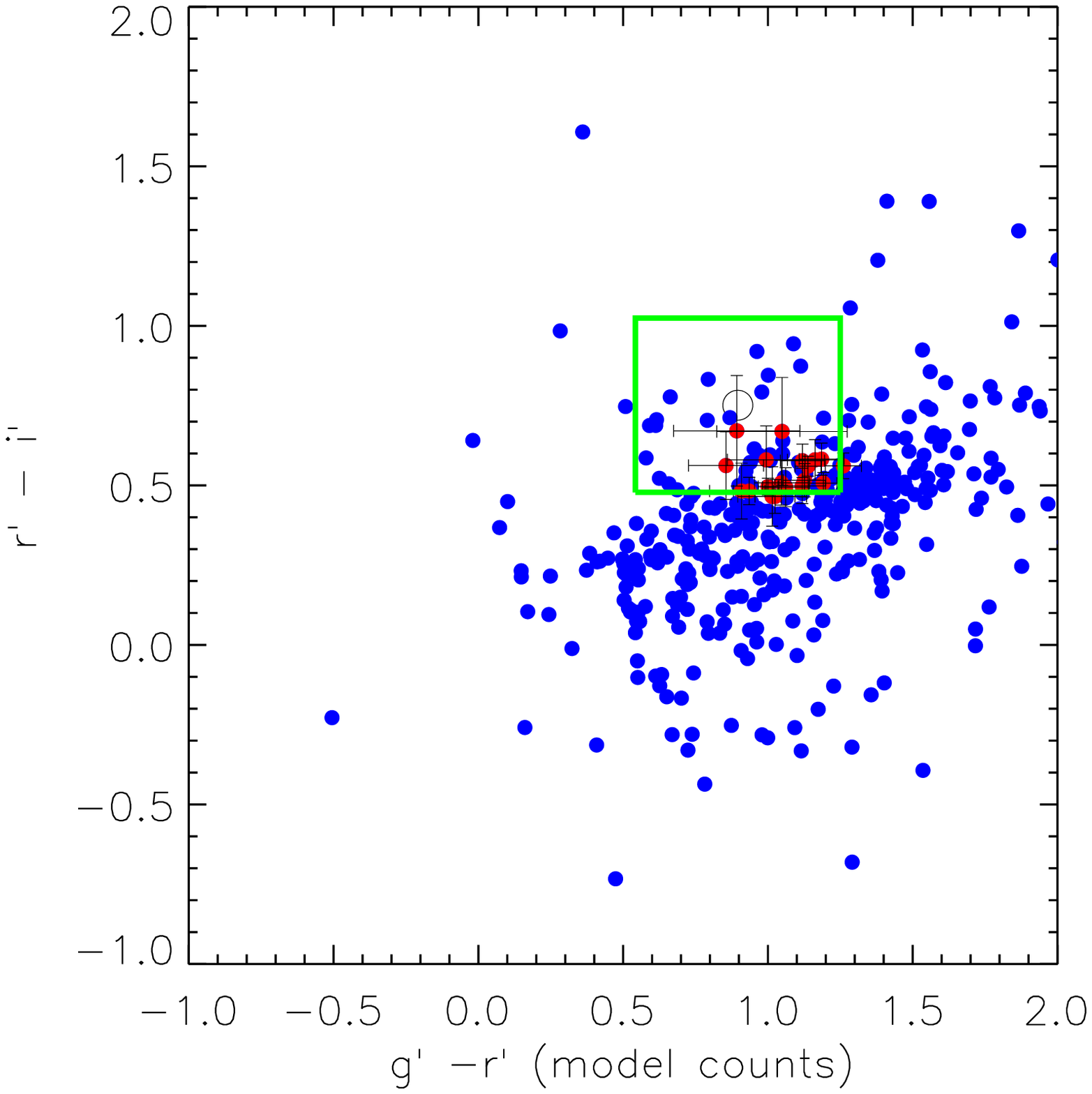}{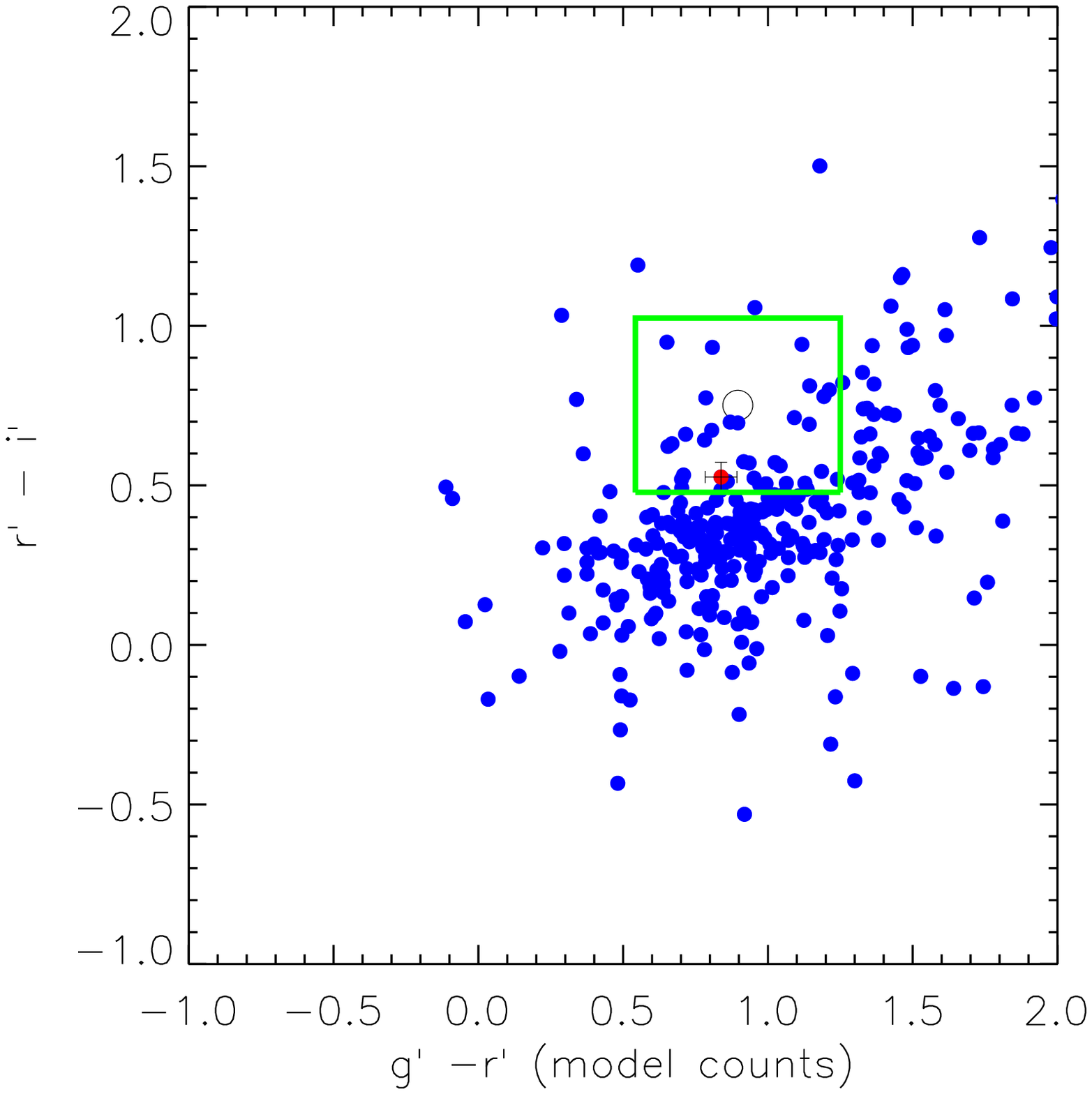} \\
\plottwo{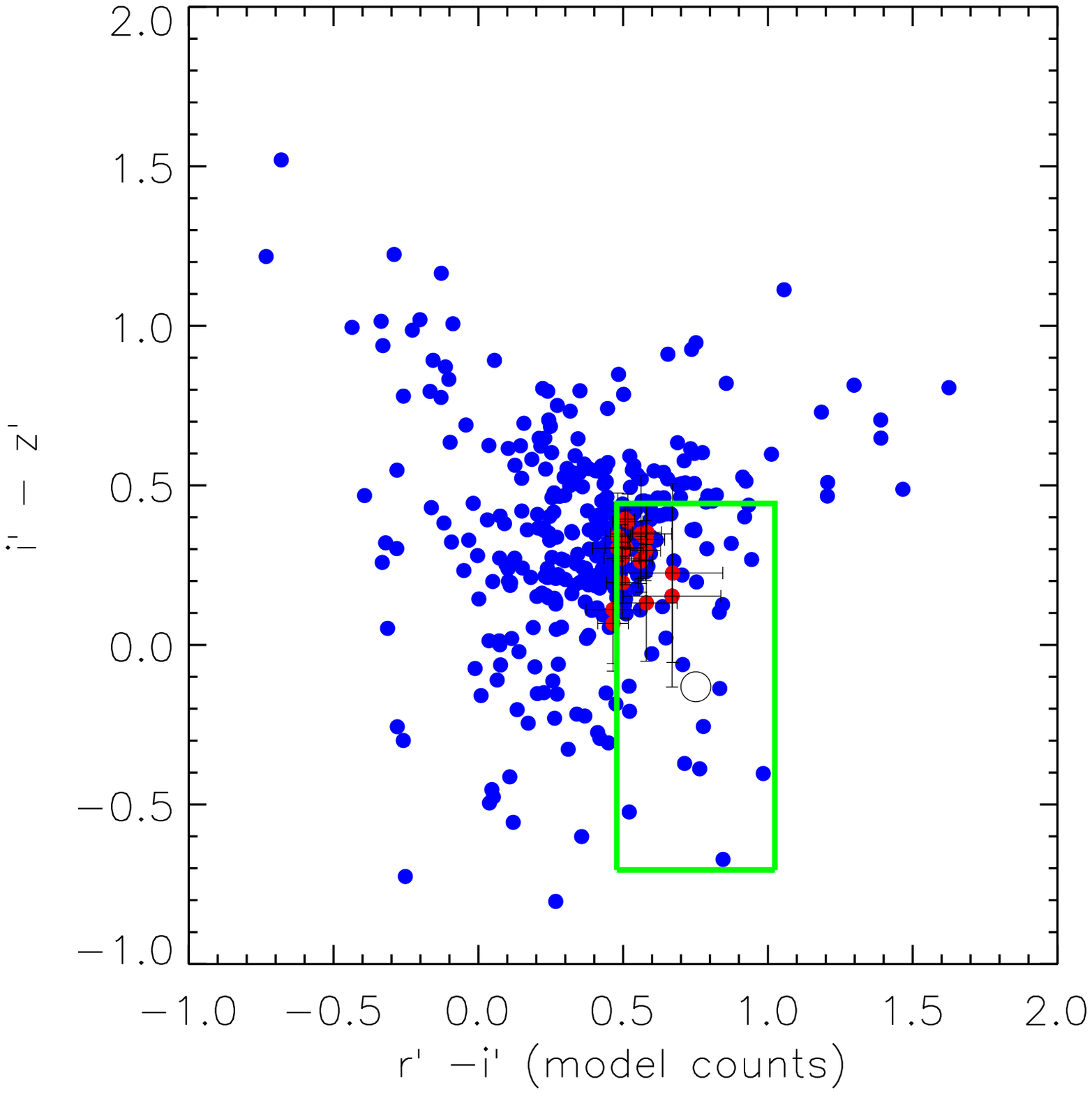}{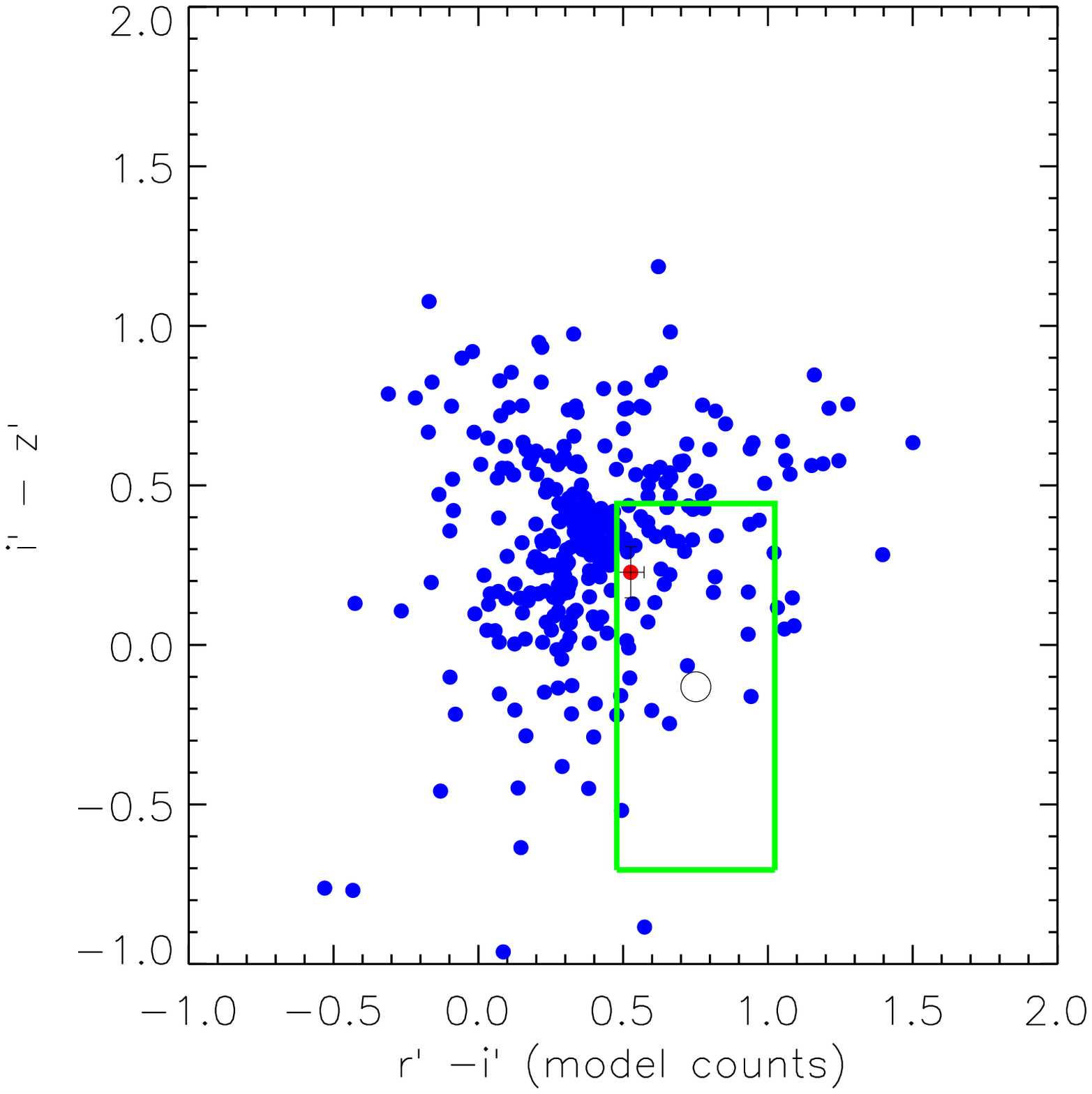}
\caption[]{
The different SDSS color-color planes for an example cluster galaxy
(left) and a randomly chosen field galaxy (right). The blue dots are all
galaxies within the spatial part of our search box (ra, dec, redshift), while
the red dots are those galaxies which are also with the color part of our
7--dimensional search aperture (shown in green).}
\label{fig:colorcolor}
\epsscale{1.0}
\end{figure*}

In Figure \ref{fig:colorcolor}, we address the first question and demonstrate
the power of using the color information in addition to the spatial
coordinates.  Here, we show the projection of the SDSS 7--dimensional search
aperture (4 color and 1 spatial coordinates) onto the different color--color
planes for both a cluster and field region. In Figure \ref{fig:colorcolor}, we
have placed the same size physical aperture over two galaxies: one in a
clustered environment (left panels in Figure \ref{fig:colorcolor}) and the
other in a field-like environment (right panels in Figure
\ref{fig:colorcolor}).  The galaxy on which the color-color plots are centered
is the target galaxy and is identified by the open circle. 

The blue dots are all galaxies within the spatial part of the search aperture.
 Visually, one might be able to detect the spatial clustering
by noticing that there are more galaxies in the
clustered versus the field environment (388 versus 327)
The red dots are galaxies that lie within the color-box in all three figures.
Note that this ``color box'' (green)
is the same, in location and size, for both the cluster and
field environments.  The over-density of galaxies in the clustered
environment now becomes much more apparent --- there are 19 cluster
galaxies (red dots) that have both similar positions and colors to the
target galaxy, while there
remains only one galaxy (with similar colors and position) in the
field-like environment.  This process increases the
signal--to--noise of the cluster over-density (compared to the field
over-density) from 388/327 to 19/1, so that the slight over-density in the
three--dimensional position--space becomes an extreme over-density in the
much sparser seven--dimensional data--space used here. 
Figure \ref{fig:colorcolor} demonstrates the elimination of
projection effects and the strength of color in galaxy clustering.
It also demonstrates the enhancement of the overdensities one can
achieve by using positions {\it and} colors in our clustering algorithm,

To address the second question, we remind the reader of our
analysis in Section \ref{strength}, where we show that four colors
does the better than just one or two. We  also stress that there are physical reasons for
wanting to use all four SDSS colors. For example, the $u$--band is an
excellent measurement of recent star--formation in galaxies (see Hopkins et
al. 2003), and is below the D4000 feature.
Therefore, the $u-g$ color allows us to discriminate between
star--forming and passive galaxies; this is demonstrated in Figure
\ref{fig:fourplot}, where we see a large color difference ($\simeq 1$ mag)
between galaxies with and without strong H$\alpha$ emission in the $u-g$
color-magnitude plot (upper left).  The $g$, $r$, $i$ passbands are the most
sensitive photometric passbands available and therefore, have the smallest
photometric errors, which results in a tight ``red sequence'' in the $g-r$,
$r-i$ color--color plane (see Figure \ref{fig:fourplot}).  Finally, the $z$
passband is useful as it provides the best measurement of the old stellar
population for these low redshift galaxies and is the least affected by
galactic reddening. The larger errors on the $z$--band photometry do not compromise
the C4 algorithm as we take the observed errors into account when constructing
the 7--dimensional search aperture. 

\begin{figure*}
\plotone{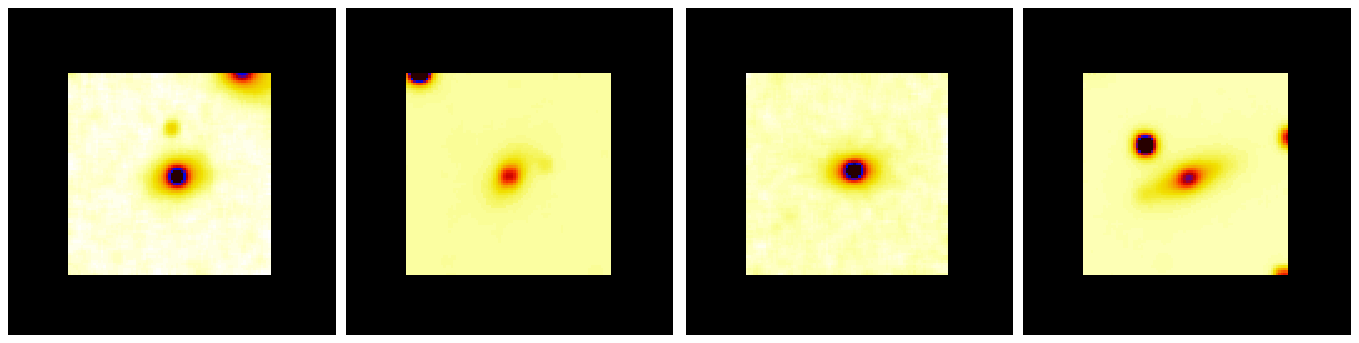}
\plotone{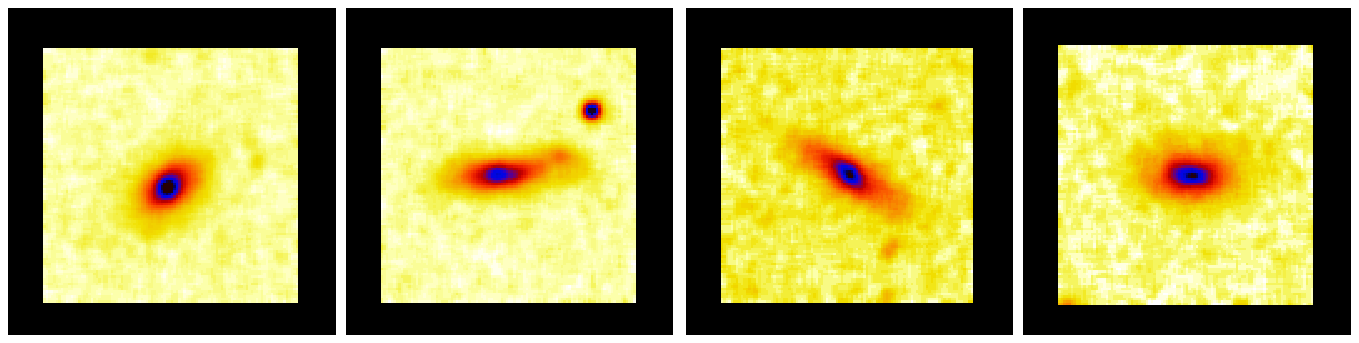}
\caption[]{We show the images of two sets of four galaxies that are clustered in
both position and color. The top four are elliptical galaxies at a redshift of $\sim 0.027$
and all lie within a projected distance of 350$h^{-1}$kpc. The bottom four are galaxies
that are diskier and have a younger stellar population (i.e., they are bluer). The bottom
four galaxies are at a redshift of $\sim 0.04$ and all lie within a circle that is 450$h^{-1}$kpc
in radius. The positions and colors of the galaxies in each of the sets are all very similar
and listed in Table 1.
}
\label{fig:sed}
\end{figure*}

With regard to using the spectra instead of the SDSS colors, we note that the
five SDSS passbands ($u, g, r, i, z$) cover a larger wavelength range than
the spectra. The central wavelengths of the SDSS photometric filters are
3550\AA, 4770\AA, 6230\AA, 7620\AA, 9130\AA~~ respectively, covering a
wavelength range from $\simeq 3300\AA$ to one micron. In comparison, the
spectra only cover a wavelength range of 3900\AA\ to 9100\AA.  From a
computational standpoint, using the spectral data instead of the photometric
colors would require working in a many thousand-dimensional data--space.  Even
with a million galaxy spectra, such a high--dimensional data--space would be
severely under-populated, leading to statistical problems in finding any
clustering in the data. In summary, the 7--dimensional data--space discussed
herein is very effective for our task of finding clusters and groups of galaxies, as
the dimensionality is sufficient to eradicate projection effects while remaining manageable in size.

Perhaps more importantly, the ability to create mock catalogs with
color clustering is currently a challenge (which we think has been met
by the catalogs used herein).
We are still a long way from having mock
catalogs that have galaxies with synthesized spectra that match the
environmental trends seen in the data.  Thus, the colors allow us to
achieve our goals of maximizing completeness and eliminating
projection effects, as tested against the mock galaxy catalogs.

\begin{deluxetable}{ccccccc}
\tablewidth{0pt}
\tablecaption{\bf Galaxy Position and Colors in Figure \ref{fig:sed}}
\tablehead{ 
\colhead{ra} & \colhead{dec} & \colhead{redshift} & \colhead{$u'-g'$} & \colhead{$g'-r'$} & \colhead{$r'-i'$} &
\colhead{$i'-z'$}}
\startdata
\cutinhead{Red Galaxies(Top)} 
219.409 & 3.946 & 0.025 & 5.01 & 0.67 & 0.31 & 0.20 \\
219.481 & 3.984 & 0.029 & 5.38 & 0.75 & 0.37 & 0.22 \\ 
219.778 & 3.999 & 0.027 & 4.00 & 0.72 & 0.37 & 0.22 \\
219.887 & 3.925 & 0.029 & 4.86 & 0.56 & 0.22 & 0.14 \\
\cutinhead{Blue Galaxies (Bottom)}
228.312 & 4.513 & 0.036 & 0.89 & 0.34 & 0.43 & 0.29 \\ 
228.445 & 4.251 & 0.041 & 1.19 & 0.32 & 0.42 & 0.00 \\
228.574 & 4.064 & 0.042 & 1.13 & 0.29 & 0.45 & 0.40 \\
228.277 & 4.195 & 0.037 & 1.09 & 0.45 & 0.40 & 0.12 \\
\enddata
\end{deluxetable}

To address the final question, in Figure \ref{fig:sed} we present two groups
of galaxies found by the C4 algorithm which possess very different galaxy
properties. The first group of galaxies in Figure \ref{fig:sed} (top) contains
galaxies which appear redder, and more elliptical--like --- as expected in a
typical group of galaxies. The second group in Figure \ref{fig:sed}
(bottom) contains much bluer, diskier galaxies.
These groups were both
detected as over--densities of galaxies with similar positions and colors (in
$u-g, g-r, r-i, i-z$) --- thus demonstrating that the C4 algorithm does not
exclude systems dominated with younger stellar populations. The
details of the galaxies in these two groups are presented in Table 1.
Likewise, in Figure \ref{fig:clust_fig2}, we show a cluster comprised of
mostly blue, star-forming galaxies at redshift $z=0.11$. Note the lack of
any prominent E/S0 ridgeline in this system. 
As discussed in Section \ref{overview},
our
algorithm is only insensitive to systems that would contain
spatially clustered galaxies that cover a broad range of spectral types.
However, galaxy types are not broadly classified, but bi-modal (spirals or elliptical, star-forming
or passive). Thus, to first order, every cluster will contain at least 50\% of
one of the two major types of galaxies and the algorithm will find such
color clustering.

\section{Measured Cluster Properties}
\label{prop}

For each C4 cluster we measure a set of quantities which include
the cluster centroid, the velocity dispersion, and the summed
$r$-band luminosity. In addition, we characterize the substructure
and local large-scale structure of each cluster.

\subsection{Cluster Centroids}
\label{centroid}
We measure three different cluster centroids: (1) the peak in the
C4 density field, (2) the luminosity weighted mean centroid, and (3)
the position of the brightest cluster galaxy. Method (1) was discussed in 
Section \ref{clustids}. Method (2) uses all galaxies within 1$h^{-1}$Mpc
of the initial centroid (Method 1) and within four velocity dispersions
(see next section). We then calculate an $r$-band luminosity weighted center.
Method (3) attempts to identify the brightest cluster galaxy (BCG). The BCG
is taken as the brightest galaxy within 500$h^{-1}$kpc of the initial
centroid (Method 1), within four velocity dispersions (see next section),
and which has no strong H${\alpha}$ emission ($<4\AA$). We then report the position
of the BCG.

The cluster redshift we report is determined via the bi-weighted
statistic of Beers et al. (1990). We use only those galaxies within 1$h^{-1}$Mpc
of the initial centroid and within $\Delta z = 0.02$ of the peak in the redshift
histogram as described in the next subsection. 

\subsection{Velocity Dispersions}
To calculate the velocity dispersion of galaxies in our clusters, we
perform an iterative technique based on the robust bi-weighted
statistic of Beers et al. (1990). Having defined the sky-positional
centroid of each cluster, we construct a redshift histogram of all
galaxies (regardless of their C4 classification) within a projected
radius of $1h^{-1}$ Mpc. We then search this histogram for a peak and
tentatively identify this peak as the velocity center of the cluster
(as a lower limit, this peak must contain at least three galaxies
within 1000 km/s of each other). We iterate by only keeping galaxies
within $1.5h^{-1}$ projected Mpc of the cluster centroid and within
$\Delta z = 0.02$ of the velocity center defined above. We then
compute the bi-weighted mean recessional velocity for these galaxies
and measure their bi-weighted velocity dispersion, $\sigma_v^{est}$.

We stress that the above procedure is performed on all available
galaxies in the SDSS spectroscopic sample. For consistency, we check
that each cluster contains a certain number of C4 galaxies, and reject
those clusters where this fraction is below 10\% of all galaxies
within 1h$^{-1}$Mpc and $\Delta z = 0.02$. Only a small fraction (2\%)
of clusters originally identified using the C4 galaxies fail this test
and are not included in the final cluster sample.
 We also reject a small number of clusters that do
not contain enough galaxies to measure an accurate velocity dispersion. 
At least eight galaxies are required to define the velocity dispersion,
consistent with previous limits to get a reliable value
(see Collins et al. 1995). Note that most
(80\%) of our clusters in the real SDSS data contain $\ge 20$ galaxies with
which to measure the velocity dispersion. Clusters which do
not meet this criteria are removed from the main sample. 
To get our final velocity dispersion measurements ($\sigma_v$), we
re-calculate it for each cluster using only galaxies within four times
the estimated velocity dispersion ($\sigma_v^{est}$) discussed
above. This is similar to the standard sigma-clipping method used in
the literature. The accuracy of these final measurements, 
which are in the observed reference frame, are discussed below.

\subsection{Summed Optical Cluster Luminosity}
To calculate the total summed $r$--band optical luminosity for each
cluster, we convert the apparent magnitudes of all cluster members
into optical luminosities, using the conversions in Fukugita et
al. (1996), and sum them. All magnitudes are $k$-corrected according
to Blanton et al. (2003a) and extinction corrected according to
Schlegel, Finkbeiner, and Davis (1998). Cluster membership is confined 
to galaxies within $4\sigma_v$ in redshift--space and a projected radius of
$1.5h^{-1}$Mpc on the sky. 
The SDSS main galaxy sample is designed to be complete ($>95\%$) to $m_r = 17.77$ and 
the C4 cluster sample is complete ($> 90\%$) to $z =0.11$. Thus, to minimize effects from the SDSS selection
function, we use an absolute magnitude limit of $M_r < -19.9$, which
is an apparent magnitude of $m_r = 17.8$ at $z \simeq 0.11$, when measuring
the optical luminosities. Clusters beyond $z = 0.11$ will need to have their
optical luminosities corrected for this incompleteness.

\subsection{Structure Contamination Flag}
We define a ``Structure Contamination Flag'' (SCF) to measure the
degree of isolation in redshift--space for each cluster.
Specifically, we examine radial variations of the velocity dispersion
for each cluster, noting large radial variations from the mean
velocity dispersion.  Clusters which are embedded in surrounding
large-scale structure can have significant velocity contamination. SCF
increases with increasing standard deviations in the velocity
dispersion profiles.  We calculate the bi-weight velocity dispersion
within 500,1000,1500,2000 and 2500$h^{-1}$kpc radial bins as described above, and
determine the standard deviation. We then assign an SCF based on the
ratio of the standard deviation of the dispersions over the mean of
the velocity dispersions. We use three bins, SCF = [0,2], in
approximating the bottom, middle, and top thirds of the distribution
of the ratio.  A cluster with SCF $=0$ has a ratio less than
15\% whereas SCF $=2$ has a ratio $> 30\%$.  The sky plot and velocity
profile are shown for a real SDSS cluster with a high SCF=2 cluster in
Figure \ref{fig:clust_fig1}, \ref{fig:clust_fig3}, and
\ref{fig:clust_fig4}.  Notice that the velocity dispersion as a
function of radius is highly erratic, producing a large standard
deviation about the mean. Figures \ref{fig:clust_fig3} and
\ref{fig:clust_fig4} show two clusters separated by less then two
tenths of one degree and $\Delta z = 0.01$. Notice that the velocity
dispersion profile increases systematically as the galaxies from the
neighboring system are picked up with increasing aperture.  These
clusters both have SCF = 2.  A clusters with SCF = 1 is shown in Figure
\ref{fig:clust_fig2} and with SCF = 0 in  \ref{fig:clust_fig5}. Notice here that the
velocity dispersion profiles are nearly constant.  This SCF flag does
not necessarily quantify the substructure of the main cluster, but
rather identifies clusters whose velocity dispersions may be
inaccurate because of nearby large-scale structure.

\subsection{Dressler-Shectman Statistic}
In addition to quantifying the local large--scale structure for each cluster,
we can use the Dressler-Shectman substructure statistic to search for local
differences in a cluster's mean recession velocity and velocity dispersion.
For each cluster member (within 1.5$h^{-1}$Mpc and $4\sigma_v$), we compute a local recession
velocity and local velocity dispersion using the ten nearest neighbors to the
galaxy which are also within  $4\sigma_v$ of the cluster redshift. We then compute the difference
between these local quantities and the mean recession velocity and velocity
dispersion for the whole cluster.  The cumulative differences are then used as
a measure of the cluster substructure. To compute the significance of this
measurement, we shuffle the galaxy velocities within the cluster and repeat
the exercise 1000 times. Using these Monte-Carlo realizations, we calculate
the probability that the observed cumulative differences would be obtained at
random, given the spatial positions of the galaxies. A low probability
indicates that the substructure is significant. For more details, see Dressler
and Shectman (1988) and Oegerle and Hill (2001).

\section{Scaling Relations and their Scatter}
\label{scaling}

Any galaxy cluster catalog will have observables that can be related
to the underlying halo dark matter mass. Typically, researchers have
used some sort of galaxy number count, i.e., richness (Abell 1958, Yee
and Ellingson 2003). While the C4 clusters certainly have galaxy
number counts, we also measure the summed $r$-band luminosity of the
galaxies within each cluster. We additionally measure a velocity
dispersion for all clusters within our spectroscopic data (using 8 or
more galaxy members within a projected radius of 1$h^{-1}$Mpc and
within four velocity dispersions).  In this section, we determine
which cluster observables scale best with the halo masses in the mock
galaxy catalogs. We stress that this section does not {\it quantify}
any absolute scaling-laws (or their scatter), which requires a
detailed analysis of the role of cosmology and of the sensitivity of
this scaling to the halo occupation.  This will be presented in an
upcoming paper.  Here, we simply study how the scatter changes as we vary
parameters of the cluster finding algorithm and measures of the local
foreground/background contamination.  In short, the scaling relations presented
in this section are used solely to guide our choice of the best
cluster observable when relating to mass, as qualified by the scatter,
and we caution the reader not to use them to draw
cosmologically-dependent conclusions.

\subsection{Structure Contamination and Velocity Dispersion}

\begin{figure*}
\plotone{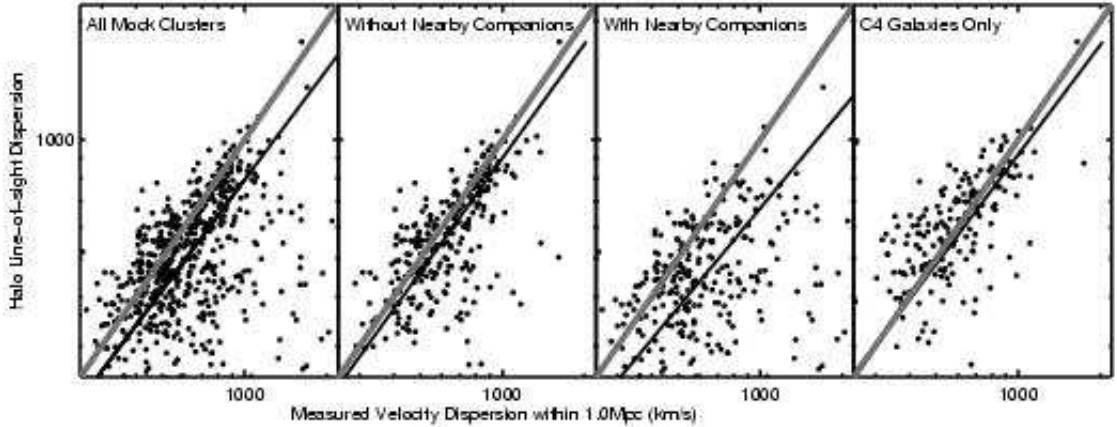}
\caption{
Using the mock clusters, we compare our measured velocity dispersions
to the line-of-sight velocity
dispersions measured directly on the halo-particles in the HV Simulations.
The thick grey line is the one-to-one line, while the black
line is the best fit when SCF = 0. We use only clusters with 10 or more
members within a projected radius of 1$h^{-1}$Mpc and within 4$\sigma_v$.
The best fit relation occurs when SCF=0 where the rms scatter is $\sim 15\%$.}
\label{fig:biwt1000_comp_allplots}
\end{figure*}

For virialized systems, the velocity dispersion is an obvious
choice when attempting to measure the mass of a cluster. So first,
we examine the validity of our method to recover the velocity dispersions. 
In Figure \ref{fig:biwt1000_comp_allplots}, we
compare our measured velocity dispersions for C4 clusters in the mock catalog
against the known particle line-of-sight velocity dispersions for the halos
in the simulations. We use only clusters with 10 or more galaxies within 1.0$h^{-1}$Mpc
and within 4$\sigma_v$.
We find an excellent one-to-one agreement with $\sim 15\%$ scatter (using
only those mock clusters with SCF = 0.
This scatter doubles if we use all
clusters. Therefore, the structure contamination flag is essential in 
recovering the true velocity dispersions of the systems. We also show the comparison
when we use only the C4 galaxies to measure the velocity dispersion. Notice that
the C4 galaxies are not strongly affected by local structure contamination. However,
there are fewer systems with enough ($>$10) galaxies to measure an accurate
dispersion. 

\begin{figure*}[]
\plotone{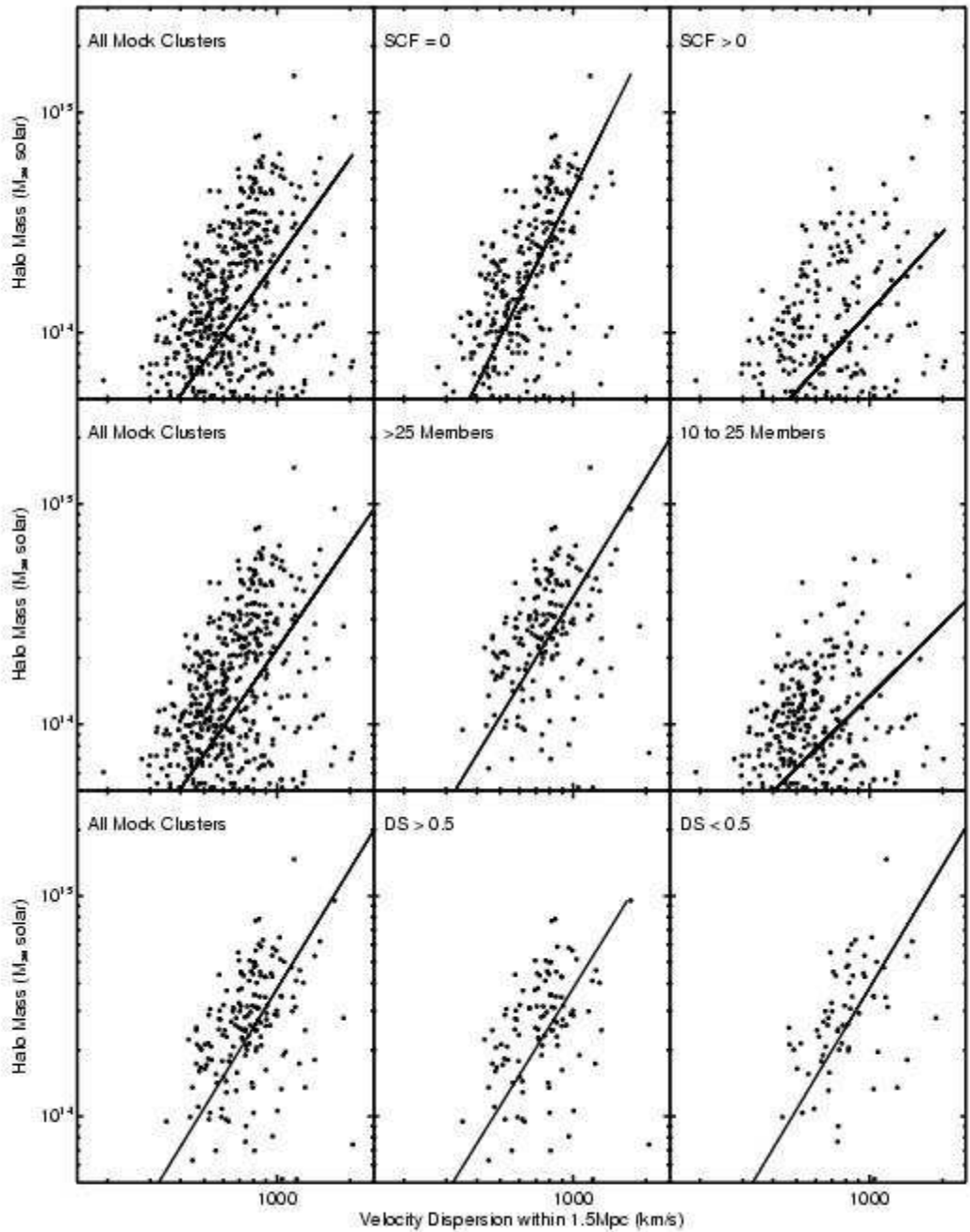}
\caption{
The relationship between  M$_{200}$ and the velocity
dispersion as a function of different methods of contamination.
{\bf Top:} The Structure Contamination Flag (SCF--defined in the text). {\bf Middle:} The number
of galaxies used to measure the dispersion. {\bf Bottom:} The Dressler-Shectman statistic.}
\label{fig:biwt1500_allplots}
\end{figure*}

In Figure \ref{fig:biwt1500_allplots} we present the correlation
between halo mass and the measured velocity dispersions of C4 clusters in the
mock SDSS catalog. We present the best (robust-fit) linear relationship
between these two physical quantities.  Here, there is also significant asymmetrical
scatter about this relationship, with many low mass systems possessing an
apparently high velocity dispersion.  Part of this asymmetry and scatter is
due to the number of galaxies used to measure the velocity dispersion. In
Figure \ref{fig:biwt1500_allplots} (middle) we split the sample into clusters with less
than 25 members and more than 25 members within 1.5$h^{-1}$Mpc. As expected,
clusters with a larger number of members have higher mass halos, but we also
observe that the scatter decreases by $\sim 50\%$ for the regression when only
the high-number systems are used.  Unfortunately, if we place a cut on the
number of cluster members to reduce the scatter in this scaling law, we also
constrain ourselves to only the higher mass systems.

As an alternative to using only clusters with the most galaxies for accurate
velocity dispersions, we plot in Figure \ref{fig:biwt1500_allplots} (top row),
the relationship between $M_{200}$ and velocity dispersion, but now 
separated on 
the Structure Contamination Flag (SCF).  The scatter in the relation is reduced by
a factor of two after we include only those clusters with SCF = 0.  Additionally,
the scatter is reduced over the whole mass range of the clusters.  This
demonstrates that a majority of the scatter in Figure
\ref{fig:biwt1500_allplots} is due to ${\rm SCF}>0$ systems, which by
definition have a known companion within $1.5\,h^{-1}\,{\rm Mpc^{-1}}$ of the
main cluster.  The end result is that velocity dispersions are typically
over-estimated when there is nearby large-scale structures. It may further
indicate that these clusters are not virialized systems due to a recent, or
on--going, merging event. Figure \ref{fig:biwt1500_allplots} (top) also shows that
for the ${\rm SCF}=0$ systems, one can obtain a tight, linear relationship
between $M_{200}$ and velocity dispersion.

Finally, we also show in
Figure \ref{fig:biwt1500_allplots} (bottom row) how the Dressler-Shectman (DS) statistic
alters the scatter in the relation between halo mass and velocity dispersion.
Here, we only use those clusters with 25 or more member galaxies to ensure
that the DS statistic is not dominated by Poisson noise. The
DS statistic does not seem to help reduce the scatter in the relation.
Since the DS statistic is really trying to measure internal velocity substructure
in clusters, this result may imply the substructure is not strongly correlated
with mass. Note that the SCF flag tends to find structure {\it outside} the cluster,
since it looks for variations in the velocity dispersion in increasing
radial bins (all the way out to 2.5Mpc beyond a cluster's center). Thus, the
SCF and the DS statistic are mapping contamination in different ways.

\subsection{Structure Contamination and Richness}

\begin{figure*}[]
\plotone{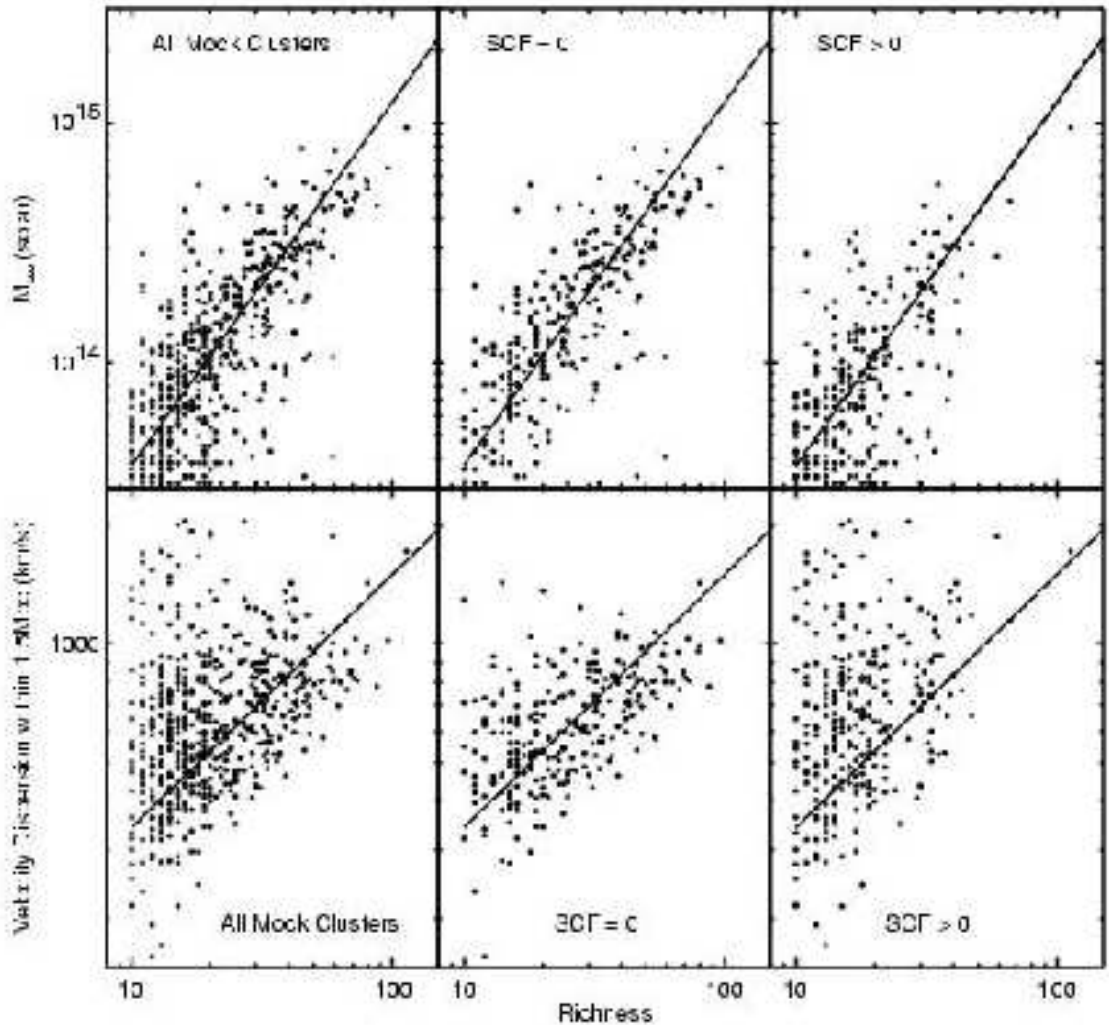}
\caption{
Using the mock clusters, we plot the relationship between the richness as measured with
1$h^{-1}$Mpc and for members within $\pm4$ velocity dispersions of the cluster,
against halo mass ({\bf top}) or velocity dispersion ({\bf bottom}). We show the effect of
the Structure Contamination Flag (SCF). The best-fit
relation is for all clusters without a nearby companion.}
\label{fig:wbiwt_allplots}
\end{figure*}

Richness (or galaxy member counts) is an often used measure
of cluster mass (see Yee and Ellingson 2003 and references therein). We examine multiple Richness measures,
including counts within metric radii (500,1000,1500,2000,2500 $h^{-1}$kpc)
and counts of only C4 galaxies (all to an absolute magnitude limit
of $M_r = -19.8$ or L$^*$). We find that the richness measured within 1 or 1.5$h^{-1}$Mpc 
to have the lowest scatter against halo mass.
Since we are working with spectroscopic data (or mock spectroscopic data),
we also apply a constraint that members be within $\pm4\times$ the cluster
velocity dispersion. Thus, our richnesses are not affected by 
foreground/background projection, which is typically the case
in optical cluster catalogs. In Figure \ref{fig:wbiwt_allplots}, we show how
halo mass scales with richness in the mock
catalogs.  Note that the scatter in mass versus richness is comparable
to that seen in mass versus velocity dispersion (Figure \ref{fig:biwt1500_allplots} top)
when SCF = 0.
It is also worth noting that when the Structure Contamination Flag is ignored,
the scatter increases by a third. 

\subsection{Structure Contamination and Summed Optical Luminosity}

While Richness (as defined here) is a fixed, integer quantity, the summed
$r$-band optical luminosity of a cluster takes into account any environmental 
dependence in luminosity, even by galaxy type (Hogg et al. 2003; Baldy et al. 2004)
We examine multiple luminosity measures,
including those within metric radii (500,1000,1500,2000,2500 $h^{-1}$kpc), using galaxies brighter
than $M_r = -19.8$ (L$^*$). We find that the cluster luminosity measured within 1 or 1.5$h^{-1}$Mpc
to have the lowest scatter against halo mass.
Since we are working with spectroscopic data (or mock spectroscopic data),
we also apply a constraint that members be within $\pm4\times$ the cluster
velocity dispersion.

\begin{figure*}[]
\plotone{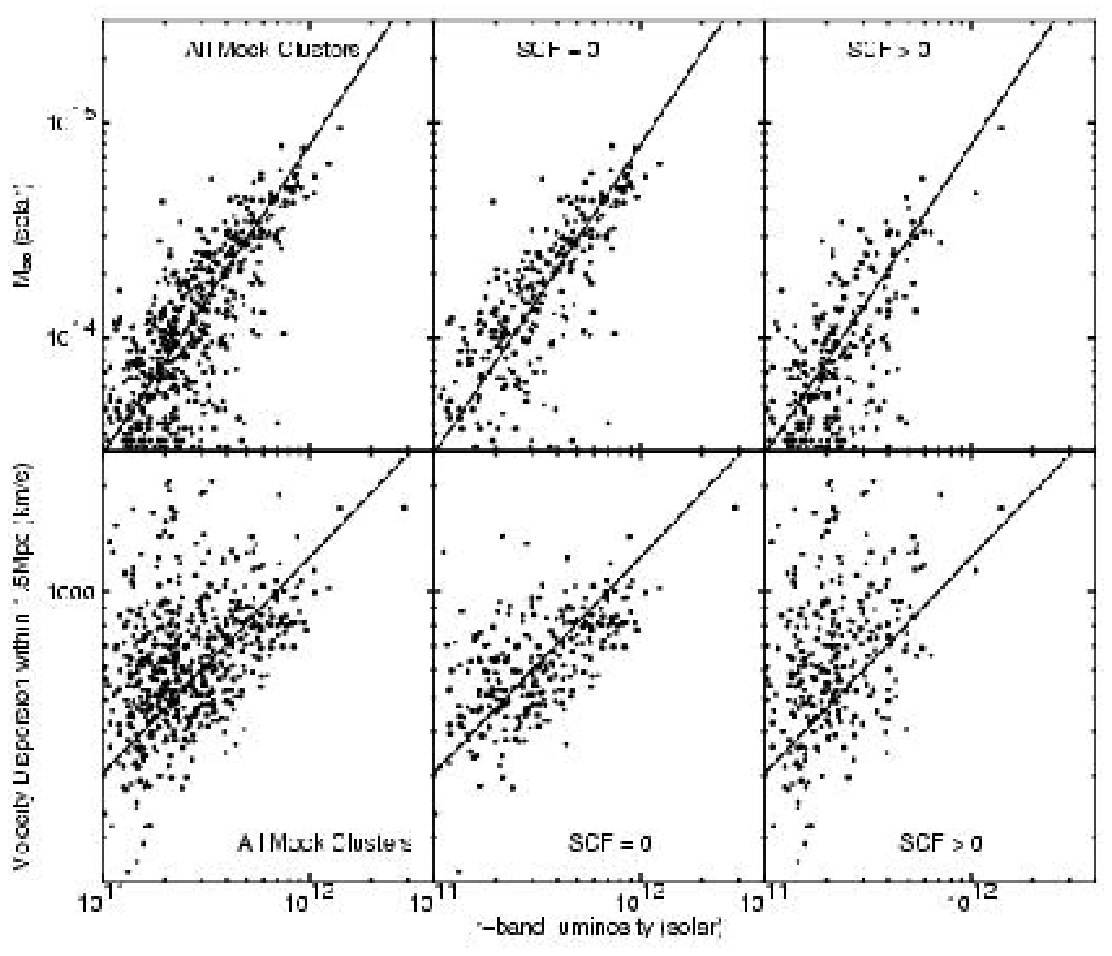}
\caption{
Using the mock clusters, we plot the relationship between total $r$-band cluster luminosity
and the halo mass ({\bf top}) or velocity dispersion ({\bf bottom}), as a function of the
Structure Contamination Flag (SCF). The best-fit
relation is for all clusters without a nearby companion
Note that unlike velocity dispersion, the scatter in mass versus luminosity is not
significantly reduced when only clusters with SCF = 0 are
used. Additionally, the scatter in the mass-luminosity relation without cuts is already as small as in
Figure \ref{fig:biwt1500_allplots} (top-middle) with SCF = 0.}
\label{fig:lumr_allplots}
\end{figure*}

In Figure \ref{fig:lumr_allplots} (top), we show the halo mass as a
function of the summed cluster $r$-band luminosity (within 1$h^{-1}$Mpc 
to $M_r = -19.8$), again separated
by the value of the structure contamination flag. We note here that the scatter on
the relation only drops by $\sim10$\% after applying the SCF cut,
demonstrating that the summed luminosity as a proxy to cluster mass is
less contaminated by nearby large-scale structure than velocity dispersion. We
also note that the scatter is smaller for mass-versus-luminosity than
it is for either mass-richness or  mass-velocity dispersion. 
Thus, we conclude that in the mock catalogs, the summed $r$-band luminosity
is the best measure of the cluster dark matter halo mass.

\begin{figure*}
\plotone{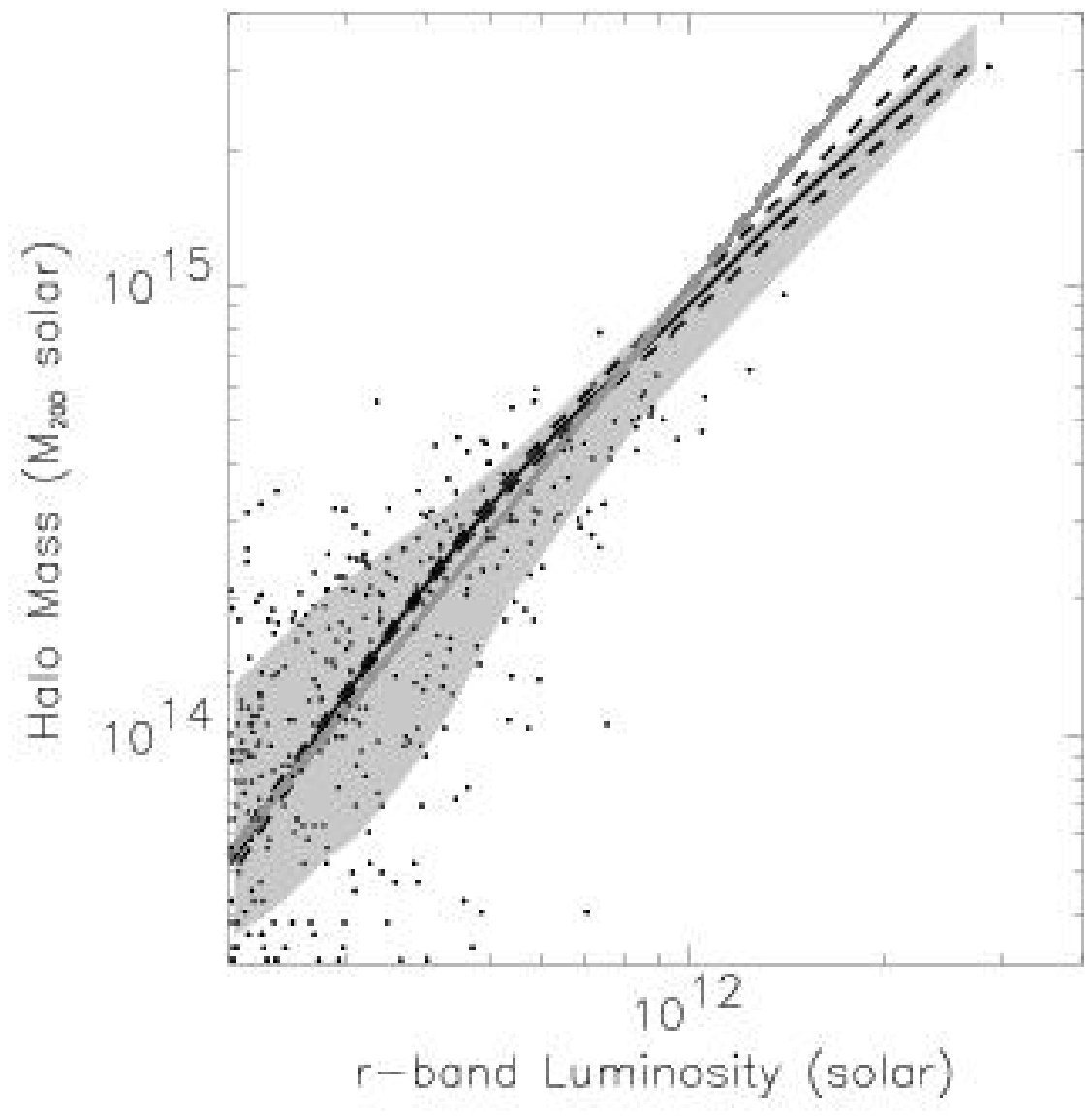}
\caption[]{Summed $r$-band cluster luminosities versus the halo mass for the
simulations.  The grey solid and dashed lines are fits to mass versus
luminosity and luminosity versus mass respectively. The solid black line is a
non-parametric fit to the data and the dashed black lines encompass the one
sigma confidence-band around this fit. The shaded grey region encompasses the
one-sigma irreducible scatter of the data about the best nonparametric fit.}
\label{fig:matchesLr}
\end{figure*}

Figure \ref{fig:matchesLr} shows the halo mass as a function of the
summed $r$-band cluster luminosity with a variety of best-fit relations and
their scatter.  We observe a linear correlation between mass and optical
cluster luminosity and the best robust-fit linear relationships (mass versus
luminosity and luminosity versus mass) are shown (light-gray solid and dashed
line respectively). One might notice that the relation is not precisely
linear, and so we have also used a nonparametric technique to more accurately
trace this scaling relation (solid black-line).
We plot a one-sigma confidence band
around our best non-parametric fit (black dashed-line).  The simultaneous
bands were determined using the method of Sun and Loader (1994).  The
gray-shaded band is the one-sigma scatter in this relation and includes
effects from matching (Section \ref{false_matches}), measurement error, and
any intrinsic scatter in the relationship.

As a consistency check, we present in Figures \ref{fig:lumr_allplots}
and Figures \ref{fig:wbiwt_allplots} (bottom) the correlation between
the velocity dispersion and richness or summed total cluster $r-$band
luminosity for clusters in our mock SDSS dataset. As seen in Figure
\ref{fig:biwt1500_allplots}, the scatter is similarly reduced by a
factor of two when we use only SCF=0 clusters.  Our conclusion is the
same as above: our velocity dispersions can be over-estimated when
there is intervening large-scale structure to interfere with the
measurement of the velocity dispersion.

In the above sections (\ref{prop} and \ref{scaling}), we have examined
how foreground/background contamination from large-scale structure (or
other clusters) affects the observables of velocity dispersion,
richness, and optical luminosity.  We pointed out a few example
clusters where the structure contamination flag (SCF) varies from 0 to
2, thus destroying any clean measure of the velocity dispersion.  We
showed the scaling laws which use the summed r-band luminosity and/or
richness are less affected by local large-scale structure than
velocity dispersion.  We conclude that the optical luminosity is the
cleanest observable measure of halo mass (using the mock catalogs). 

\section{Application to SDSS DR2 Data}
\label{data}

In this Section, we discuss the application of the C4 cluster--finding
algorithm to the Sloan Digital Sky Survey (SDSS) spectroscopic galaxy data (Strauss et
al. 2002). Technical details of the SDSS instrumentation and operations can be
found in Gunn et al. (1998), York et al. (2000), Hogg et al. (2001), Stoughton et al. (2002),
Smith et al. (2002) and Pier et al. (2003), as well as at \verb+http://www.sdss.org.+
In this paper, we focus on the photometric and spectroscopic data of the Second
Data Release (DR2) of the SDSS (Abazajian et al. 2004), {\it i.e.}, we apply
the C4 algorithm to $\simeq$ 250,000 spectroscopically observed galaxies,
which cover $\sim 2600\,\, {\rm deg^2}$ of the sky. We apply some quality
constraints to the sample. Specifically, we take all objects spectroscopically identified
as galaxies (including the deeper Luminous Red Galaxy sample; Eisenstein et al. 2001),
excluding those objects with SDSS warning flags set for low-confidence redshift, no
currently available spectrum, no red end, or no blue end. Also, the zConf variable
must be greater than 0.7. This results in 249725 unique galaxies.

Using the methodology outlined above, we find $748$
clusters and groups of galaxies in the SDSS DR2 data. This is a relatively
small number of systems compared to other published SDSS photometric cluster catalogs
(Goto et al. 2002; Kim et al. 2002; Bahcall et al. 2003), but we stress that
our sample is confined to systems with at least eight {\it spectroscopic} redshifts within
$1h^{-1}$Mpc of the cluster center. This limits our sample to nearby
clusters, in the redshift range $0.02<z<0.17$. Other SDSS cluster surveys are based only on the
SDSS photometry and therefore extend to higher redshift and contain many more
clusters, at the expense of not having measured velocity dispersions or
precisely determined membership for all of
the clusters.

\subsection{Fiber Collisions}
\label{targeting}
To this point,
we have made no correction in our analysis for the problem of ``fiber
collisions''. As discussed in Strauss et al. (2002), no two spectroscopic
fibers can be placed within 55 arcseconds of each other in any given SDSS
spectroscopic plate. These fiber collisions have been minimized by overlapping
adjacent plates (i.e. tiling),  and are $<20\%$ effect at all surface densities of galaxies
(see Blanton et al. 2003b). We study fiber collisions in two ways: (1) for the mock
catalogs, we run the C4 algorithm before and after fiber collisions are
applied; (2) for the real data, we run the algorithm on the spectroscopy (i.e. after fiber collisions) and
add in SDSS photometric cluster galaxies that were missed by the tiling algorithm.
Since we do not know the redshifts of these
missed SDSS targets, it
is impossible to run the algorithm on the real data prior to fiber collisions --unless
estimated redshifts are used. Unfortunately, even very good photo-z's (i.e. with $\Delta z \sim 0.05$)
have errors larger than the redshift box used in our algorithm.

The mock catalogs allow us to study the affect of fiber collisions on
completeness and purity, while both the mocks and the real data
provide an independent analysis of the effect fiber collisions have on cluster luminosity.
In the mock catalogs, we find clusters both before and after fiber collisions, and so
the measured properties can change in many ways. For instance, the cluster centers
and velocity dispersions can change, both of which would add or subtract to
cluster membership.
However in the real data, the cluster centers and velocity dispersions are determined solely through the spectroscopic
sample.
We can only correct the total cluster luminosity by adding in
the missed photometric galaxies that are within the projected radius. To minimize foreground/background
contamination, we also constrain these to have colors
and magnitudes near the E/S0 ridgeline of the clusters.
Thus unlike the mock catalogs where the fiber corrected luminosities can be higher or lower,
in the real SDSS data, the fiber corrected summed $r$-band luminosities 
will always be the same or higher.

\subsubsection{Fiber Collisions in the Mock Catalogs}
To create our mock catalogs with fiber collisions,
we find all mock galaxy pairs within 55 arcseconds and
randomly choose the brightest galaxy in each pair. We do this such that 70\%
of all galaxies with nearest-neighbor separations of less than 55 arcseconds
are retained. This mimics (albeit conservatively) the targeting algorithm
described in Blanton et al. (2003b). We then re-ran our cluster finder on
these revised  mock catalogs. 

We find that regardless of fiber collisions, the completeness and purity remain unchanged for the more
massive systems. As the mass (or luminosity) of the clusters decreases,
the completeness after applying fiber collisions drops slightly ($\sim 5\%$).
As a consequence of finding fewer systems, the purity increases slightly
after fiber collisions are applied (again by 5\%). This can be seen in
Figure \ref{fig:purity} where the solid (dashed) lines are before (after) fiber collisions.
Thus, fiber collisions play a very minimal
role in finding clusters.

\begin{figure*}[tp]
\plotone{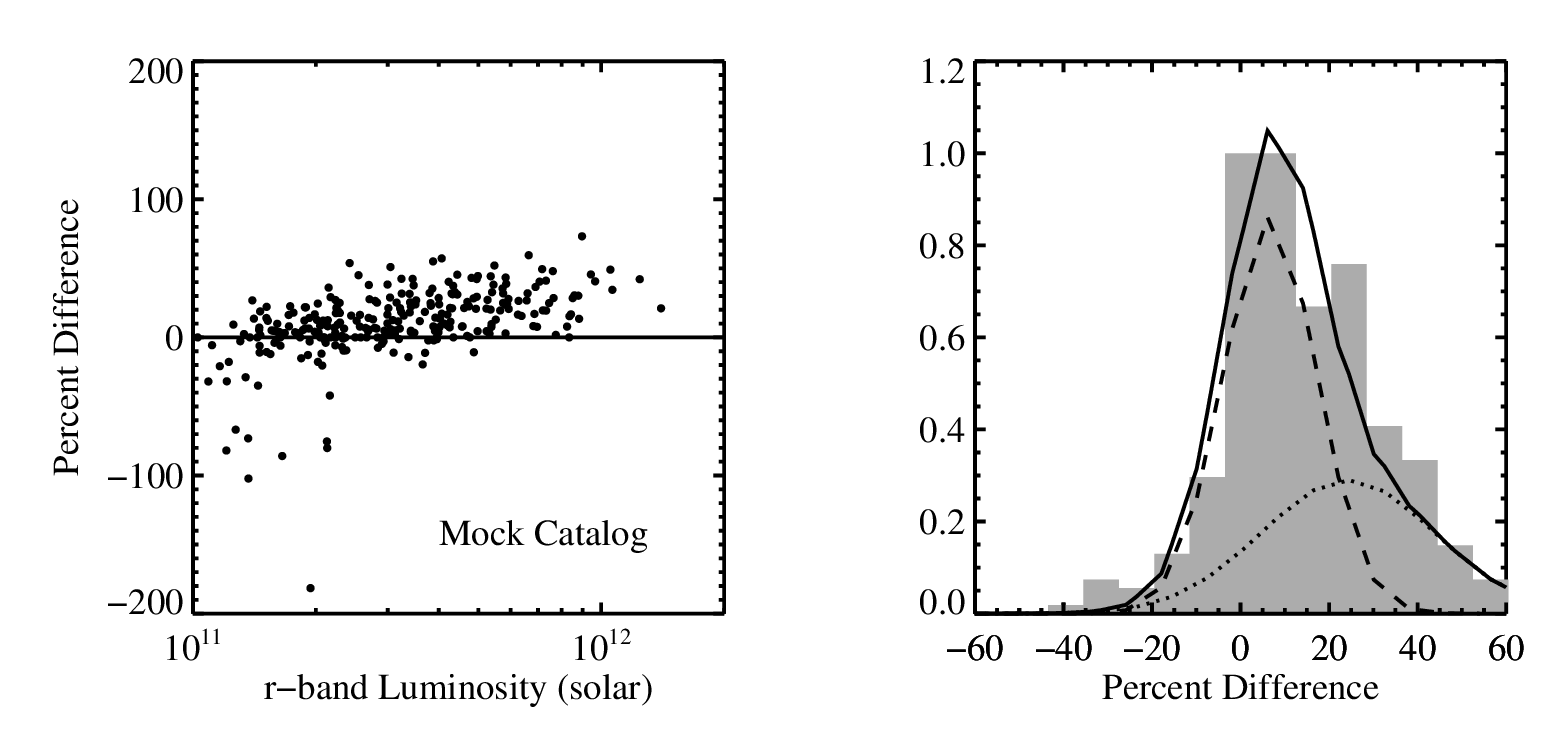}
\plotone{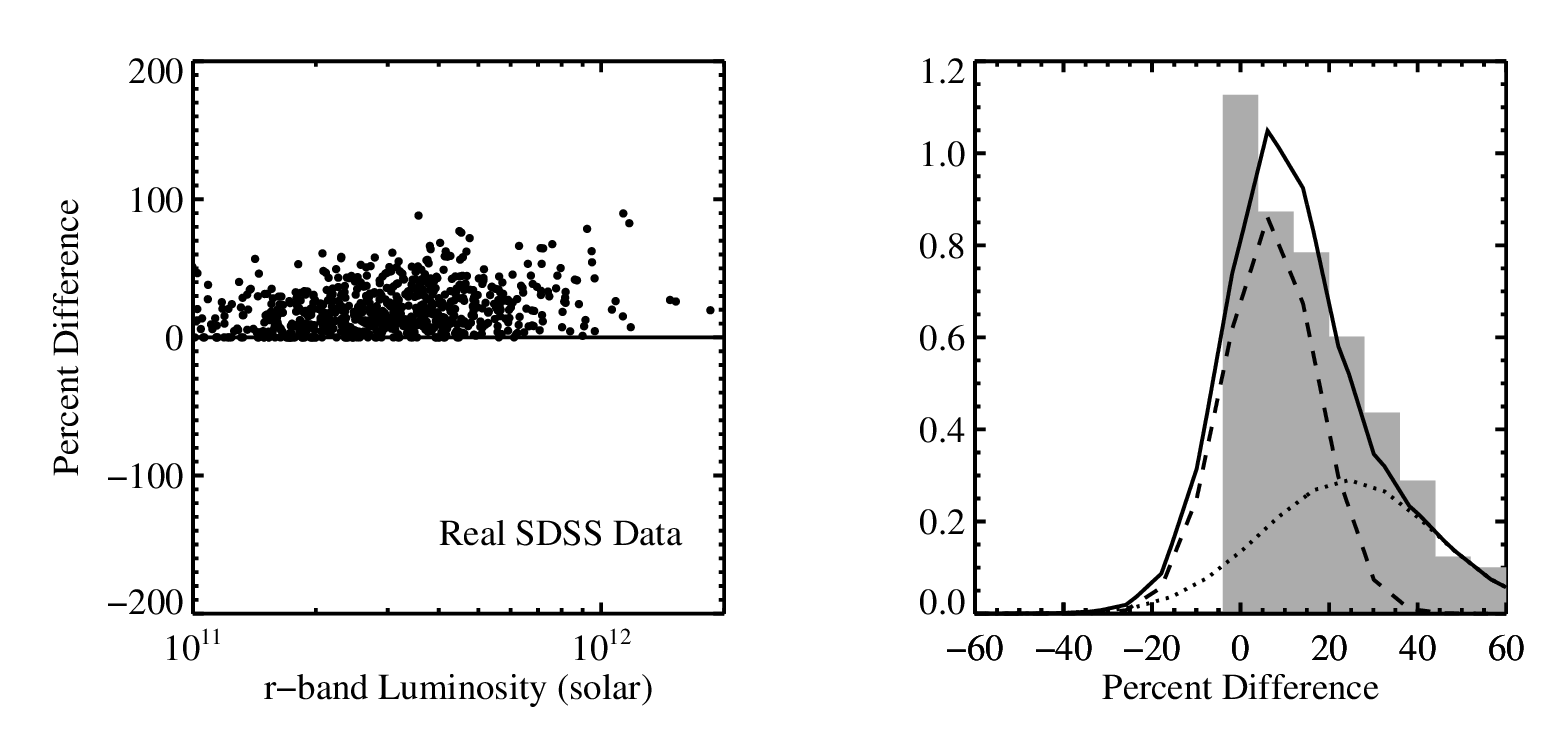}
\caption{Using the mock clusters, we compare our summed $r$-band luminosities
before and after we remove mock galaxies due to fiber collisions. The clusters
are systematically brighter (by 5\%) before we make the fiber collision corrections.
The Gaussian is a fit to those clusters with revised luminosities after the
fiber collisions were put in and has a width of $\sim 10\%$.}
\label{fig:diff_lumr}
\end{figure*}

However, the summed optical
$r$-band luminosities are systematically underestimated as seen in Figure \ref{fig:diff_lumr} Top.
In this figure we show the difference in the measured mock
cluster luminosities before and after (i.e. minus) the fiber collision algorithm is applied.
Bright massive clusters are more affected then small dim clusters.
The histogram has been normalized to the average number of clusters with differences between zero and ten percent.
As indicated by the peak of the histogram, many of the cluster luminosities remain
unchanged. Most
are systematically brighter before we remove mock galaxies due to fiber collisions (as
expected). A few are dimmer {\it before} galaxies are removed from fiber collisions. 
As described above, the clusters found after the fiber collision algorithm is applied
have different membership criteria than those before (due to a different centroid or
velocity dispersion). Thus we expect a few clusters to be brighter after
galaxies are removed from the mocks.

The distribution of the histogram in Figure \ref{fig:diff_lumr} is not a simple Gaussian.
We have investigated this and find that the sum of two Gaussians is statistically a good fit.
The two Gaussians are explained by two different populations of missed targets: typical
cluster galaxies and brightest cluster galaxies (BCG). In the mock catalogs, we find that the BCG
would be missed by fiber collisions $\sim 38\%$ of the time. These BCGs are significantly
brighter than typical cluster galaxies and they account for the tail in the histogram of
Figure \ref{fig:diff_lumr}. To confirm this, we have fit the histograms of the BCGs and non-BCGs
in Figure \ref{fig:diff_lumr} separately (as shown by the dotted (BCG) and dashed (non-BCG) curves.
The sum of these two Gaussians (the solid line) is then a good fit to the overall distribution.
The centers of these Gaussians tell us (on average) how much fiber collisions lower the 
cluster luminosities. When the BCG is missed, clusters are dimmer by 24\%, while when the
BCG is spectroscopically observed, clusters are dimmer by only 6\%.

\subsubsection{Fiber Collisions in the SDSS Data}
Since we cannot find clusters in the real SDSS spectroscopic data before fiber collisions,
we attempt to correct the SDSS C4 clusters by adding galaxies that were targeted
for observation, but missed due to the tiling algorithm.
We extract all galaxies within a projected $1h^{-1}$Mpc of the cluster center and brighter
than $m_r = 19.6$ and $M_r = -19.8$, where we assume the galaxies to be at the
redshift of the cluster. We then add this additional galaxy light to the cluster luminosities and
look for a new BCG.  To avoid adding too much noise
from foreground/background galaxies, we require these additional
galaxies to have colors that lie within the cluster's E/S0 ridgeline
and also be no more than two magnitudes dimmer than the orginally identified brightest
cluster galaxy (BCG). 

In Figure \ref{fig:diff_lumr} bottom, we show the differences in the cluster
luminosities before and after the fiber corrections. Recall that by design, the
corrected cluster luminosities will always be as bright or brighter than before
the correction (since we only add in galaxies from the photometry). As in the
mock catalogs described above, the BCGs are missed $\sim 39\%$ of the time. Also
similar to the mock catalogs, the peak of the histogram is at zero difference.
For comparison, we over plot the Gaussians from the mock catalog histograms
(Figure \ref{fig:diff_lumr} top). The close match betweem the mock catalogs and 
the real SDSS data in terms of the tails of these distributions as well as in the fraction
of missed BCGs
indicate that, (a) the fiber collision algorithm we use on the mock catalogs
is close to the tiling of the SDSS, and (b) that the cluster luminosity differences
have two components (BCG and non-BCG) and can be corrected.

\subsection{Galaxy Deblending and Targeting}

To be targeted for spectroscopy by the SDSS (Blanton et al. 2003b), a galaxy
must first be identified in the photometric data. Since we're only looking at
the brightest galaxies in the SDSS, we assume that star-galaxy separation is
a non-issue (see Scranton et al. 2003). However for various reasons, the photometric
deblender occasionally fails to correctly identify a galaxy. An extreme example
is for Abell 1539 (ra, dec = 186.5813, 62.5563), where the multiple identification of a satellite trail caused the deblender
to reach its maximum number of objects within a given field. Thus, many galaxies within the
cluster were not identified.

While a paper presenting the SDSS deblender algorithm and tests is 
currently under preparation (Lupton et al. 2005), we made a minimal effort
to quantify any SDSS deblender issues, specifically with regards to
galaxy clusters. To do so, we identified 110 Abell clusters within the
sky coverage of the SDSS DR2 spectroscopic survey and visually examined
each one using the SDSS Sky Server Cutout tool. This web-based tool allows
the user to look at any SDSS photometric field and overlay photometric or
spectroscopic SDSS galaxies (those both targeted and observed). One can also
click on the galaxies to determine their redshifts, magnitudes, etc. 

The purpose of this exercise was to determine whether SDSS deblending or targeting
problems could affect the C4 algorithm.
To perform this analysis, we took the following steps:
\begin{enumerate}
\item{Using the SDSS Finding Chart cutout tool, look at a $7\times7$ arcminute field around each Abell cluster center.}
\item{Identify the BCG, its coordinates and r-band magnitude.}
\item{Determine the fraction of spectroscopic targets that actually have spectroscopy.}
\item{Determine whether the BCG was targeted and observed spectroscopically.}
\item{Count the number of galaxies that were missed by the deblender.}
\item{Count the number of galaxies that are brighter than the SDSS spectroscopic limit ($m_r = 17.77$) but not targeted.}
\end{enumerate}
In a few cases, no cluster was obvious to the eye and so the BCG parameters and other measures were skipped
(Abell 130, 630,  682, 685, 796, 2195, 2433, 2703).

We find that 32\% of the Abell clusters are not affected by fiber collisions at all and that
73\% of cluster cores have more than 70\% of their targeted galaxies actually observed spectroscopically
by the SDSS tiling algorithm. The worst case is Abell 2644 where only four of the ten targeted galaxies
were observed. The cluster BCGs were all targeted and 70\% were observed, which is slightly better than the
global C4 average of 61\% as discussed above. The SDSS deblender seemed to miss objects in less
than 10\% of the clusters. In most of these few cases (i.e., except Abell 1539 mentioned above), 
only one obvious galaxy object was missed in the
entire field (which usually contains hundreds of objects).

Our conclusions from this visual examination of over 100 Abell clusters are that
the SDSS deblender works extremely well for these low redshift systems. Likewise,
the targeting and tiling algorithms perform as expected.

\subsection{The SDSS-C4 Clusters}

\begin{figure}
\plotone{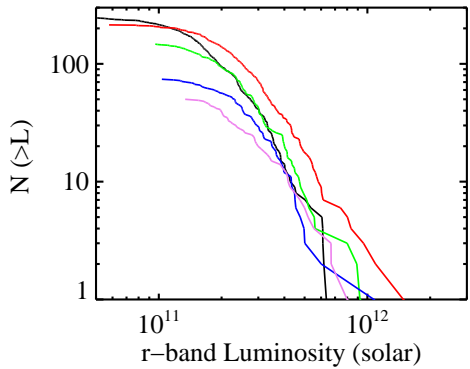}
\caption[]{The number of SDSS C4 clusters brighter than $L_r$ for
equal volume shells, each at increasing redshift. The shells are described in the
text. They go from the lowest to highest redshift as black,red, green, blue, violet.
The red line covers a volume which includes the SDSS Great Wall (Gott et al. 2004) where
we find additional massive clusters compared to the other four equal volumes.}
\label{fig:zlimit_data}
\end{figure}

In Figure \ref{fig:zlimit_data}, we present the number of clusters as a function
of $r$-band luminosity for the same equal volume shells as described in Section \ref{maglimit}.
In a comparison with the bottom panel of Figure \ref{fig:zlimit}, we see the same
trend of finding fewer low-mass systems at higher redshifts. A major difference between
Figure \ref{fig:zlimit_data} and Figure \ref{fig:zlimit} (bottom) is for the redshift
shell z = [0.075,0.093], which contains the Sloan Great Wall (Gott et al. 2004) at redshift $z = 0.08$ (the
red line in Figures \ref{fig:zlimit} and \ref{fig:zlimit_data}). In this one bin of volume,
we seem to find many more systems compared to the other equal volumes. Additionally, there
appears to be more massive clusters in this bin then in the other equal volume bins,
while the number of
low mass systems is little changed.

In Table \ref{DR2clus}, we present the $748$ clusters detected using
the C4 algorithm on the DR2 dataset. For each cluster, we present an
SDSS C4 ID (Column 1); the (J2000) right ascension and declination of the
centroid via Method (1) in Section \ref{centroid} (Columns 2,3); 
the ra and dec via Method (2) in Section \ref{centroid}  (Columns 4,5);
the right ascension
and declination of the Brightest Cluster
Galaxy (Columns 6,7);. the mean redshift of the cluster
(Column 8); the velocity dispersion (to convert to the rest-frame of the cluster, divide by $1+z$);
(Column 9); the Richness 
(Column 10); the summed $r$-band cluster luminosity (corrected for missed targets) (Column 11);
the Structure Contamination Flag (SCF; Column 12); the
Dressler-Shectman substructure probability (Column
13); and the names of other clusters matched to within 10 arcminute of the cluster centroid (Column 14).
The names are as follows-- Abell: Abell (1958); RBS: Rosat Bright Source Catalog (Voges et al. 1999); RX: The Hamburg/RASS catalog of optical
identifications (Bade et al. 1998); ZwCl: Zwicky compact groups and clusters (Zwicky and others
1961,1963,1965,1966,1968a,b); WXD: Wei et al. (1999); RXC: Oppenheimer et al. (1997); WBL: Catalog
of Nearby Poor Clusters
White et al. 1999); CAN: Wegman et al. (1996); SCC: S{\" o}chting et al. (2002); MACS/VMF:
Vikhlinin et al.(1998); KDCS: Deep Range Survey (Postman et al. 2002). Note that the
match is not performed against redshift, thus some of these matches are in projection only.
More information on these names and additional matches can be found on the C4 webpage.

In Figures \ref{fig:clust_fig1} through \ref{fig:clust_fig6}, we show the sky plots,
color--magnitude relations, velocity histograms, and velocity dispersion profiles for
six C4 clusters detected in the DR2. 
We have made available
these plots for all 748 C4 DR2 clusters via the world-wide web. These plots provide for
quality assurance, so users of this catalog can inspect the individual
systems and determine whether they suit their scientific requirements.

An electronic version of the catalog given in Table \ref{DR2clus} can be
downloaded from \verb+http://www.ctio.noao.edu/~chrism/C4/+ We intend to update this C4 catalog as
more SDSS data becomes available and therefore, we encourage the reader to
check this webpage periodically.

\subsection{Comparison to Other Cluster Catalogs}

In this subsection we take a closer look at the C4 cluster sample by comparing it
to two different samples from the literature: the Abell (1958)
catalog and the RASS-SDSS catalog (Popesso et al. 2004). These two have
remarkable different selection algorithms (which are fully described in their
respective references). In short, the Abell catalog was created by the eyeball
examination of the {\it optical} Palomar All Sky Survey plates. On the other hand, the RASS-SDSS sample
identifies the optical counterparts to {\it X-ray discovered} clusters taken from the literature.
Only 36\% of the RASS-SDSS clusters are also Abell clusters. 

As opposed to the C4 catalog, neither the
Abell nor the RASS-SDSS cluster catalogs have selection functions and so we can
make no quantitative claims about their completeness or purity. However, since the C4 catalog is
$>90\%$ pure regardless of redshift and/or mass and $>90\%$ complete
for masses above 2$\times10^{14}$ solar masses and redshifts below $z = 0.12$ (see Figure \ref{fig:zlimit}),
we expect to be able to recover the more massive clusters from other samples.

\subsubsection{The C4-Abell Clusters}
We identified those Abell clusters that lie in the area of sky coverage for
the SDSS DR2. 
We then match the two
cluster samples and find any C4 clusters that are within 10 arcminutes of an Abell cluster.
We find 346 Abell
clusters that meet these criteria and 123 are uniquely matched to C4 clusters (only 5 Abell clusters are matched to two different C4 clusters).
However, to more fairly
compare the two samples, we must place a redshift constraint on the two samples so that
some minimum level of completeness is met by the algorithms. The Abell cluster
catalog is thought to be fairly complete within $0.03 \le z \le 0.12$ (see Miller et al. 2002). And as shown
in Figure \ref{fig:zlimit}, the C4 sample is also $>90\%$ complete in this redshift range for masses
greater than $2\times10^{14}$ solar masses. When we apply this redshift
constraint, there are 104 Abell clusters and
71\% of them are also found by the C4 algorithm. We took a closer examination of the 30 missing
Abell clusters (see Table \ref{abell_c4}) and found that six of them are actually too deep
for the C4 algorithm, eight have too few spectra, and six are not clusters at all (as determined
through a visual examination of the photometry and spectra around their Abell centers).
So if we
exclude these 20 from the list of Abell clusters the C4 algorithm should be able to find, than the
recovery rate goes up to 88\%. If you include another three Abell clusters that were
found, but blended into nearby clusters, the recovery rate is 92\%. 
Thus, the the C4 algorithm
can recover $\sim 90\%$ of Abell clusters and the ones that are missed are
typically found at redshifts where the C4 algorithm is less than 90\% complete ($z > 0.12$).

\begin{deluxetable*}{cccccl}
\tablewidth{0pt}
\tablecaption{\bf Abell Clusters Not Recoverd by the C4 Algorithm}
\tablehead{
\colhead{Abell ID} & \colhead{ra} & \colhead{dec} & \colhead{redshift} & \colhead{Abell Richness} & \colhead{Note}}
\startdata
    87 &   10.75653 &   -9.79327 &    0.05500  &      1    &    Blended with Abell 85\\
   116 &   13.96094 &    0.63749 &    0.06650 &       0   &     Too few spectra\\
   190 &   20.91811 &   -9.85647 &    0.10150 &       0   &     \\
   237 &   25.24066 &    0.26942 &    0.10320 &       0   &     Not a cluster\\
   605 &  119.07229 &   27.39939 &    0.10960 &       0   &     Not a cluster\\
   733 &  135.32976 &   55.62065 &    0.11590 &       1   &     \\
   756 &  138.10278 &   48.47721 &    0.11760 &       1   &     Too few spectra\\
   869 &  146.55203  &   2.35163 &    0.11740 &       0   &     Too few spectra\\
   912 &  150.29539 &   -0.10799 &    0.04460 &       0   &     \\
   919 &  151.23901 &   -0.69325 &    0.09540 &       1   &     Too deep (z=0.2)\\
   975 &  155.65785 &   64.63030 &    0.11860 &       1   &     \\
  1032 &  157.57814 &    4.01031 &    0.07940  &      0   &     Blended with Abell 1024\\
  1215 &  169.87431 &    4.34328 &    0.04940  &      1   &     Too deep  (z=0.156)\\
  1289 &  172.90517 &   60.75701 &    0.11180  &      0   &     Not a cluster\\
  1322 &  174.24931 &   63.22324 &    0.11140  &      0   &     Too deep (z=0.245)\\
  1389 &  177.34004 &   -1.37796 &    0.10320  &      0   &     Not a cluster\\
  1402 &  178.13921 &   60.42187 &    0.10620  &      0   &     Not a cluster\\
  1404 &  178.08986 &   -2.81113 &    0.10320  &      0   &     Too few spectra\\
  1406 &  178.30972 &   67.88883 &    0.11780  &      1   &     Too few spectra\\
  1432 &  179.92520 &   68.10468 &    0.11350  &      0   &     Too deep (z=0.265)\\
  1459 &  181.06020 &    2.50469 &    0.08070  &      1    &    \\
  1477 &  182.22612 &   64.07182 &    0.11090  &      1    &    Too deep (z=0.24)\\
  1630 &  192.93535 &    4.56138 &    0.06480  &      1    &    \\
  1729 &  201.00092 &   -3.36041 &    0.11440  &      1    &    Too few spectra\\
  2023 &  226.45502 &    2.85714 &    0.05470  &      1    &    Too few spectra\\
  2053 &  229.31754 &   -0.68262 &    0.11270  &      1    &    Blended with A2051\\
  2356 &  323.93530 &    0.12408 &    0.11610  &      1    &    \\
  2433 &  334.61554 &   14.01783 &    0.08800  &      0    &    Too deep (z=0.12)\\
  2644 &  355.28537 &    0.09426 &    0.06930  &      1    &    Too few spectra\\
  2705 &    1.49696 &   15.79528 &    0.11470  &      1    &    No cluster\\
\enddata
\label{abell_c4}
\end{deluxetable*}

\subsubsection{The RASS-SDSS-C4 Clusters}
We applied the same area constraints to the Popesso et al. (2004) sample
as we did to the Abell clusters. Of the 97 RASS-SDSS clusters that lie within
the area of the SDSS DR2 spectroscopic survey, we find 53 using the C4 algorithm.
However, when we apply the same volume constraints as for the Abell sample, the C4 algorithm finds
39 out of the 43 RASS-SDSS clusters within $0.03 \le z \le 0.12$. Five missing
RASS-SDSS clusters (one has a mistaken redshift) are listed
in Table \ref{rass_c4}, along with notes made after additional examination. The RASS-SDSS recovery rate 
goes up to 98\% if we account for a slightly larger matching radius at very low redshifts,
edge effects, and the occasional cluster de-blending issue.

\begin{deluxetable*}{cccccl}
\tablewidth{0pt}
\tablecaption{\bf RASS-SDSS Clusters Not Recoverd by the C4 Algorithm}
\tablehead{
\colhead{RASS-SDSS ID} & \colhead{ra} & \colhead{dec} & \colhead{redshift} &  \colhead{X-ray Luminosity} & \colhead{Note}}
\startdata
RXCJ0736.4+3925 & 114.10400  &  39.43290 &   0.11700&  4.81$\times10^{37}$W &                            \\ 
RXCJ0953.6+0142 & 148.42310  &  1.71180  &  0.09800 &  0.97$\times10^{37}$W & Blended with SDSSDR2-C4-1304\\
RXCJ1303.9+6731 & 195.98540 &  67.51770  &  0.10600 &  0.35$\times10^{37}$W & Too deep (z=0.22)\\
RXCJ1511.5+0145 & 227.88969 &   1.76420  &  0.03700 &  0.07$\times10^{37}$W & Too close to survey edge\\
RXCJ2324.3+1439 & 351.08771  & 14.66450 &   0.04200 &  0.97$\times10^{37}$W & SDSSDR2-C4-2001 is 14 arcminutes away.
\enddata
\label{rass_c4}
\end{deluxetable*}

\section{Discussion}
\label{summary}

As mentioned in the introduction, our goal with the C4 catalog was to
construct an optical cluster catalog with a well--determined selection
function, {\it i.e.}, in which we know both the completeness and
purity of the catalog as a function of the underlying dark halo mass
and the cluster observables.  To achieve this, we have used a new
breed of mock galaxy catalogs
that are designed to match the distribution of galaxy colors and
luminosities {\em and their environmental dependence} in the real SDSS
data.  We have then run the exact same cluster-finding algorithm on
the mock catalog as we run on the data; this allows us to calculate
the completeness and purity of the cluster catalog, and examine how
these quantities change as we adjust different parameters in the
cluster--finding algorithm. 

We have thus used
the mock galaxy catalogs to fine--tune our algorithm such that it maximizes both our
completeness and purity. Using the W05 mock SDSS catalogs, we find
that the C4 catalog has less than 10\% contamination 
over all cluster masses/luminosities
and redshifts examined.  The catalog has less than $5\%$ contamination
for clusters with masses greater than $2\times 10^{14}{\rm M_{\odot}}$
 --- and is more than 90\% complete for clusters with
masses larger than $ 2\times 10^{14}{\rm M_{\odot}}$ and within $z =
0.12$.

We have also used the W05 mock catalog to study key cluster scaling
relations and the scatter about these relations,
to determine which cluster observable-to-mass scaling has the
smallest scatter.  For the first time,
we are able to directly relate cluster observables (velocity
dispersion, summed cluster optical luminosities, and richness) to the halo masses
used to derive theoretical mass functions (Jenkins et al. 2001; Evrard
et al. 2002). The scaling relation with the smallest scatter (measured
perpendicular to the best-fit) is halo mass versus velocity dispersion
--- when only systems with a Structure Contamination
Flag, SCF = 0, are
considered ({\it i.e.}, only those with no surrounding large--scale
structure) (see Figure \ref{fig:biwt1500_allplots} Middle). 
The relation between halo mass and summed $r$-band luminosity has only
slightly more scatter than the best-case for velocity dispersion. 
However, this scatter is not
significantly affected by limiting to systems without close
companions --- i.e., the sample is less affected by
intervening large-scale structure than the velocity dispersion is.  Thus,
while the velocity dispersion (which has a simple physical
relationship to mass via the virial theorem) 
correlates well with halo mass,
the optical luminosity may be more appropriate when the velocity
dispersion is contaminated or cannot be accurately measured.

A much more thorough analysis of the scaling relations and how well
they can be calibrated against simulations will be presented in W05; a
detailed comparison of various observational scaling relations for C4
and maxBCG clusters (Annis et al. 2001) will be presented in McKay et al (in preparation).

We present this C4 catalog to
demonstrate the power of the SDSS spectroscopic data in finding nearby
clusters of galaxies, and to advance our understanding of cluster
scaling relations and galaxy evolution in dense environments (for
which the sample has been used in the studies of Gomez et al. 2003;
Miller et al. 2003; Balogh et al. 2004). In future work, we will use
the C4 catalog to constrain cosmological parameters and to
statistically understand cluster properties like cluster profiles and
luminosity functions. By the end of the SDSS, we estimate that the C4
catalog will contain $\simeq 2500$ clusters, making it one of the
largest, most homogeneous catalogs of nearby clusters in existence.

\begin{figure*}[tp]
\epsscale{0.7}
\plotone{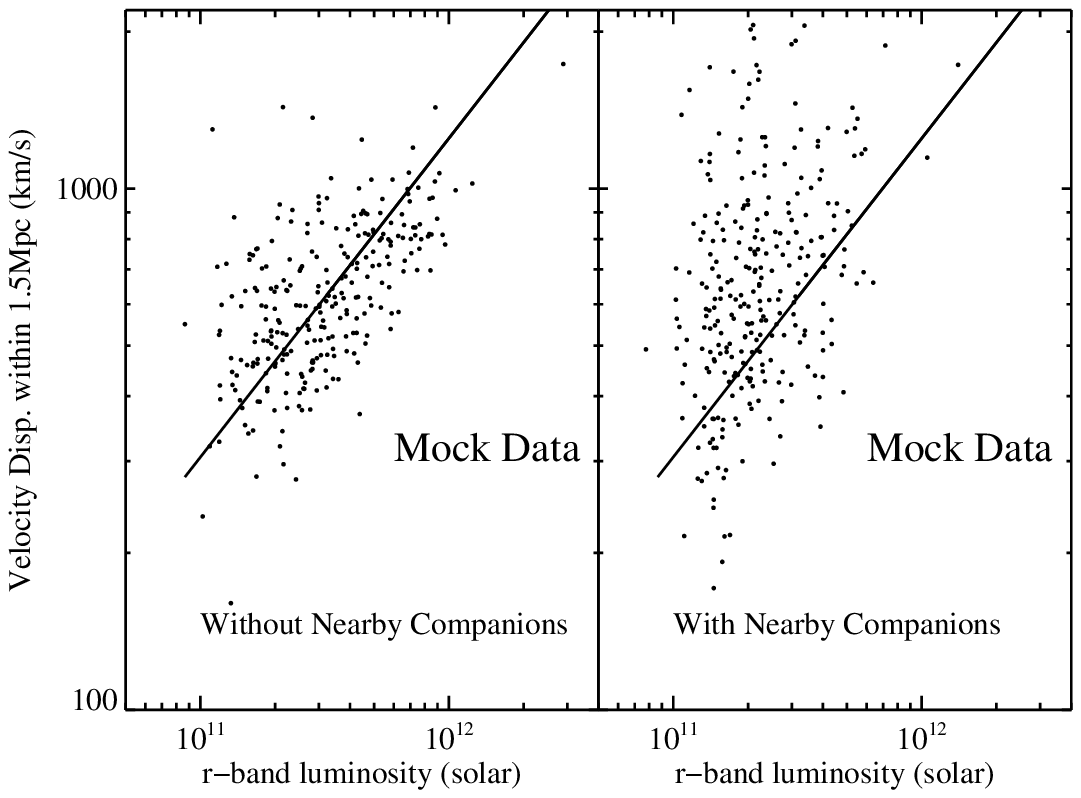}

\plotone{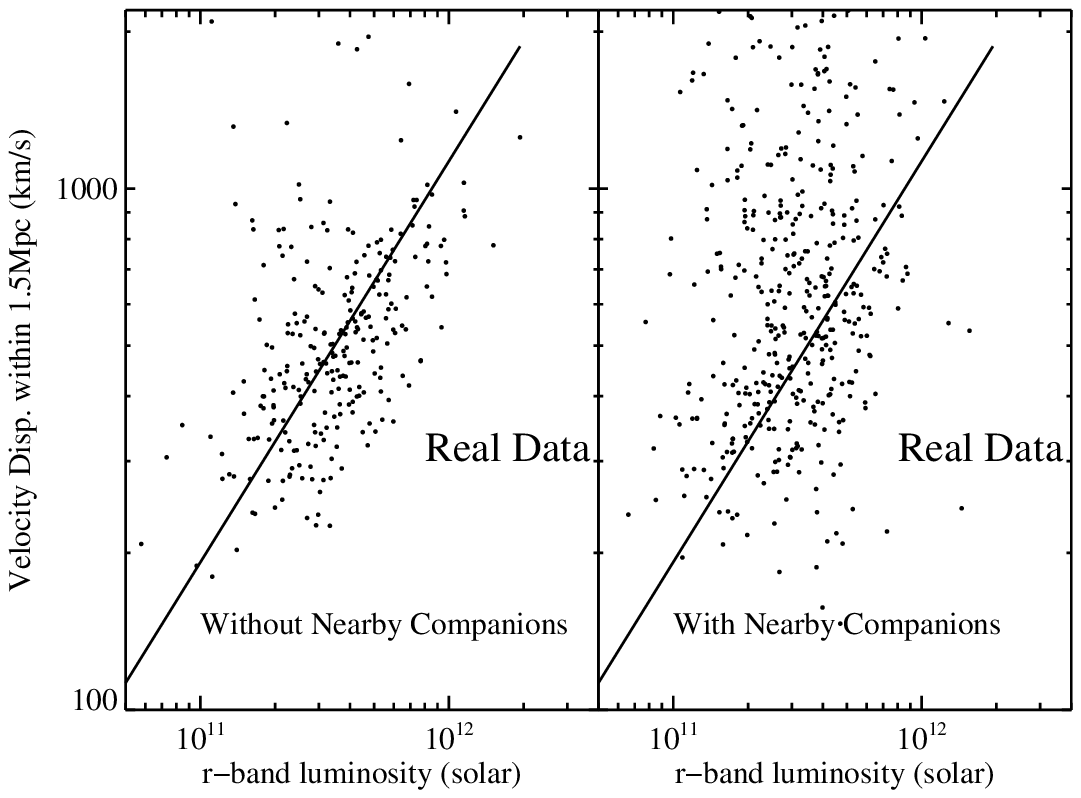}
\caption{A plot of the relationship between our main cluster observables, the
velocity dispersion, $r$-band cluster luminosity, and the Structure Contamination Flag.  The
top plots are for the mock data, while bottom plot is for the real SDSS
data. The scatter in the scaling relation between velocity dispersion and
$r$-band cluster luminosity is reduced by a factor of $\sim 2$ once nearby
large-scale structure is accounted for. We show the best-fit linear relation
to the data without nearby companions (left--hand plot) for both the real data
and the mock catalog.}
\epsscale{1.0}
\label{fig:biwt1500_companion_real}
\end{figure*}

As an illustration of the quality of the C4 catalog, we present in
Figure \ref{fig:biwt1500_companion_real} the scaling relation between
the total $r$-band cluster luminosity and the velocity dispersion,
separated by the value of the companion flag, for both the real SDSS
data (bottom) and the mock SDSS catalog (top).  These cluster
properties are our main observables, and in the real data, their
interpretation does not rely on the details of the mock catalog generation. This
plot again demonstrates the power of the Structure Contamination Flag in reducing
the scatter on these scaling relations.  Note the striking similarity
between the scaling relations in the mock data and the real
data. Within the uncertainties, the only difference between the two is
in the number of clusters available, which is higher in the mock
catalog. The best fits to these scaling relations are almost identical
between the simulations and real data. In summary, the clusters in our
mock Universe are statistically identical to those in the
Universe, suggesting that the underlying cosmological model of our mock
catalogs --- $\Omega_{matter}=0.3$, $\Omega_{\lambda}=0.7$ and
$\sigma_8=0.9$ --- is a good representation of reality.  Detailed comparison of these
scaling relations will be presented in a future work.

RHW acknowledges support provided by the National Aeronautics and
Space Administration (NASA) through Hubble Fellowship grant
HST-HF-01168.01-A awarded by the Space Telescope Science Institute
(STScI).  The authors acknowledge the hospitality of the Aspen Center
for Physics, where some of this work was completed.

Funding for the creation and distribution of the SDSS Archive has been
provided by the Alfred P. Sloan Foundation, the Participating
Institutions, the National Aeronautics and Space Administration, the
National Science Foundation, the U.S. Department of Energy, the
Japanese Monbukagakusho, and the Max Planck Society. The SDSS Web site
is http:\/\/www.sdss.org/.

The SDSS is managed by the Astrophysical Research Consortium (ARC) for
the Participating Institutions. The Participating Institutions are The
University of Chicago, Fermilab, the Institute for Advanced Study, the
Japan Participation Group, The Johns Hopkins University, Los Alamos
National Laboratory, the Max-Planck-Institute for Astronomy (MPIA),
the Max-Planck-Institute for Astrophysics (MPA), New Mexico State
University, University of Pittsburgh, Princeton University, the United
States Naval Observatory, and the University of Washington.

This research has made use of the NASA/IPAC Extragalactic Database
(NED) which is operated by the Jet Propulsion Laboratory, California
Institute of Technology, under contract with the National Aeronautics
and Space Administration.

\clearpage

\begin{figure}
\caption[]{This figure is only included in the online edition. \\
In Figures \ref{fig:clust_fig1} through \ref{fig:clust_fig6}, we show the sky plots (top left), the
velocity histograms (top right), and $g-r$ vs. $r$ color-magnitude relation (CMR) (bottom left), and the velocity dispersion profile (bottom
right). In all figures, red and green circles are for cluster members within 1.5$h^{-1}$Mpc and within $\pm4\sigma_v$, with
red having weak  H$\alpha$  emission and green having strong H$\alpha$ emission. Blue circles are the Brightest
Cluster Galaxy.
Yellow filled circles are the C4 galaxies. Black filled circles are: all galaxies in the sky plot; all galaxies within
1.5$h^{-1}$Mpc in the CMR; and all galaxies within 1.5$h^{-1}$Mpc  and  within $\pm4\sigma_v$ in the velocity dispersion profile.
{\bf Cluster ID: SDSSDR2-C4-1028}. This specific cluster is highly elliptical and the galaxy velocity dispersion profile has a high level
of scatter in the outer radii, causing the SCF flag to be set to 2. Notice that the C4 galaxy velocity dispersion
profile has much less scatter, but there is evidence that the C4 galaxies underestimate the true velocity dispersions
(see Figure \ref{fig:biwt1000_comp_allplots}).}
\label{fig:clust_fig1}
\end{figure}
\begin{figure}

\caption[]{
This figure is only included in the online edition. \\
{\bf Cluster ID: SDSSDR2-C4-1088}. The panels and symbols are the same as in Figure \ref{fig:clust_fig1}. The core of this cluster is dominated by a galaxy population with strong H$\alpha$ emission and blue-colors, indicative of star-formation.
This cluster has SCF=0.}
\label{fig:clust_fig2}
\end{figure}
\begin{figure}

\caption[]{
This figure is only included in the online edition. \\
{\bf Cluster ID: SDSSDR2-C4-3001}. The panels and symbols are the same as in Figure \ref{fig:clust_fig1}. This cluster is the southern component to a binary system (notice the concentration of galaxies at dec = 34.1). Notice how
the velocity dispersion profile increases (by a factor of 2)
as one probes to larger and larger radii, thus bringing in galaxies from the northern cluster.
This cluster has SCF=2}
\label{fig:clust_fig3}
\end{figure}
\begin{figure}

\caption[]{
This figure is only included in the online edition. \\
{\bf Cluster ID: SDSSDR2-C4-3004}. The panels and symbols are the same as in Figure \ref{fig:clust_fig1}. This is the northern counterpart to Figure \ref{fig:clust_fig3}.}
\label{fig:clust_fig4}
\end{figure}
\begin{figure}


\caption[]{This figure is only included in the online edition. \\
{\bf Cluster ID: SDSSDR2-C4-3269}. The panels and symbols are the same as in Figure \ref{fig:clust_fig1}. This is one of our higher redshift clusters (z $\sim 0.168$). This is a relatively low-mass system (400km/s). Notice how the C4 galaxies
(yellow) are have a smaller dispersion then the red galaxies (red). We show in Figure \ref{fig:biwt1000_comp_allplots} that the
C4 galaxies are a biased measure of the true velocity dispersion (on average). This cluster has SCF=0.}
\label{fig:clust_fig5}
\end{figure}
\begin{figure}

\caption[]{This figure is only included in the online edition. \\
The panels and symbols are the same as in Figure \ref{fig:clust_fig1}. This is one of the lowest redshift systems (z $\sim 0.027$) found by the C4 algorithm.
This cluster has SCF = 0. Notice in the sky plot how the C4 galaxies are always cluster members (i.e., there are no
yellow filled circles without a green or red outline). This figure shows how well the C4 algorithm does in spearating
cluster members from the background (black dots). Due to its low redshift, this cluster is not included in the final catalog.}
\label{fig:clust_fig6}
\end{figure}

\clearpage
\LongTables
\begin{landscape}
\begin{deluxetable}{cccccccccccccl}
\tablecolumns{14}
\tablewidth{0pt}
\tablecaption{\bf The SDSS-C4 Cluster Catalog (DR2)}
\tablehead{
\colhead{ID} & \colhead{RA } & \colhead{Dec} & 
\colhead{RA } & \colhead{Dec } & \colhead{BCG RA } & \colhead{BCG Dec } &
\colhead{redshift} & \colhead{$\sigma_v$} & \colhead{Richness} & 
\colhead{Lum$_r$} & \colhead{SCF} & \colhead{DS} & \colhead{Other Names} \\ 
\colhead{} & \colhead{(deg)} &  \colhead{(deg)} & \colhead{(deg)} & \colhead{(deg)} & \colhead{(deg)} & \colhead{(deg)} & \colhead{} &
\colhead{ km/s} & \colhead{} & \colhead{M$_{\odot}$} & \colhead{} & \colhead{} & \colhead{}}
\startdata
3002 & 258.1272 &  64.0166 & 258.1362 &  64.0183 & 258.1200 &  64.0608 & 0.08078 & 1355 &  150 & 1.933993e+12 & 0 &  0.02 & RBS 1630,ABELL 2255                                \\
1031 & 213.7458 &  -0.3874 & 213.7301 &  -0.4146 & 213.7405 &  -0.3496 & 0.14037 &  931 &   48 & 1.584852e+12 & 1 &  0.99 & ABELL 1882                                         \\
3153 & 133.9645 &  42.1373 & 133.9973 &  42.1182 & 133.9380 &  42.0673 & 0.13867 &  607 &   21 & 1.558037e+12 & 1 &  0.91 & no matches                                         \\
1047 & 229.2174 &  -0.9029 & 229.1859 &  -0.9629 & 229.2174 &  -0.9029 & 0.11856 &  870 &   84 & 1.511027e+12 & 0 &  0.00 & ABELL 2051                                         \\
3238 & 257.6715 &  56.7729 & 257.7072 &  56.7694 & 257.5823 &  56.7573 & 0.12508 &  274 &   13 & 1.448392e+12 & 2 &  0.18 & no matches                                         \\
3077 & 240.2678 &  53.9485 & 240.3145 &  53.9002 & 240.3670 &  53.9474 & 0.10725 &  611 &   30 & 1.282911e+12 & 2 &  0.77 & RXC J1601.3+5354,ABELL 2149                        \\
3001 & 255.6148 &  33.4949 & 255.6578 &  33.5058 & 255.6381 &  33.5167 & 0.08847 & 1599 &  104 & 1.232508e+12 & 1 &  0.95 & RX J1702.5+3330,ABELL 2245                         \\
1039 & 228.8332 &   4.3846 & 228.8090 &   4.3502 & 228.8088 &   4.3862 & 0.09817 &  971 &   56 & 1.159921e+12 & 0 &  0.94 & ABELL 2048                                         \\
1055 & 202.7347 &  -1.7053 & 202.7637 &  -1.7449 & 202.7960 &  -1.7273 & 0.08470 & 1112 &   82 & 1.149299e+12 & 0 &  0.00 & ABELL 1750                                         \\
2002 & 358.5570 & -10.4192 & 358.5397 & -10.3877 & 358.5570 & -10.4190 & 0.07612 &  976 &   85 & 1.147930e+12 & 0 &  1.00 & ABELL 2670                                         \\
1043 & 234.1718 &  -2.0237 & 234.1550 &  -2.0297 & 234.1242 &  -1.9634 & 0.14477 &  832 &   41 & 1.123676e+12 & 0 &  0.95 & ABELL 2094                                         \\
1135 & 180.2013 &   3.4281 & 180.1875 &   3.3906 & 180.2303 &   3.4488 & 0.13586 & 1595 &   35 & 1.069101e+12 & 0 &  0.18 & ABELL 1437                                         \\
3130 & 186.8255 &  63.3870 & 186.8714 &  63.4071 & 186.9634 &  63.3848 & 0.14477 &  556 &   16 & 1.055957e+12 & 1 &  0.12 & ABELL 1544,RX J1227.8+6323                         \\
1140 & 176.3583 &  -2.3925 & 176.3784 &  -2.4031 & 176.3790 &  -2.4531 & 0.12399 & 2182 &   36 & 1.033189e+12 & 1 &  0.94 & ABELL 1373,ABELL 1373E                             \\
2005 &  19.8822 &  14.8913 &  19.8902 &  14.8660 &  19.9095 &  14.8980 & 0.12850 &  773 &   40 & 9.780194e+11 & 0 &  0.08 & ABELL 0175                                         \\
3030 & 126.3232 &  47.1691 & 126.3386 &  47.1648 & 126.3710 &  47.1335 & 0.12721 &  813 &   43 & 9.738780e+11 & 0 &  0.32 & ABELL 0655,ZwCl 0822.8+4722                        \\
3004 & 255.6803 &  34.0804 & 255.6607 &  34.0462 & 255.6898 &  34.0611 & 0.09945 & 1372 &   88 & 9.644175e+11 & 1 &  0.99 & ABELL 2244,ABELL 2244B                             \\
1041 & 194.6444 &  -1.7209 & 194.6891 &  -1.7280 & 194.6729 &  -1.7615 & 0.08392 &  864 &   68 & 9.573658e+11 & 0 &  0.28 & ABELL 1650                                         \\
3075 & 254.9274 &  32.5790 & 254.9603 &  32.5765 & 254.9331 &  32.6153 & 0.09855 & 1607 &   54 & 9.350994e+11 & 1 &  0.90 & RX J1659.7+3236,ABELL 2241,ABELL 2241B             \\
1035 & 175.8631 &  -1.7308 & 175.9095 &  -1.7533 & 175.8722 &  -1.6680 & 0.10611 &  599 &   39 & 9.341238e+11 & 0 &  0.19 & ABELL 1364                                         \\
1014 & 175.2992 &   5.7348 & 175.2856 &   5.7342 & 175.2992 &   5.7347 & 0.09824 &  852 &   73 & 9.289833e+11 & 0 &  0.50 & ABELL 1346,ABELL 1346N                             \\
2088 & 321.2136 &  -6.9301 & 321.2015 &  -6.9369 & 321.2933 &  -6.9636 & 0.11729 &  766 &   37 & 8.767952e+11 & 1 &  0.51 & no matches                                         \\
3057 & 176.4409 &  67.4865 & 176.4169 &  67.4827 & 176.4491 &  67.5495 & 0.11569 &  788 &   46 & 8.649162e+11 & 1 &  0.38 & ABELL 1366                                         \\
2116 & 327.6470 &  -7.8728 & 327.6525 &  -7.9011 & 327.6604 &  -7.7888 & 0.12204 &  696 &   23 & 8.553782e+11 & 0 &  0.32 & no matches                                         \\
\multicolumn{14}{c}{The rest of the table can be found in the July online edition of the Astronomical Journal, or at www.ctio.noao.edu/~chrism/c4} \\
\enddata

\label{DR2clus}
\end{deluxetable}
\clearpage
\end{landscape}


\begin{thebibliography}{}
\bibitem[Abazajian et al.(2004)]{2004AJ....128..502A} Abazajian, K., et 
al.\ 2004, \aj, 128, 502 
\bibitem[Abell(1958)]{1958ApJS....3..211A} Abell, G.~O.\ 1958, \apjs, 3,
211
\bibitem[Abell, Corwin, \& Olowin(1989)]{1989ApJS...70....1A} Abell, G.~O.,
Corwin, H.~G., \& Olowin, R.~P.\ 1989, \apjs, 70, 1
\bibitem[Adami et al.(2000)]{2000ApJS..131..391A} Adami, C., Ulmer, M.~P., 
Romer, A.~K., Nichol, R.~C., Holden, B.~P., \& Pildis, R.~A.\ 2000, \apjs, 
131, 391 
\bibitem[Annis et al.(1999)]{1999AAS...195.1202A} Annis, J., et al.\ 1999, 
Bulletin of the American Astronomical Society, 31, 1391 
\bibitem[Bahcall, Fan, \& Cen(1997)]{1997ApJ...485L..53B} Bahcall, N.~A.,
Fan, X., \& Cen, R.\ 1997, \apjl, 485, L53
\bibitem[Bahcall et al.(2003)]{2003ApJS..148..243B} Bahcall, N.~A.~et al.\
2003, \apjs, 148, 243
\bibitem[Bade et al.(1998)]{1998A&AS..127..145B} Bade, N., et al.\ 1998, 
\aaps, 127, 145 
\bibitem[Baldry et al.(2004)]{2004ApJ...600..681B} Baldry, I.~K., 
Glazebrook, K., Brinkmann, J., Ivezi{\' c}, {\v Z}., Lupton, R.~H., Nichol, 
R.~C., \& Szalay, A.~S.\ 2004, \apj, 600, 681 
\bibitem[Balogh et al.(2004)]{2004MNRAS.348.1355B} Balogh, M., et al.\ 
2004, \mnras, 348, 1355 
\bibitem[Beers, Flynn, \& Gebhardt(1990)]{1990AJ....100...32B} Beers,
T.~C., Flynn, K., \& Gebhardt, K.\ 1990, \aj, 100, 32
\bibitem[(Benjamini \& Hochberg 1995)]{BH} Benjamini, Y., Hochberg, Y. 1995, 
J. R. Stat. Soc. B, 57, 289
\bibitem[Blakeslee et al.(2003)]{2003ApJ...596L.143B} Blakeslee, J.~P., et 
al.\ 2003, \apjl, 596, L143 
\bibitem[Blanton et al.(2003)]{2003AJ....125.2348B} Blanton, M.~R.~et al.\
2003a, \aj, 125, 2348
\bibitem[Blanton et al.(2003)]{2003AJ....125.2276B} Blanton, M.~R., Lin,
H., Lupton, R.~H., Maley, F.~M., Young, N., Zehavi, I., \& Loveday, J.\
2003b, \aj, 125, 2276
\bibitem[Baum(1959)]{1959IAUS...10...23B} Baum, W.~A.\ 1959, IAU Symp.~ 10:
The Hertzsprung-Russell Diagram, 10, 23
\bibitem[B{\" o}hringer et al.(2000)]{2000ApJS..129..435B} B{\" o}hringer,
H.~et al.\ 2000, \apjs, 129, 435
\bibitem[Bower, Lucey, \& Ellis(1992)]{1992MNRAS.254..589B} Bower, R.~G.,
Lucey, J.~R., \& Ellis, R.~S.\ 1992, \mnras, 254, 589
\bibitem[Carlstrom, Holder, \& Reese(2002)]{2002ARA&A..40..643C} Carlstrom,
J.~E., Holder, G.~P., \& Reese, E.~D.\ 2002, \araa, 40, 643
\bibitem[Chester \& Roberts(1964)]{1964AJ.....69..635C} Chester, C.~\&
Roberts, M.~S.\ 1964, \aj, 69, 635
\bibitem[Cole, Hatton, Weinberg, \& Frenk(1998)]{1998MNRAS.300..945C} Cole,
S., Hatton, S., Weinberg, D.~H., \& Frenk, C.~S.\ 1998, \mnras, 300, 945
\bibitem[Collins, Guzzo, Nichol, \& Lumsden(1995)]{1995MNRAS.274.1071C}
Collins, C.~A., Guzzo, L., Nichol, R.~C., \& Lumsden, S.~L.\ 1995, \mnras,
274, 1071
\bibitem[Coorway and Sheth 2002]{} Cooray, A. and Sheth R. 2002, Phys. Reports, 372, 1

\bibitem[Dalton, Efstathiou, Maddox, \&
Sutherland(1992)]{1992ApJ...390L...1D} Dalton, G.~B., Efstathiou, G.,
Maddox, S.~J., \& Sutherland, W.~J.\ 1992, \apjl, 390, L1
\bibitem[Diaferio, Kauffmann, Colberg, \& White(1999)]{1999MNRAS.307..537D} 
Diaferio, A., Kauffmann, G., Colberg, J.~M., \& White, S.~D.~M.\ 1999, 
\mnras, 307, 537 
\bibitem[Dressler \& Shectman(1988)]{1988AJ.....95..985D} Dressler, A.~\&
Shectman, S.~A.\ 1988, \aj, 95, 985
\bibitem[Ebeling et al.(1997)]{1997ApJ...479L.101E} Ebeling, H., Edge,
A.~C., Fabian, A.~C., Allen, S.~W., Crawford, C.~S., \& Boehringer, H.\
1997, \apjl, 479, L101
\bibitem[Edge \& Stewart(1991)]{1991MNRAS.252..414E} Edge, A.~C.~\&
Stewart, G.~C.\ 1991, \mnras, 252, 414
\bibitem[Eisenstein et al.(2001)]{2001AJ....122.2267E} Eisenstein, D.~J., 
et al.\ 2001, \aj, 122, 2267
\bibitem[Eke et al.(2004)]{2004MNRAS.tmp..483E} Eke, V.~R., et al.\ 2004, 
\mnras, 483 
\bibitem[Evrard et al.(2002)]{2002ApJ...573....7E} Evrard, A.~E.~et al.\
2002, \apj, 573, 7
\bibitem[Frenk, White, Efstathiou, \& Davis(1990)]{1990ApJ...351...10F}
Frenk, C.~S., White, S.~D.~M., Efstathiou, G., \& Davis, M.\ 1990, \apj,
351, 10
\bibitem[Frenk et al.(1999)]{1999ApJ...525..554F} Frenk, C.~S., et al.\ 
1999, \apj, 525, 554 
\bibitem[Fukugita et al.(1996)]{1996AJ....111.1748F} Fukugita, M.,
Ichikawa, T., Gunn, J.~E., Doi, M., Shimasaku, K., \& Schneider, D.~P.\
1996, \aj, 111, 1748
\bibitem[Gioia et al.(1990)]{1990ApJ...356L..35G} Gioia, I.~M., Henry,
J.~P., Maccacaro, T., Morris, S.~L., Stocke, J.~T., \& Wolter, A.\ 1990,
\apjl, 356, L35
\bibitem[Gladders(2002)]{2002PhDT.........3G} Gladders, M.\ 2002,
Ph.D.~Thesis
\bibitem[Gladders \& Yee(2000)]{2000AJ....120.2148G} Gladders, M.~D.~\&
Yee, H.~K.~C.\ 2000, \aj, 120, 2148
\bibitem[G{\' o}mez et al.(2003)]{2003ApJ...584..210G} G{\' o}mez, P.~L.~et
al.\ 2003, \apj, 584, 210
\bibitem[Goto et al.(2002)]{2002AJ....123.1807G} Goto, T.~et al.\ 2002,
\aj, 123, 1807
\bibitem[Gott III et al.(2003)]{2003astro.ph.10571G} Gott III, J.~R., 
Juri{\' c}, M., Schlegel, D., Hoyle, F., Vogeley, M., Tegmark, M., Bahcall, 
N., \& Brinkmann, J.\ 2003, ArXiv Astrophysics e-prints, astro-ph/0310571 
\bibitem[Gunn et al.(1998)]{1998AJ....116.3040G} Gunn, J.~E.~et al.\ 1998,
\aj, 116, 3040
\bibitem[Henry \& Arnaud(1991)]{1991ApJ...372..410H} Henry, J.~P.~\&
Arnaud, K.~A.\ 1991, \apj, 372, 410
\bibitem[Henry et al.(1995)]{1995ApJ...449..422H} Henry, J.~P.~et al.\
1995, \apj, 449, 422 
\bibitem[Hogg, Finkbeiner, Schlegel, \& Gunn(2001)]{2001AJ....122.2129H} 
Hogg, D.~W., Finkbeiner, D.~P., Schlegel, D.~J., \& Gunn, J.~E.\ 2001, \aj, 
122, 2129 
\bibitem[Hogg et al.(2003)]{2003ApJ...585L...5H} Hogg, D.~W., et al.\ 2003, 
\apjl, 585, L5 
\bibitem[Hopkins et al.(2002)]{2002AJ....123.1086H} Hopkins, A.~M., Miller, 
C.~J., Connolly, A.~J., Genovese, C., Nichol, R.~C., \& Wasserman, L.\ 
2002, \aj, 123, 1086 
\bibitem[Hopkins et al.(2003)]{2003ApJ...599..971H} Hopkins, A.~M., et al.\ 
2003, \apj, 599, 971 
\bibitem[Jenkins et al.(2001)]{2001MNRAS.321..372J} Jenkins, A., Frenk, 
C.~S., White, S.~D.~M., Colberg, J.~M., Cole, S., Evrard, A.~E., Couchman,
H.~M.~P., \& Yoshida, N.\ 2001, \mnras, 321, 372 
\bibitem[Kauffmann, Colberg, Diaferio, \& White(1999)]{1999MNRAS.303..188K}
Kauffmann, G., Colberg, J.~M., Diaferio, A., \& White, S.~D.~M.\ 1999,
\mnras, 303, 188 
\bibitem[Kepner et al.(1999)]{1999ApJ...517...78K} Kepner, J., Fan, X.,
Bahcall, N., Gunn, J., Lupton, R., \& Xu, G.\ 1999, \apj, 517, 78 
\bibitem[Kochanek et al.(2003)]{2003ApJ...585..161K} Kochanek, C.~S., 
White, M., Huchra, J., Macri, L., Jarrett, T.~H., Schneider, S.~E., \&
Mader, J.\ 2003, \apj, 585, 161 
\bibitem[Kim et al.(2002)]{2002AJ....123...20K} Kim, R.~S.~J.~et al.\ 2002,
\aj, 123, 20 
\bibitem[Lacey \& Cole(1994)]{1994MNRAS.271..676L} Lacey, C.~\& Cole, S.\ 
1994, \mnras, 271, 676 
\bibitem[Lasker(1970)]{1970AJ.....75...21L} Lasker, B.~M.\ 1970, \aj, 75,
21 
\bibitem[Lee et al.(2004)]{2004AJ....127.1811L} Lee, B.~C., et al.\ 2004, 
\aj, 127, 1811 
\bibitem[Lin, Mohr, \& Stanford(2003)]{2003ApJ...591..749L} Lin, Y., Mohr, 
J.~J., \& Stanford, S.~A.\ 2003, \apj, 591, 749 
\bibitem[Lucey(1983)]{1983MNRAS.204...33L} Lucey, J.~R.\ 1983, \mnras, 204,
33 
\bibitem[Lumsden, Nichol, Collins, \& Guzzo(1992)]{1992MNRAS.258....1L} 
Lumsden, S.~L., Nichol, R.~C., Collins, C.~A., \& Guzzo, L.\ 1992, \mnras,
258, 1 
\bibitem[Lupton et al.(2002)]{2002SPIE.4836..350L} Lupton, R.~H., Ivezic, 
Z., Gunn, J.~E., Knapp, G., Strauss, M.~A., \& Yasuda, N.\ 2002, \procspie, 
4836, 350 
\bibitem[Mandolesi, Villa, \& Valenziano(2002)]{2002AdSpR..30.2123M} 
Mandolesi, N., Villa, F., \& Valenziano, L.\ 2002, Advances in Space 
Research, 30, 2123 
\bibitem[McClure \& van den Bergh(1968)]{1968AJ.....73..313M} McClure,
R.~D.~\& van den Bergh, S.\ 1968, \aj, 73, 313 
\bibitem[Miller, Batuski, Slinglend, \& Hill(1999)]{1999ApJ...523..492M} 
Miller, C.~J., Batuski, D.~J., Slinglend, K.~A., \& Hill, J.~M.\ 1999, 
\apj, 523, 492 
\bibitem[Miller, Nichol, \& Batuski(2001)]{2001ApJ...555...68M} Miller, 
C.~J., Nichol, R.~C., \& Batuski, D.~J.\ 2001a, \apj, 555, 68 
\bibitem[Miller, Nichol, \& Batuski(2001)]{2001Sci...292.2302M} Miller, 
C.~J., Nichol, R.~C., \& Batuski, D.~J.\ 2001b, Science, 292, 2302 
\bibitem[Miller et al.(2001)]{2001AJ....122.3492M} Miller, C.~J.~et al.\
2001c, \aj, 122, 3492 
\bibitem[Miller, Krughoff, Batuski, \& Hill(2002)]{2002AJ....124.1918M} 
Miller, C.~J., Krughoff, K.~S., Batuski, D.~J., \& Hill, J.~M.\ 2002, \aj, 
124, 1918 
\bibitem[Miller et al.(2003)]{2003ApJ...597..142M} Miller, C.~J., Nichol, 
R.~C., G{\' o}mez, P.~L., Hopkins, A.~M., \& Bernardi, M.\ 2003, \apj, 597, 
142 
\bibitem[Oegerle \& Hill(2001)]{2001AJ....122.2858O} Oegerle, W.~R.~\&
Hill, J.~M.\ 2001, \aj, 122, 2858 
\bibitem[Ostrander, Nichol, Ratnatunga, \& 
Griffiths(1998)]{1998AJ....116.2644O} Ostrander, E.~J., Nichol, R.~C.,
Ratnatunga, K.~U., \& Griffiths, R.~E.\ 1998, \aj, 116, 2644
\bibitem[Oukbir \& Blanchard(1992)]{1992A&A...262L..21O} Oukbir, J.~\&
Blanchard, A.\ 1992, \aap, 262, L21 
\bibitem[Oppenheimer, Helfand, \& Gaidos(1997)]{1997AJ....113.2134O} 
Oppenheimer, B.~R., Helfand, D.~J., \& Gaidos, E.~J.\ 1997, \aj, 113, 2134 
\bibitem[Perlmutter \& Schmidt(2003)]{2003sgrb.conf..195P} Perlmutter, 
S.~\& Schmidt, B.~P.\ 2003, LNP Vol.~598: Supernovae and Gamma-Ray 
Bursters, 195 
\bibitem[Petrosian(1976)]{1976ApJ...209L...1P} Petrosian, V.\ 1976, \apjl, 
209, L1 
\bibitem[Pier et al.(2003)]{2003AJ....125.1559P} Pier, J.~R., Munn, J.~A.,
Hindsley, R.~B., Hennessy, G.~S., Kent, S.~M., Lupton, R.~H., \& Ivezi{\'
c}, {\v Z}.\ 2003, \aj, 125, 1559 
\bibitem[Popesso et al.(2004)]{2004A&A...423..449P} Popesso, P., B{\" 
o}hringer, H., Brinkmann, J., Voges, W., \& York, D.~G.\ 2004, \aap, 423, 
449 
\bibitem[Postman et al.(1996)]{1996AJ....111..615P} Postman, M., Lubin,
L.~M., Gunn, J.~E., Oke, J.~B., Hoessel, J.~G., Schneider, D.~P., \&
Christensen, J.~A.\ 1996, \aj, 111, 615 
\bibitem[Postman, Lauer, Oegerle, \& Donahue(2002)]{2002ApJ...579...93P} 
Postman, M., Lauer, T.~R., Oegerle, W., \& Donahue, M.\ 2002, \apj, 579, 93 
\bibitem[Reichart et al.(1999)]{1999ApJ...518..521R} Reichart, D.~E., 
Nichol, R.~C., Castander, F.~J., Burke, D.~J., Romer, A.~K., Holden, B.~P.,
Collins, C.~A., \& Ulmer, M.~P.\ 1999, \apj, 518, 521 
\bibitem[Romer et al.(2000)]{2000ApJS..126..209R} Romer, A.~K.~et al.\
2000, \apjs, 126, 209 
\bibitem[Romer \& et al.(2004)]{2004cgpc.sympE..47R} Romer, A.~K.~\& et 
al.\ 2004, Clusters of Galaxies: Probes of Cosmological Structure and 
Galaxy Evolution
\bibitem[Sand, Treu, \& Ellis(2002)]{2002ApJ...574L.129S} Sand, D.~J.,
Treu, T., \& Ellis, R.~S.\ 2002, \apjl, 574, L129
\bibitem[Schlegel, Finkbeiner, \& Davis(1998)]{1998ApJ...500..525S} 
Schlegel, D.~J., Finkbeiner, D.~P., \& Davis, M.\ 1998, \apj, 500, 525
\bibitem[Scranton et al.(2002)]{2002ApJ...579...48S} Scranton, R., et al.\ 
2002, \apj, 579, 48 
\bibitem[Smith et al.(2002)]{2002AJ....123.2121S} Smith, J.~A.~et al.\
2002, \aj, 123, 2121  
\bibitem[S{\" o}chting, Clowes, \& Campusano(2002)]{2002MNRAS.331..569S}
S{\" o}chting, I.~K., Clowes, R.~G., \& Campusano, L.~E.\ 2002, \mnras,
331, 569
\bibitem[Stoughton et al.(2002)]{2002AJ....123..485S} Stoughton, C.~et al.\
2002, \aj, 123, 485 
\bibitem[Strauss et al.(2002)]{2002AJ....124.1810S} Strauss, M.~A.~et al.\
2002, \aj, 124, 1810 
\bibitem[Sutherland(1988)]{1988MNRAS.234..159S} Sutherland, W.\ 1988,
\mnras, 234, 159 
\bibitem[Tully, Mould, \& Aaronson(1982)]{1982ApJ...257..527T} Tully,
R.~B., Mould, J.~R., \& Aaronson, M.\ 1982, \apj, 257, 527  
\bibitem[Viana \& Liddle(1996)]{1996MNRAS.281..323V} Viana, P.~T.~P.~\&
Liddle, A.~R.\ 1996, \mnras, 281, 323 
\bibitem[Vikhlinin et al.(1998)]{1998ApJ...502..558V} Vikhlinin, A., 
McNamara, B.~R., Forman, W., Jones, C., Quintana, H., \& Hornstrup, A.\ 
1998, \apj, 502, 558 
\bibitem[Visvanathan \& Sandage(1977)]{1977ApJ...216..214V} Visvanathan,
N.~\& Sandage, A.\ 1977, \apj, 216, 214 
\bibitem[Voges et al.(1999)]{1999A&A...349..389V} Voges, W., et al.\ 1999, 
\aap, 349, 389 
\bibitem[Wechsler(2004)]{2004cgpc.sympE..53D} Wechsler, R.~H.\ 2004, 
Clusters of Galaxies: Probes of Cosmological Structure and Galaxy 
Evolution
\bibitem[Wegner et al.(1996)]{1996ApJS..106....1W} Wegner, G., Colless, M., 
Baggley, G., Davies, R.~L., Bertschinger, E., Burstein, D., McMahan, R.~K., 
\& Saglia, R.~P.\ 1996, \apjs, 106, 1 
\bibitem[Wei, Xu, Dong, \& Hu(1999)]{1999A&AS..139..575W} Wei, J.~Y., Xu, 
D.~W., Dong, X.~Y., \& Hu, J.~Y.\ 1999, \aaps, 139, 575 
\bibitem[White et al.(1999)]{1999AJ....118.2014W} White, R.~A., Bliton, M., 
Bhavsar, S.~P., Bornmann, P., Burns, J.~O., Ledlow, M.~J., \& Loken, C.\ 
1999, \aj, 118, 2014 
\bibitem[White(2002)]{2002ApJS..143..241W} White, M.\ 2002, \apjs, 143, 241 
\bibitem[Wittman(2002)]{2002glat.conf...55W} Wittman, D.\ 2002, LNP
Vol.~608: Gravitational Lensing: An Astrophysical Tool, 55 
\bibitem[Yang et al.(2004)]{2004MNRAS.350.1153Y} Yang, X., Mo, H.~J., Jing, 
Y.~P., van den Bosch, F.~C., \& Chu, Y.\ 2004, \mnras, 350, 1153 
\bibitem[Yee and Ellingson (2003)]{} Yee, H. and Ellingson, E. 2003, \apj, 585, 215
\bibitem[York et al.(2000)]{2000AJ....120.1579Y} York, D.~G.~et al.\ 2000,
\aj, 120, 1579 
\bibitem[Zehavi et al.(2002)]{2002ApJ...571..172Z} Zehavi, I., et al.\ 
2002, \apj, 571, 172 
\bibitem[Zwicky(1952)]{1952PASP...64..247Z} Zwicky, F.\ 1952, \pasp, 64,
247 
\bibitem[Zwicky, Herzog, \& Wild(1961)]{1961CGCG1.C...0000Z} Zwicky, F., 
Herzog, E., \& Wild, P.\ 1961, CGCG1, 0 
\bibitem[Zwicky \& Herzog(1963)]{1963CGCG2.C...0000Z} Zwicky, F.~\& Herzog, 
E.\ 1963, CGCG2, 0 
\bibitem[Zwicky, Karpowicz, \& Kowal(1965)]{1965CGCG5.C...0000Z} Zwicky, 
F., Karpowicz, M., \& Kowal, C.~T.\ 1965, CGCG5, 0 
\bibitem[Zwicky \& Herzog(1966)]{1966CGCG3.C...0000Z} Zwicky, F.~\& Herzog, 
E.\ 1966, CGCG3, 0 
\bibitem[Zwicky \& Herzog(1968)]{1968CGCG4.C...0000Z} Zwicky, F.~\& Herzog, 
E.\ 1968a, CGCG4,0
\bibitem[Zwicky \& Kowal(1968)]{1968CGCG6.C...0000Z} Zwicky, F.~\& Kowal, 
C.~T.\ 1968b, CGCG6, 0 


\end{thebibliography}
\end{document}